\documentclass[prd,twocolumn,showpacs,showkeys,preprintnumbers,floatfix,
  nofootinbib,superscriptaddress]{revtex4-1}
\usepackage{amsfonts} 
\usepackage{amssymb} 
\usepackage{amsmath} 
\usepackage{subfigure} 
\usepackage{graphicx} 
\usepackage{array} 
\usepackage{dcolumn} 
\usepackage{bm} 
\usepackage{latexsym} 
\usepackage{longtable} 
\usepackage{hyperref} 
\newcommand{\TB}{\textrm{B}}
\newcommand{\TI}{\textrm{I}}
\newcommand{\TP}{\textrm{P}}
\newcommand{\TV}{\textrm{V}}
\newcommand{\TA}{\textrm{A}}
\newcommand{\TT}{\textrm{T}}

\newcommand{\CO}{{\cal O}}

\newcommand{\CV}{{\cal V}}

\newcommand{\CM}{{\cal M}}

\newcommand{\GeV}{\mathop{\rm GeV}\nolimits}

\begin{document}


\title{ $B_K$ using HYP-smeared staggered fermions in $N_f=2+1$
  unquenched QCD }
\author{Taegil Bae}
\affiliation{
  Lattice Gauge Theory Research Center, FPRD, and CTP, \\
  Department of Physics and Astronomy,
  Seoul National University, Seoul, 151-747, South Korea
}
\author{Yong-Chull Jang}
\affiliation{
  Lattice Gauge Theory Research Center, FPRD, and CTP, \\
  Department of Physics and Astronomy,
  Seoul National University, Seoul, 151-747, South Korea
}
\author{Chulwoo Jung}
\email[E-mail at ]{chulwoo@bnl.gov}
%
%
\affiliation{
Physics Department, Brookhaven National Laboratory,
Upton, NY11973, USA
}
\author{Hyung-Jin Kim}
\affiliation{
  Lattice Gauge Theory Research Center, FPRD, and CTP, \\
  Department of Physics and Astronomy,
  Seoul National University, Seoul, 151-747, South Korea
}
\author{Jangho Kim}
\affiliation{
  Lattice Gauge Theory Research Center, FPRD, and CTP, \\
  Department of Physics and Astronomy,
  Seoul National University, Seoul, 151-747, South Korea
}
\author{Jongjeong Kim}
\altaffiliation[Current address: ]{
  Physics Department,
  University of Arizona,
  Tucson, AZ 85721, USA
}
\affiliation{
  Lattice Gauge Theory Research Center, FPRD, and CTP, \\
  Department of Physics and Astronomy,
  Seoul National University, Seoul, 151-747, South Korea
}
\author{Kwangwoo Kim}
\affiliation{
  Lattice Gauge Theory Research Center, FPRD, and CTP, \\
  Department of Physics and Astronomy,
  Seoul National University, Seoul, 151-747, South Korea
}
\author{Weonjong Lee}
\email[E-mail at ]{wlee@snu.ac.kr}
\homepage[Home page at ]{http://lgt.snu.ac.kr/}
\altaffiliation[Visiting professor at ]{
  Physics Department,
  University of Washington,
  Seattle, WA 98195-1560, USA
}
\affiliation{
  Lattice Gauge Theory Research Center, FPRD, and CTP, \\
  Department of Physics and Astronomy,
  Seoul National University, Seoul, 151-747, South Korea
}
\author{Stephen R. Sharpe}
\email[E-mail at ]{sharpe@phys.washington.edu}
\homepage[Home page at ]{http://www.phys.washington.edu/users/sharpe/}
\affiliation{
  Physics Department,
  University of Washington,
  Seattle, WA 98195-1560, USA
}
\author{Boram Yoon}
\affiliation{
  Lattice Gauge Theory Research Center, FPRD, and CTP, \\
  Department of Physics and Astronomy,
  Seoul National University, Seoul, 151-747, South Korea
}
\collaboration{SWME Collaboration}
\date{\today}
\begin{abstract}
  We present results for kaon mixing parameter $B_K$ 
  calculated using HYP-smeared improved
  staggered fermions on the MILC asqtad lattices.  We use three
  lattice spacings ($a\approx 0.12$, $0.09$ and $0.06\;$fm),
  ten different valence quark masses ($m\approx m_s/10-m_s$),
  and several light sea-quark masses in order to control the
  continuum and chiral extrapolations. We derive the next-to-leading order
  staggered chiral perturbation theory (SChPT) results necessary to fit
  our data, and use these results to do extrapolations based both on
  SU(2) and SU(3) SChPT. The SU(2) fitting is particularly straightforward
  because parameters related to taste-breaking and matching errors appear
  only at next-to-next-to-leading order. We match to the continuum 
  renormalization scheme (NDR) using one-loop perturbation theory.
  Our final result is from the SU(2) analysis, with the SU(3) result
  providing a (less accurate) cross check. We find 
  $B_K(\text{NDR}, \mu = 2 \text{ GeV}) = 0.529 \pm 0.009 \pm 0.032$ 
  and $\widehat{B}_K =B_K(\text{RGI})= 0.724 \pm 0.012 \pm 0.043$,
  where the first error is statistical and the second systematic. 
  The error is dominated by the truncation error in the matching factor.
  Our results are consistent with those obtained using
  valence domain-wall fermions on lattices generated with
  asqtad or domain-wall sea quarks.
\end{abstract}
\maketitle

\section{Introduction \label{sec:intr}}

CP violation was first observed, long ago, in neutral kaon mixing.
The long-lived eigenstate $K_L$ has a small CP-even impurity, allowing
it to decay into two pions. The amount of this impurity is
parametrized by $\varepsilon$, whose experimental
value is~\cite{pdg-2009}.
\begin{equation} 
\varepsilon = (2.228 \pm 0.011)\times 10^{-3} \,.
\end{equation}
In the standard model, 
CP violation is induced by box diagrams involving
virtual W and Z bosons, which lead to the prediction
\begin{eqnarray}
\varepsilon &=& \exp(i\phi_\varepsilon) \
      \sqrt{2} \sin(\phi_\varepsilon) \ C_\varepsilon \
                  {\rm Im}\lambda_t \ X \ \widehat{B}_K  + \xi
\end{eqnarray}
where
\begin{subequations}
\begin{eqnarray}
      X &=& {\rm Re} \lambda_c [ \eta_1 S_0(x_c) - \eta_3 S_3(x_c,x_t) ]
      - {\rm Re} \lambda_t \eta_2 S_0(x_t) \nonumber \\ \\
      \lambda_i &=& V_{is}^* V_{id} \\
      x_i &=& m_i^2 / M_W^2 \\
      C_\varepsilon &=& \frac{G_F^2 F_K^2 m_K M_W^2}
      {6 \sqrt{2} \pi^2 \Delta M_K}  \\
      \xi &=& \exp(i\phi_\varepsilon) \sin(\phi_\varepsilon)
      \frac{{\rm Im}A_0}{{\rm Re}A_0}
\end{eqnarray}
\end{subequations}
For the values of the parameters, see, e.g., Ref.~\cite{buras-2008}.
The key point is that, aside from known kinematic factors and
QCD Wilson coefficients, $\varepsilon$ is given
in terms of elements of the CKM matrix (occurring in the
$\lambda_i$) and the hadronic matrix element
$\widehat B_K$. This matrix element has the form
\begin{eqnarray}
  \widehat{B}_K &=& C(\mu) B_K(\mu)
  \\
  B_K(\mu) &=& \frac{\sum_{\nu}
     \langle \bar{K}_0 | [\bar{s} \gamma_\nu (1\!-\!\gamma_5) d]
    [\bar{s} \gamma_\nu (1\!-\!\gamma_5) d] | K_0 \rangle }{
    \frac{8}{3} \langle \bar{K}_0 | \bar{s} \gamma_0\gamma_5 d | 0 \rangle
    \langle 0 | \bar{s} \gamma_0\gamma_5 d | K_0 \rangle }
  \label{eq:bk-def}
\end{eqnarray}
where $C(\mu)$ is the Wilson coefficient which makes 
$\widehat{B}_K$ RG (renormalization group) invariant
(as discussed further in Sec.~\ref{ssec:matching}),
$\mu$ is the renormalization scale of the operator,
and we have assumed that the kaons are at rest.
Thus if one can determine the value of $\widehat B_K$, one can
constrain the CP-violating part of the CKM matrix.

It has been a long-standing goal of lattice QCD to calculate
$B_K$. The challenging part of the required ratio of matrix elements, 
Eq.~(\ref{eq:bk-def}), is the numerator. This is, at first sight,
relatively straightforward to evaluate, because  it involves only
a single, stable particle in both initial and final states.
A complication
arises, however, from the left-handed structure of the four-fermion
operator. This constrains its behavior in and around the chiral limit.
In particular, at leading order in chiral perturbation theory,
the matrix element in the numerator is proportional to $m_K^2$.
This vanishing is not guaranteed if
one uses a formulation of lattice fermions in which chiral symmetry
is broken. Because of this, most recent lattice calculations of
$B_K$ have used fermions in which one has at least some remnant of
chiral symmetry---domain-wall, staggered, overlap and twisted-mass fermions.

After a long history of quenched calculations, advances in algorithms,
methodology and computers have allowed calculations of $B_K$ using light
dynamical quarks. Particularly noteworthy are two 
``$2+1$ flavor'' calculations in which all three of the physical light
quarks are dynamical, although the up  and down quarks are kept 
degenerate.\footnote{%
We also compare to results from calculations with two degenerate
dynamical quarks in the concluding section.}
These calculations both use valence domain-wall fermions, with one
using asqtad staggered sea quarks~\cite{ALV-09}, and the other
domain-wall sea quarks~\cite{rbc-uk-08,Kelly:2009fp}.
Both calculations have produced results in
which all sources of systematic error are controlled,
and the total error is less than 5\%.
In particular, both calculations use two lattice spacings to 
allow a continuum extrapolation.
The results of these two calculations are consistent.

It is important for a parameter such as $B_K$, which is used to
constrain the standard model, to have multiple calculations using
different lattice methods. In particular, we think it is important
to use different valence fermions. Thus we have undertaken a
calculation using valence staggered fermions. In particular, we 
use HYP-smeared improved staggered fermions, since these are known
to reduce taste-breaking, and are computationally inexpensive to implement.
We use the same asqtad lattices as in Ref.~\cite{ALV-09}, except that
we add a third, smaller lattice spacing ($a\approx 0.12$, $0.09$
and $0.06\;$fm).
An important feature of our approach is that the four-fermion
operators are constructed using HYP-smeared (rather than thin) links.
This is known to reduce the size of perturbative corrections~\cite{wlee-02}
and scaling violations~\cite{wlee-06-1}.

There has been one previous calculation
of $B_K$ using valence staggered fermions on
the asqtad, $N_f = 2+1$ lattices~\cite{gamiz-06}.
This work differs from ours in that they used
thin links to construct the operators, and only worked on the
$a\approx 0.12\;$fm lattices.
As a consequence, the result contains potentially large
discretization and matching errors.

A drawback of using staggered fermions is that each flavor
comes with 4 tastes, and one needs to
use to the rooting prescription to reduce the number
of sea quarks to the physical complement. 
Rooting leads to unphysical effects at non-zero lattice spacing.
We assume, following the arguments of
Refs.~\cite{Bernard:2006zw,Sharpe:2006re,Bernard:2007ma,Shamir:2006nj}, 
that the effects of rooting vanish in the continuum limit,
and are taken into account at non-zero lattice spacing
by our use of extrapolations based on staggered chiral
perturbation theory (SChPT). 
The phenomenological successes of rooted staggered 
fermions~\cite{milc-rmp-09}
provide support to this assumption.

This paper is organized as follows.
In Section \ref{sec:sxpt}, we derive the necessary next-to-leading order (NLO)
SChPT results,
first for SU(3) ChPT including the effects of our use of
a mixed action, and then for SU(2) ChPT.
In Section \ref{sec:data-anal}, we explain our methodology,
describe how we obtain results for $B_K$,
and then present our fits using both SU(3) and SU(2) SChPT.
In the SU(3) case we must use constrained, Bayesian fitting,
given the complexity of the functional form.
We also discuss the light sea-quark mass dependence and
the continuum extrapolation.
We close the section by presenting final results and error
budgets for both SU(3) and SU(2) analyses.
We conclude in Section \ref{sec:conclude} with a
comparison of our result with previous work and some
comments on our future plans.
In four appendices we present, respectively, the functional forms
needed for SU(3) fitting, the functional forms needed for SU(2)
fitting, a derivation of several key results needed to obtain
the SU(2) functional forms, and tables of a subset of our
numerical results.

Preliminary results from this project have been presented 
in Refs.~\cite{wlee-09-1,wlee-09-2,wlee-09-3,wlee-09-4}.

\section{Staggered chiral perturbation theory \label{sec:sxpt}}
To extrapolate our data to the physical (average) light quark mass,
and to the continuum limit, we use staggered chiral perturbation 
theory (SChPT).
We have carried out fits using the functions
predicted by both SU(3) (strange quark treated
as light) and SU(2) (strange quark treated as ``heavy'') SChPT.
The purpose of this section, and the
associated appendices, is to derive these functions and present them
in an explicit form suitable for fitting.

For SU(3), a next-to-leading order (NLO) SChPT analysis
of $B_K$ 
has been given in Ref.~\cite{steve-06}. That work, however,
considered the theory in which the sea and valence quarks had
the same action. In our set-up, by contrast, we have asqtad sea quarks
(since we use configurations generated by the MILC 
collaboration~\cite{milc-04})
and HYP-smeared valence quarks (as described in the following section).
Thus we must generalize the results of Ref.~\cite{steve-06} to
a mixed-action theory. We also present the results
in an explicit notation rather than using the compact but dense
notation of Ref.~\cite{steve-06}.

The use of SU(2) ChPT has been advocated in Ref.~\cite{rbc-uk-08-3},
and the continuum partially quenched NLO predictions 
for a variety of quantities, including $B_K$, have been presented.
These results are directly applicable to domain-wall fermions,
and have been used for extrapolating $B_K$ in Ref.~\cite{rbc-uk-08-1}. 
Here we generalize the results to staggered ChPT, 
finding that the complications of the SU(3) NLO SChPT form
are greatly reduced.

Before presenting the derivations we describe our notation for
the quantities that appear in the SChPT expressions,
and explain how we obtain the numerical values for these
quantities.
We work throughout in the isospin-symmetric limit, and label
the up and down sea quarks collectively as $l$.
Thus $m_l$ is the common light sea-quark mass, while $m_s$
is the mass of the strange sea quark.
Following Ref.~\cite{steve-06},
valence down (anti-)quarks are labeled $x$ while 
valence strange (anti-)quarks are labeled $y$.
These have masses $m_x$ and $m_y$, respectively. 
As noted above, the sea quarks use the asqtad action, while
the valence quarks use the HYP-smeared staggered action
(more precisely the HYP (II) choice in the convention
of Refs.~\cite{wlee-03,wlee-02}).

Many pseudo-Goldstone bosons (PGB) appear in the NLO chiral expressions.
The mass of the ``kaon'' composed of $x$-quarks and $y$-antiquarks
(or vice-versa) is denoted $m_{xy:\TB}$, where $\TB$ is the taste.
The taste can be I, P, V, A or T, standing, respectively for
taste scalar [${\bf 1}$], pseudo-scalar [$\xi_5$], vector
[$\xi_\mu$], axial [$\xi_{\mu 5}$] and tensor [$\xi_{\mu\nu}$].
Here we use the result that at leading order (LO)
in SChPT there is an accidental SO(4) taste symmetry~\cite{wlee-99}.
Similarly, the masses of ``pions'' composed of $x$-quarks and $x$-antiquarks
are labeled $m_{xx:\TB}$, 
while those of PGBs composed of $y$-quarks and antiquarks are
denoted $m_{yy:\TB}$.
In the following, PGB masses are always in physical units, 
converted from the measured lattice masses using the scales determined
by the MILC collaboration~\cite{milc-rmp-09,milc-private}.
These scales are given in Table~\ref{tab:r1/a} below.

To simplify the chiral expressions, we also use the following
shorthand notations:
\begin{itemize}
  \item The squared masses of the valence kaons are denoted by
    $K_\text{B} \equiv m^2_{xy:\text{B}}$. 
    We also use a special shorthand $G \equiv K_\TP$ 
    for the Goldstone-taste kaon. This is the particle which
    we use for the external states in our calculation.

  \item The squared masses of the valence pions are
    denoted: $X_\text{B} = m^2_{xx:\text{B}}$.
    Similarly, we use   $Y_\text{B} = m^2_{yy:\text{B}}$.
    Note that for tastes B = I, V and A 
    the pion masses have LO contributions from quark-disconnected diagrams
    involving ``hairpin'' vertices. These are, by definition, 
    {\em not included} in $X_B$ and $Y_B$. In other words, these mass-squareds
    are those of the pions in which the quark and antiquark are taken 
    (implicitly) to have different flavors.

  \item We also need the squared masses of 
    analogous flavor non-singlet pions composed
    of $\bar ll$ and $\bar ss$ quarks. We call these $L_\text{B} \equiv
    m^2_{ll:\text{B}}$ and $S_\text{B} \equiv m^2_{ss:\text{B}}$,
    respectively, for B = I, V and A.

  \item Finally, we need the squared masses of the $\eta_\text{B}$ and
    $\eta'_\text{B}$ pions for B = V, A and I. We call these masses simply
    $\eta_\text{B}$ and $\eta'_\text{B}$.
    The $\eta_\text{B}$ and $\eta'_\text{B}$ 
    are the flavor singlet, sea-quark mesons that
    result from the inclusion of mixing between $\bar l l$ and $\bar ss$
    pions due to hairpin diagrams~\cite{bernard-03}.

\end{itemize}

We now address how we obtain the numerical values of the various
masses discussed above. 
The valence kaon and non-singlet pion masses 
$K_\text{B}$, $X_\text{B}$, and $Y_\text{B}$ are obtained
from our simulations in an ancillary calculation.
In fact, we find to an extremely good approximation that,
for our range of valence quark masses,
the expected LO forms hold:\footnote{%
SChPT predicts that $\Delta_\TP = 0$, although in
practice the missing NLO terms together with finite-volume
effects can lead to a small non-zero value for $\Delta_\TP$.}
\begin{subequations}
\label{eq:pion-mass-sq-1}
\begin{eqnarray}
K_\text{B} &=& b_1(m_x + m_y) + \Delta_\TB \\
X_\text{B} &=& b_1(2 m_x) + \Delta_\TB \\
Y_\text{B} &=& b_1(2 m_y) + \Delta_\TB\,.
\end{eqnarray}
\end{subequations}
In practice we do a fit to these equations and
extract $b_1$ and the $\Delta_\TB$, and then use the
latter to reconstruct the mass-squareds. 
More details on these fits are given below in Sec.~\ref{ssec:bk-comp}.

\begin{table}[htb]
\caption{Masses of flavor non-singlet,
taste P sea-quark masses, in lattice units~\cite{milc-04,milc-private}.
Ensemble labels are from
Table~\protect\ref{tab:milc-lat} below.
  \label{tab:LPandSP}}
\begin{ruledtabular}
\begin{tabular}{l  l  l  l }
Ensemble & $a \sqrt{L_P}$ & $a \sqrt{S_P}$ \\
\hline
C1 & 0.3779 & 0.4875 \\
C2 & 0.3113 & 0.4889 \\
C3 & 0.2245 & 0.4944 \\
C4 & 0.1889 & 0.4923 \\
C5 & 0.1597 & 0.4913 \\
\hline
F1 & 0.1479 & 0.3273 \\
F2 & 0.1046 & 0.3273 \\
\hline
S1 & 0.09371 & 0.2075 \\
\end{tabular}
\end{ruledtabular}
\end{table}

\begin{table}[htb]
\caption{Taste splittings, $r_1^2 \Delta_B$, for flavor
non-singlet pions composed of sea quarks~\cite{milc-private}.
These are for the ensembles C3, F1 and S1 
(see Table~\protect\ref{tab:milc-lat} below).
  \label{tab:Delta_B:sea}}
\begin{ruledtabular}
\begin{tabular}{l  l  l  l }
taste & coarse & fine & superfine \\
\hline
$r_1^2 \Delta_A$ & 0.2053 & 0.0706 & 0.0253 \\
$r_1^2 \Delta_T$ & 0.3269 & 0.1154 & 0.0413 \\
$r_1^2 \Delta_V$ & 0.4391 & 0.1524 & 0.0552 \\
$r_1^2 \Delta_S$ & 0.5370 & 0.2062 & 0.0676 \\
\end{tabular}
\end{ruledtabular}
\end{table}

In the case of pions composed of sea quarks, we obtain the
flavor non-singlet masses
$L_\text{B}$ and $S_\text{B}$ from MILC collaboration 
results~\cite{milc-04,milc-private}.
The sea pion and ``ss'' meson masses are quoted in Table~\ref{tab:LPandSP},
while taste splittings are collected in Table~\ref{tab:Delta_B:sea}.
We do not include the errors on these splittings in our analysis
as their impact is much smaller than other sources of error.
This is because the errors are very small, and furthermore
enter into a numerically small part of the one-loop correction.
We use the same taste-splittings for all the coarse ensembles,
and similarly for both the fine ensembles,
which we expect to be a very good approximation based on our results
for valence pions.

For the flavor singlets sea-quark mesons, the taste-singlet is simplest,
since the $\eta'_\TI$ (which corresponds to the $\eta'$ of QCD)
is not a PGB and can be integrated out.
This leads to the following result for the $\eta_\TI$ mass:~\cite{bernard-03}:
\begin{equation}
\eta_\TI = 
\frac{1}{3}L_\TI + \frac{2}{3} S_\TI\,.
\end{equation}
For tastes B=V and A, we use the result of Ref.~\cite{bernard-03}:
\begin{eqnarray}
\eta_\text{B} &=& m^2_{\eta:\text{B}} = 
\frac{1}{2} \bigg( L_\text{B} + S_\text{B} + 
\frac{3}{4} a^2 {\delta'}_\text{B}^{ss} - Z_\text{B} \bigg)\,,
\label{eq:etaB}
\\
\eta'_\text{B} &=& m^2_{\eta:\text{B}} = 
\frac{1}{2} \bigg( L_\text{B} + S_\text{B} + 
\frac{3}{4} a^2 {\delta'}_\text{B}^{ss} + Z_\text{B} \bigg)\,,
\label{eq:etapB}
\\
Z_\text{B} &=& \sqrt{H_\text{B}^2 
- \frac{1}{2} a^2 {\delta'}_\text{B}^{ss} H_\text{B}
+ \frac{9}{16} (a^2 {\delta'}_\text{B}^{ss})^2 }\,,
\label{eq:ZB}
\\
H_\text{B} &=& S_\text{B} - L_\text{B}\,.
\label{eq:HB}
\end{eqnarray}
The superscript on the hairpin vertices $a^2 {\delta'}_{\TA,\TV}^{ss}$ 
will be explained in the following section.
Their numerical values are taken from the fits of Ref.~\cite{milc-04}. 
Note that their dimensions are given by
${\delta'}_\TB^{ss}\propto \Lambda_{\rm QCD}^4$.

The taste violating parameters on the coarse lattices
are~\cite{milc-private}
\begin{eqnarray}
& & r_1^2 a^2 \delta'_A = -0.30(1)(3) 
\\
& & r_1^2 a^2 \delta'_V = -0.05(2)(3)\,.
\end{eqnarray}
These come from fits using the ``smoothed'' $r_1$
values that we adopt, and that are given in Table~\ref{tab:r1/a} below.
As for the sea-quark taste splittings, we fix these parameters to
their central values and ignore the errors.
For the other lattice spacings we scale the parameters assuming an
$a^2\alpha_s^2$ dependence.

\subsection{Mixed action effects in SU(3) SChPT\label{ssec:mixed}}

The inclusion of mixed-action effects into SChPT can be accomplished
by a straightforward generalization of the methods worked out
in Refs.~\cite{Bar:2002nr,Bar:2005tu,Golterman:2005xa,Chen:2009su}.
A mixed action can be described field-theoretically at the lattice level
in the same way as a partially quenched (PQ) theory: there are valence
quarks, corresponding ghosts, and sea quarks. The difference is that
there is no longer a symmetry interchanging valence and sea quarks.
This symmetry only emerges in the continuum limit, 
where the choice of discretization does not matter.
What this means is that one has a standard PQ set up when $a\to 0$, but, for
$a\ne 0$, the corrections introduced by discretization errors 
in the valence, sea and mixed valence-sea sectors
[parametrized by low-energy coefficients (LECs)] are unrelated.
At first sight (and in light of the complexity of the analysis
of $B_K$ in an unmixed context~\cite{steve-06}),
the required generalization appears rather daunting.
It turns out, however, that this is not the case for our particular
application, since nearly all of the new LECs
do not enter at NLO in $B_K$.

\begin{figure}[tbp]
\begin{tabular}{ccccc}
	\includegraphics[width=0.9in]{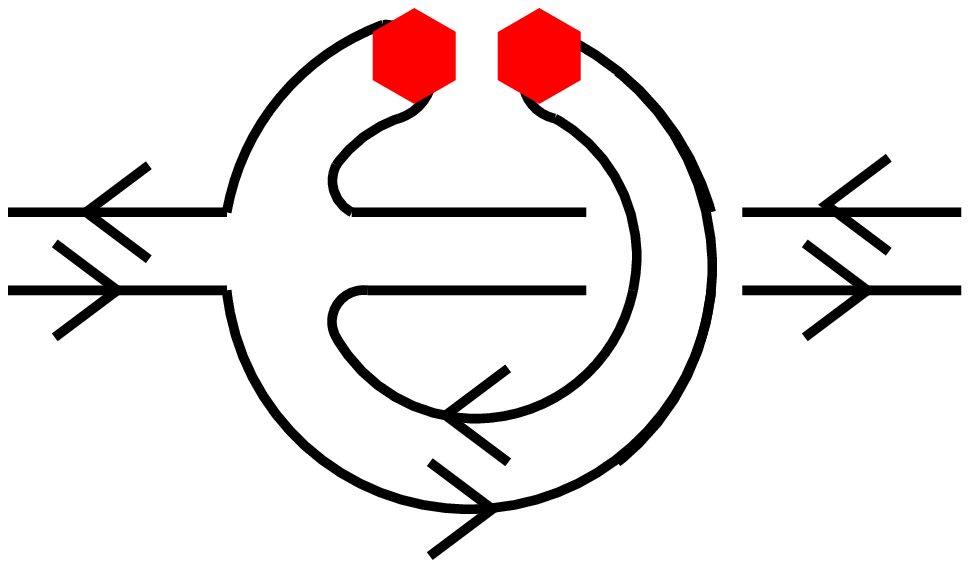} &
	$\;\;\;\;\;\;$ & 
	\includegraphics[width=0.9in]{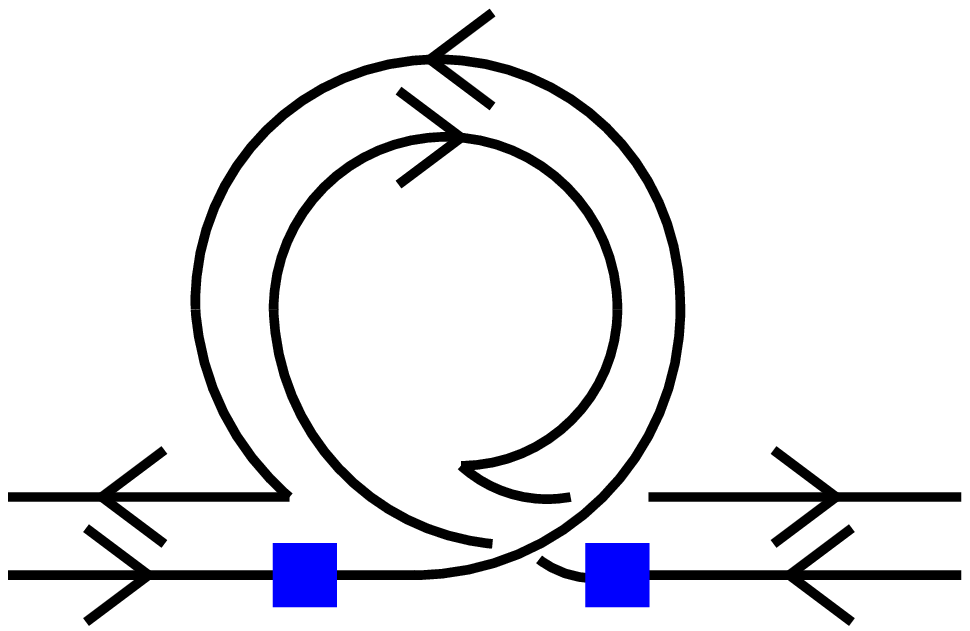} &
        $\;\;\;\;\;\;\;\;$ & 
	\includegraphics[width=0.9in]{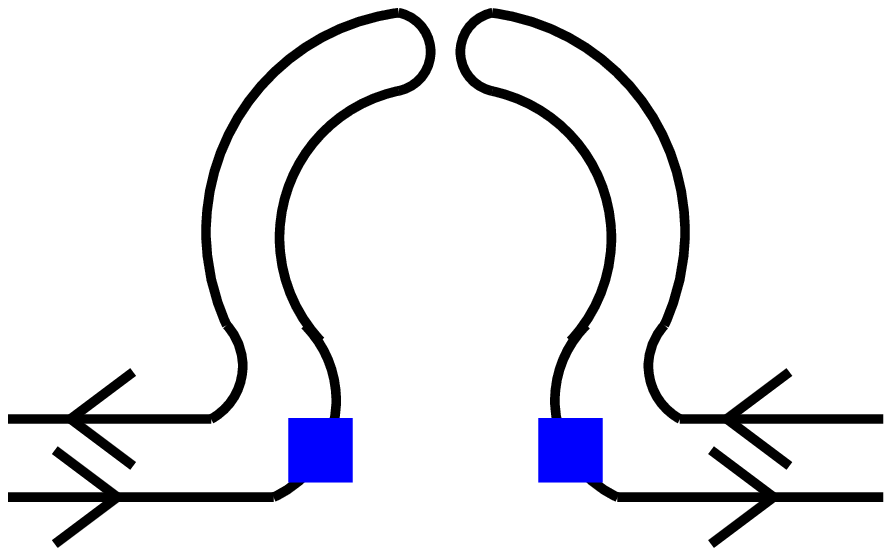} 
\\ 
        (a) &	$\;\;\;\;\;\;$ & (b) & $\;\;\;\;\;\;$ & (c) \\
	\end{tabular}
	\caption{Classes of quark-line diagrams 
contributing to $B_K$ at 1-loop order. 
The four-quark operator is represented as two bilinears which
change a valence $d$-quark into a valence $s$-quark. These bilinears
are shown as either (blue) squares (if the bilinear connects
to both quark and antiquark from an external kaon) or as (red)
hexagons (if the bilinear connects a quark from one kaon to an
antiquark from the other). For each of the three diagrams there
are others (not shown) related by interchanging the roles of
quark and antiquark. In (c) the ``hairpin vertex''
contains, implicitly, a sum of ``bubbles'' composed of sea quarks.
  \label{fig:quarkflow}}
\end{figure}

The easiest way to see this is to look at the quark-flow diagrams
which contribute. These have been presented in Ref.~\cite{steve-06},
and we display them in Fig.~\ref{fig:quarkflow}.
Only valence-quark lines occur in Figs.~\ref{fig:quarkflow}(a) and (b), 
so that the PGBs entering in loop diagrams are all of
the ``valence-valence'' type. The {\em form} 
of the contributions from these diagrams
will thus be unchanged by the use of a mixed action, and we can simply
use the results of Ref.~\cite{steve-06}.
The contributions of these diagrams give what is called the ``connected''
part of the matrix element, ${\cal M}_{\rm conn}$, where connected means
that the loops do not involve hairpin vertices.
We present a fully explicit form for ${\cal M}_{\rm conn}$
in the next section.
We stress that the {\em values}
of the LECs in ${\cal M}_{\rm conn}$ are altered by the change from
unmixed to mixed action, but this has no practical consequences
as these are {\em a priori} unknown parameters that are either
determined from the spectrum or by the fits to $B_K$ itself.

Another feature of the quark-line diagrams is that no ``valence-sea''
mesons appear at NLO. Thus the new LECs associated with such mixed mesons
do not enter into our result.

New LECs do appear in the ``disconnected'' matrix element,
${\cal M}_{\rm disc}$, for which the quark-line diagram is
shown in Fig.~\ref{fig:quarkflow}(c). 
The hairpin vertices in the loop implicitly
contain a sum over any number of insertions of sea-quark loops,
corresponding to the propagation of ``sea-sea'' PGBs.
This means that three types of hairpin vertex enter, instead of
the single type for an unmixed action:
$a^2 {\delta'}_\TB^{vv}$, for the hairpin connecting
valence-valence pions to themselves;
$a^2 {\delta'}_\TB^{vs}$, for the hairpin connecting
valence-valence pions to sea-sea pions, and
$a^2 {\delta'}_\TB^{ss}$, for the hairpin connecting sea-sea pions
to themselves. The last of these has already appeared in
Eqs.~(\ref{eq:etaB}-\ref{eq:ZB}).
These hairpins occur as LO vertices only for tastes B$=$V and A,
so that, for $B_K$, the change to a mixed action introduces only
four new LECs:
${\delta'}_\TV^{vv}$, ${\delta'}_\TA^{vv}$,
${\delta'}_\TV^{vs}$ and ${\delta'}_\TA^{vs}$.
We stress that these hairpin vertices are independent
of the quark masses.

Hairpins also occur at LO in the taste singlet channel, but here the
dominant contribution is that which leads to the bulk of the physical
$\eta'$ mass.
This is a continuum contribution,
which  is therefore independent of the fermion action, 
and thus is common to $vv$, $vs$ and $ss$ vertices.
 Small differences between these vertices
of $O(a^2)$ do not contribute to $B_K$ until 
next-to-next-to-leading order (NNLO).
Thus the three types of vertices are the same at LO in the
taste-singlet channel, and the result of Ref.~\cite{steve-06} for
this channel is unchanged.

In light of the preceding discussion, we see
that the ingredients that change when moving from a PQ
(unmixed action) theory to a mixed-action (MA) set-up are the
hairpin propagators, $D_{xx}^{\TB,\rm PQ}(q)$,
$D_{xy}^{\TB,\rm PQ}(q)$, and $D_{yy}^{\TB,\rm PQ}(q)$, where
B=V or A labels the taste.
We first recall the form of these propagators in
a theory with an unmixed action~\cite{steve-06}.
The ``$xy$'' hairpin propagator is
\begin{eqnarray}
D_{xy}^{\TB,\rm PQ}(q) &=& - a^2 {\delta'}_\TB
\frac{1}{(q^2+X_\TB)(q^2+Y_\TB)} R_\TB(q) \,,
\label{eq:DxyPQ}
\\
R_\TB(q) &=&
\frac{(q^2+L_\TB)(q^2+S_\TB)}{(q^2+\eta_\TB)(q^2+\eta'_\TB)}
\,,
\end{eqnarray}
[this is Eq.~(42) of Ref.~\cite{steve-06} restricted to
the case of $2+1$ flavors].
The expressions for
$\eta_\TB$ and $\eta'_\TB$ are those given in (\ref{eq:etaB}) and
(\ref{eq:etapB}) above except that ${\delta'}_\TB^{ss}$ is
replaced by the single, common, hairpin vertex ${\delta'}_\TB$. 
The results for $D_{xx}^{\TB,\rm PQ}$ are simply
obtained by the replacement $Y_\TB\to X_\TB$,
while the opposite replacement yields the $D_{yy}^{\TB,\rm PQ}$.
Similar replacements hold for the mixed-action theory, and
so we discuss only $D_{xy}$ below.

The result (\ref{eq:DxyPQ}) can be understood as follows:
$a^2 {\delta'}_\TB$ is the hairpin vertex, $R_\TB(q)$ is the
result of the sum over sea-sea mesons ``within'' the hairpin vertex,
and the remaining two propagators are those of the ``external''
$xx$ and $yy$ pions.
One can then see how the result changes when the hairpin vertex splits
into three types. If there are no sea-quark loops, one
has the $vv$ hairpin vertex and the external propagators, but no 
factor of $R_\TB(q)$.
If there are sea-quark loops, then one has a common factor
of $({\delta'}_\TB^{vs})^2$ (from $vv\to ss$ and $ss\to vv$),
and the sum $R_\TB(q)$ except for the first term, with
one factor of ${\delta'}_\TB^{ss}$ removed to avoid double-counting.
Thus one finds that 
$D_{xy}^{\TB,\rm PQ}(q)$ should be replaced with
the mixed-action hairpin propagator
\begin{widetext}
\begin{eqnarray}
D_{xy}^{\TB,\rm MA}(q) &=& -\frac{1}{(q^2+X_V)(q^2+Y_V)}
\left[ a^2 {\delta'}_\TB^{vv} 
+
\left(a^2{\delta'}^{vs}_\TB\right)^2
\frac{(R_\TB(q)-1)}{a^2{\delta'}^{ss}_\TB}
\right]
\\
&=&
\left(\frac{(\delta^{'vs}_\TB)^2}{\delta^{'ss}_\TB\delta^{'}_\TB}\right)
D_{xy}^{\TB,\rm PQ}(q)
- a^2
\left(\frac{\delta^{'vv}_V\delta^{'ss}_\TB-(\delta^{'vs}_\TB)^2}
                 {\delta^{'ss}_\TB}\right)
\frac{1}{(q^2+X_\TB)(q^2+Y_\TB)}\,.
\label{eq:D_MA}
\end{eqnarray}
\end{widetext}
From the second line, we see that the impact of using a mixed
action is twofold.
\begin{enumerate}
\item
One must replace the overall factor multiplying the loop contribution
calculated in the unmixed case as follows:
\begin{equation}
{\delta'}_\TB \longrightarrow \delta^{\rm MA1}_\TB =
\frac{(\delta^{'vs}_\TB)^2}{\delta^{'ss}_\TB}\,,\quad
{\TB=\TV,\TA}\,.
\end{equation}
The form of the contribution will be otherwise unchanged,
since the $q$ dependence in the first term of (\ref{eq:D_MA}) is
the same as that of $D^{\TB, \rm PQ}_{xy}(q)$.
We also note that our lack of knowledge of the two $\delta_\TB^{\rm MA1}$
does not introduce new fit parameters since these hairpin vertices
multiply unknown operator coefficients from the
chiral theory [as shown explicitly in Eq.~(\ref{eq:app:b13_14})].
\item
In addition, one must evaluate a new loop integral,
in which propagator has the same form as for a quenched hairpin, 
and in which the overall constant is
\begin{equation}
\delta^{\rm MA2}_\TB =
\frac{{\delta'}^{vv}_\TB{\delta'}^{ss}_\TB-
({\delta'}^{vs}_\TB)^2}
                 {{\delta'}^{ss}_\TB}\,,\quad {\TB=\TV,\TA}\,.
\end{equation}
The result is given as $F^{(2)}_\TB$ in Appendix~\ref{app:def-F12}.
This does introduce two new fit parameters.
\end{enumerate}

We close this section with a comment on
the expected magnitudes of the four new LECs.
The hairpin vertices of SChPT are the result of
mapping four-quark operators in the Symanzik effective
theory into the chiral theory. The structure of the
underlying four-quark operators mirrors that of the
chiral operators they produce. For example, 
the hairpin vertex proportional to ${\delta'}^{vs}_\TB$
arises from an operator containing a valence-quark bilinear
and a sea-quark bilinear, while that proportional to
${\delta'}^{vv}_\TB$ arises from an operator
containing two valence-quark bilinears.
The important observation is these four-quark operators also give
non-hairpin taste-dependent contributions to the pion masses
(viewed as having the gluons exchanged in the t-channel rather
than the s-channel). The operators leading to ${\delta'}^{vv}_\TB$,
${\delta'}^{vs}_\TB$ and ${\delta'}^{ss}_\TB$
lead, respectively, to taste splittings for valence-valence,
valence-sea and sea-sea pions.
But we know that the valence HYP-smeared
action reduces taste splittings by a factor of $\approx 3$ 
compared to that for the sea-quark asqtad action~\cite{wlee-08-2}.
Assuming that this is a universal reduction, applicable to
all four-quark operators, we conclude that 
${\delta'}^{vv}_\TB/{\delta'}^{ss}_\TB \approx 1/3$.
Since the improvement in taste-splittings arises from a reduction
of the coupling of quarks to high-momentum gluons, we expect that
it ``factorizes'', so that
${\delta'}^{vs}_\TB/{\delta'}^{ss}_\TB \approx 1/\sqrt{3}$.
Given these assumptions, we are led to the expectations
\begin{equation}
\delta^{\rm MA1}_\TB \approx {\delta'}^{ss}_\TB/3
\qquad{\rm and}\qquad
\delta^{\rm MA2}_\TB \approx 0
\,.
\end{equation}
If these expectations are close to correct, then the taste V and A
hairpin contributions are significantly reduced by the use
of the mixed action: the ``old'' contributions 
from $D^{\TB, \rm PQ}_{xy}(q)$
are reduced by a factor of $\sim 3$, while the ``new''
$\delta^{\rm MA2}_\TB$ terms are suppressed.
We use these expectations to simplify our
fitting strategy.

\subsection{SU(3) SChPT result \label{ssec:su3}}
In this subsection we summarize the NLO result for $B_K$
in SU(3) SChPT---obtained from Ref.~\cite{steve-06} with
the generalization explained in the previous subsection. 
The utility of this result depends on the
extent to which one can treat the strange quark mass as
light compared to $\Lambda_{\rm QCD}$. We will attempt to judge this
{\em a posteriori} based on how
successful we are in fitting with the SU(3) form.

Discretization errors in $B_K$ arise from the action
and from the four-quark operator. If they involve taste-breaking
then they are, for both HYP-smeared and asqtad actions, of $O(a^2\alpha^2)$,
while non taste-breaking errors are of $O(a^2)$ for HYP-smeared
quarks and of $O(a^2\alpha)$ for asqtad quarks.
The different factors of $\alpha$ are, however, misleading, since numerical
results indicate that taste-breaking effects are enhanced.
Furthermore, as noted above,
different actions can lead to significantly different numerical
sizes of taste-breaking effects, although formally these effects
are of the same order.
Thus, following Ref.~\cite{steve-06}, we adopt the 
phenomenologically-based power-counting scheme in
which all discretization errors are treated as being of $O(p^2)$,
irrespective of the number of powers of $\alpha$.
In light of this choice, and for the sake of brevity,
we refer below to all discretization errors as being
simply of ``$O(a^2)$'', except where the associated
powers of $\alpha$ are important.

Additional errors arise if one uses
perturbatively calculated matching factors to relate the lattice
operator to its continuum counterpart.
We use one-loop matching, and furthermore
keep only the subset of the one-loop induced
operators involving bilinears with taste P.
This results in ``mixing errors'' proportional to both
$\alpha_s$---from one-loop mixing with operators with tastes other 
than P---and $\alpha_s^2$---from the unknown two-loop mixing with
all operators, including those with taste P.
As explained in Ref.~\cite{steve-06}, it is reasonable phenomenologically
to treat both these effects as also of $O(p^2)$.
The justification for treating $O(\alpha)\sim O(\alpha^2)$ is that the
one-loop coefficients of the operators we drop are known to be small,
while two-loop coefficients are unknown and might not be small.
Indeed, with HYP-smeared fermions, we find that the relevant
one-loop coefficients are of typical size 
$\alpha_s/(4\pi) \approx 0.08 \alpha_s$
\cite{KimLeeSharpe,wlee-03},
and thus expect that their effects are noticeably smaller than those
of $O(\alpha_s^2)$.
In light of this, we refer to truncation errors below as simply
of $O(\alpha^2)$ unless we need to be more specific.

Using this power-counting scheme, Ref.~\cite{steve-06}
determined all the operators in SChPT that must be included at NLO,
and carried out the required one-loop calculation.
This involves using LO propagators in the loops, which implies
that the accidental SO(4) taste symmetry~\cite{wlee-99,bernard-03} 
holds for their masses.
As already  noted, we determine these masses from an ancillary spectrum
calculation on the same configurations, or from published
MILC results, so that they are not unknowns in the fits.

The resulting expression for $B_K$ can be written as
\begin{equation}
  B_K^{\text{SU(3)}} = \sum_{j=1}^{16} b_j H_j\,.
\label{eq:bk-su3-ma}
\end{equation}
Here the coefficients $b_j$ are related to
unknown SChPT LECs, while the $H_j$ are known functions of 
the (numerically determined) PGB masses.
In the following we run through these 16 terms in turn,
explaining their origin and their properties.
The explicit expressions for the $H_j$ are collected
in Appendix \ref{app:defs}.

The dominant contribution to $B_K$ comes from the function
\begin{equation}
H_1  = 1 + \frac{1}{8 \pi^2 f_\pi^2 G} 
\left[ \CM_{\textrm{conn}}^0 + \CM_{\textrm{disc}}^0 \right]\,.
\label{eq:H1def}
\end{equation}
This contains the LO contribution (the ``$1$'') together
with the ``continuum-like'' loop contributions to
$\CM_{\textrm{conn}}$ and $\CM_{\textrm{disc}}$.
By ``continuum-like'' we mean that the logarithms have the continuum
form (thus the superscript $0$, indicating $a\to 0$),
except that the pions in the loops
have taste-splittings included in their masses.
If these taste-splittings are set to zero, then
$\CM_{\textrm{conn}}^0$ and $\CM_{\textrm{disc}}^0$.
go over to the continuum chiral logarithms.
Assuming that $f_\pi$ is known, the coefficient of
the chiral logarithms is predicted.\footnote{%
As discussed below,
our data is insufficient to determine $f_\pi$ from the fits. 
Thus we use a fixed value, which we vary among reasonable
choices, as discussed below.}
Since we know the values of the required PGB masses, 
we know the value of the function $H_1$ for each choice
of valence and sea-quark masses.
For this reason it makes sense
to lump these terms together into a single function, $H_1$.
The explicit forms for 
$\CM_{\textrm{conn}}^0$ and $\CM_{\textrm{disc}}^0$
are given in eqs.~(\ref{eq:app:Mconndef}) and (\ref{eq:app:Mdiscdef}), 
respectively.
It turns out that $\CM_{\textrm{disc}}^0$ vanishes for
degenerate valence quarks.

The chiral logarithms in (\ref{eq:H1def}) 
have the generic form $m_K^2 \ln m_K$.
Since an explicit $1/G$ factor is taken out,
the functions $\CM_{\textrm{conn}}^0$ 
and $\CM_{\textrm{disc}}^0$ have generic form $m_K^4\ln m_K$.

The coefficient of $H_1$ is given by
\begin{equation}
b_1 = B_0 + O(a^2, \alpha^2) \,.
\label{eq:b1res}
\end{equation}
Here $B_0$ is the value of $B_K$ in the combined 
$SU(3)$ chiral and continuum
limits. $B_0$ is a quantity that is of interest for continuum ChPT.
The remaining terms indicate the manner in which $b_1$
approaches the continuum limit: there are contributions from
both non-taste breaking and taste-breaking
discretization errors, and from the truncation of 
perturbation theory.\footnote{%
In this case there are no terms linear in $\alpha$~\cite{steve-06}.}
When fitting at a single lattice spacing, one can treat $b_1$ as
a single constant. The scaling violations must be accounted for
when extrapolating to the continuum limit. Alternatively
one can attempt a combined continuum-chiral fit, as has been
done by the MILC collaboration~\cite{milc-04}.
We have not done so here.

The next four terms are the analytic NLO
(and, in the case of $H_3$, NNLO) corrections:
\begin{align}
H_2 & = G / \Lambda_\chi^2
\label{eq:H2def}
\\
H_3 & = ( G / \Lambda_\chi^2 )^2
\label{eq:H3def}
\\
H_4 & = \frac{(X_\TP - Y_\TP)^2}{ G \Lambda_\chi^2 }
\label{eq:H4def}
\\
H_5 & = \frac{ L_\TP + S_\TP/2 }{ \Lambda_\chi^2 }
\label{eq:H5def}
\end{align}
All of these terms are present in the continuum limit,
and to the order we work their coefficients are 
independent of $a$ and the quark masses.
We have chosen to write them in terms of the PGB masses
rather than the quark masses. This allows us to set the
scale of these terms by the chiral scale $\Lambda_\chi$
which we take to be $1\;$GeV in our fitting.
Then, assuming naive dimensional analysis, we expect the
coefficients of these terms $b_2-b_4$ to be of $O(1)$.

We find that the NNLO analytic term, $b_3 H_3$, is necessary to
obtain reasonable fits. The need for a NNLO term
is not surprising given relatively large kaon masses
included in our SU(3) fits.
It is, perhaps, surprising that we can get away with 
only a single term beyond NLO,
given that the MILC collaboration requires
NNNLO and NNNNLO terms in their fits to pion and kaon
properties~\cite{milc-04}. The difference appears to be that
the errors in $B_K$ are larger than those in the quantities
studied by the MILC collaboration---PGB masses and
decay constants.

We stress that we are not including the complete NNLO expression
(which is not known). In particular, we are not including non-analytic
NNLO terms, and are keeping only one of the NNLO analytic terms.
Our approach here is phenomenological---we are keeping the
minimum number of NNLO analytic terms needed to obtain
reasonable fits.

The remaining 11 $H_j$ are pure lattice artifacts,
caused either by discretization errors or the truncation
of matching factors, or both. They all arise from taste-violating
interactions.\footnote{%
There are also taste-conserving discretization and truncation errors;
these lead to the corrections in the expression for the
coefficient $b_1$, Eq.~(\protect\ref{eq:b1res}).}
The first 7 are corrections to $\CM_{\textrm{conn}}$,
and are labeled by
\begin{align}
H_6 & = F^{(4)}_{\TI}
\label{eq:H6def}
\\
H_7 & = F^{(4)}_{\TP}
\label{eq:H7def}
\\
H_8 & = F^{(4)}_{\TV}
\label{eq:H8def}
\\
H_{9} & = F^{(4)}_{\TA}
\label{eq:H9def}
\\
H_{10} & = F^{(4)}_{\TT}
\label{eq:H10def}
\\
H_{11} & = F^{(5)}
\label{eq:H11def}
\\
H_{12} & = F^{(6)}\,.
\label{eq:H12def}
\end{align}
The explicit functional forms of $F^{(4)}_{\TB}$, $F^{(5)}$,
and $F^{(6)}$ are given in eqs.~(\ref{eq:app:F4def}),
(\ref{eq:app:F5def}) and (\ref{eq:app:F6def}), respectively.
The subscript on $F^{(4)}_{\TB}$ indicates the taste
of the pion in the loop. 
The functions $F^{(5)}$ and $F^{(6)}$ vanish in the limit
of degenerate valence quarks.

The chiral logarithms in the functions
$H_{6-12}$ are enhanced over the continuum-like logarithms
in $H_1$. While the latter are of standard form $m_K^2 \ln m_K$,
the former are simply proportional to $\ln m_K$.
This enhancement of the chiral logarithm is balanced by the fact
that the coefficient, $b_j$, is small:
\begin{equation}
b_{6-12} \sim \Lambda_{\rm QCD}^2 \times
\left(a^2 \Lambda_{\rm QCD}^2 \ {\rm or}\ \alpha^2 \right)\,.
\label{eq:b6-12est}
\end{equation}

The final four functions $H_j$ with $j=13-16$ arise from
corrections to $\CM_{\textrm{disc}}$ due to discretization 
and truncation errors. These are the functions which are 
affected by our use of a mixed action, 
as explained above in Sec.~\ref{ssec:mixed}.
$H_{13}$ and $H_{14}$ are the loop integrals
involving the hairpin propagator $D_{xy}^{\TB,\rm PQ}(q)$
[i.e. the first term in Eq.~(\ref{eq:D_MA})].
They are, for B$=$V and A, respectively,
\begin{align}
H_{13} & = F^{(1)}_\TV\,,
\\
H_{14} & = F^{(1)}_\TA\,,
\end{align}
and are multiplied by coefficients proportional
to $\delta_\TV^{\rm MA1}$ and $\delta_\TA^{\rm MA1}$.
This leaves the two functions multiplied by the two
additional LECs introduced by our use of a mixed action,
i.e. $\delta_\TV^{\rm MA2}$ and $\delta_\TA^{\rm MA2}$.
These are, respectively,
\begin{align}
H_{15} & = F^{(2)}_\TV\,,
\\
H_{16} & = F^{(2)}_\TA\,.
\end{align}

The functions $F^{(1)}_\TB$ and $F^{(2)}_\TB$  are given in 
Appendix~(\ref{app:def-F12}).
They are yet more divergent in the chiral limit than $F^{(4-6)}$, behaving 
generically as $\ln m_K/m_K^2$ (although they vanish when $m_x=m_y$).
The power-counting is restored by coefficients
[whose explicit expressions in terms of chiral coefficients
are given in eqs.~(\ref{eq:app:b13_14}) and (\ref{eq:app:b15_16})]
which are proportional to the product of two factors arising from
discretization or truncation errors:
\begin{eqnarray}
b_{13-16} &\sim& \Lambda_{\rm QCD}^2 \times
\left(a^2\delta_\TB^{\rm MA1}\ {\rm  or}\ a^2\delta_\TB^{\rm MA2}\right)
\nonumber\\
&&
\ \ \times\left(a^2 \Lambda_{\rm QCD}^2 \ {\rm or}\ \alpha^2 \right)\,.
\label{eq:b13-16est}
\end{eqnarray}
(We recall that $\delta_\TB^{\rm MA1,2}\propto \Lambda_{\rm QCD}^4$.)
As discussed at the end of Sec.~\ref{ssec:mixed}, we expect
$\delta_\TB^{\rm MA1}$ to be suppressed by a factor of 3 from its natural
size, and for $\delta_\TB^{\rm MA2}$ to be close to zero. Thus we expect
the contributions from $H_{13}$ and $H_{14}$ to be small, while
those from $H_{15}$ and $H_{16}$ to be negligible.

\subsection{SU(2) SChPT result\label{ssec:su2}}

In the recent literature there has been much
discussion of the convergence and reliability of SU(3) ChPT when
extrapolating lattice results to the physical kaon mass.
As examples, we contrast the apparently successful extrapolations 
of Refs.~\cite{milc-04,milc-rmp-09,ALV-09} (the latter being for 
$B_K$ using a domain-wall valence/staggered sea mixed action),
with the lack of convergence found in Ref.~\cite{rbc-uk-08,jlqcd-chipt-08}.
Recent reviews of the situation are given in Refs.~\cite{Lellouchlat08,Boyle09},
and a related discussion in continuum ChPT is in Ref.~\cite{Donoghue09}.
Our view is that whether SU(3) ChPT can be used reliably
depends on the quantity considered (the rate of convergence is not
universal) and on whether one includes NNLO and higher order terms (these
are needed to represent data in the region of the physical kaon mass).
Thus we have attempted SU(3) fits using the theoretical
form described in the previous subsection (which includes one NNLO term),
and find the reasonably successful results to be described below.

It is nevertheless true that, for the non-degenerate lattice kaons
which lie closest to the physical kaon, and which thus have the dominant
effect on the extrapolation, the non-analytic dependence comes
dominantly from chiral logarithms involving {\em only} the light quarks
($x$ and $\ell$ in our notation).
Chiral logarithms of mesons containing strange quarks
(i.e. of $K_\TB$, $S_\TB$, $Y_\TB$, $\eta_\TI$ and $\eta'_\TB$\footnote{%
We do not include $\eta_V$ or $\eta_A$ in this list because
the strange-quark component is small and these mesons are light.
This can be seen from Eq.~(\protect\ref{eq:etaB}), which in the limit that
$a^2\delta'_\TB \ll |S_\TB-L_\TB|$, which is the case in practice,
gives $\eta_\TB\approx L_\TB+ a^2\delta'_\TB$.
}%
) 
can be expanded about their values when the strange quark mass
equals its physical value and represented accurately by analytic 
terms. This means that, in effect, one is using an (approximate)
SU(2) ChPT expression for those kaons closest to the physical kaon
even when nominally doing an SU(3) extrapolation.

The RBC collaboration were led, by
considerations along these lines, together with the poor convergence
they found for SU(3) ChPT, to propose
the use of SU(2) ChPT for the extrapolation of kaon (and pion) 
properties~\cite{rbc-uk-08}. In this approach, first considered
systematically in Ref.~\cite{Roessl:1999iu}
and extended in Refs.~\cite{rbc-uk-08,rbc-uk-08-1}, the
kaon is treated as a heavy, static source, and
no expansion is made in the mass of the (valence or sea) strange quark.
The expansion in powers of $(m_\pi/\Lambda_\chi)^2$ is
supplemented by an expansion in $(m_\pi/m_K)^2$.
In our calculation, this means that the SU(2) ChPT result
will apply only to the subset of our non-degenerate masses in which
$m_x$ is small and $m_y$ is large.

This methodology has been applied successfully 
to a calculation of $B_K$ using domain-wall fermions
by the RBC collaboration~\cite{rbc-uk-08-1}.
Since they use lattice fermions with an almost exact chiral symmetry,
the corresponding ChPT result is that of the continuum, except that
the LECs can depend on $a^2$. For our calculation, however, what is
needed is an SU(2) {\em staggered} ChPT result.
At first sight, this would appear to require a generalization of
the rather involved enumeration of operators performed in 
Ref.~\cite{steve-06} to the case of SU(2) chiral symmetry
with a heavy kaon source.
It turns out, however, that a simpler approach suffices.
One can show that it is sufficient to consider
the $m_x, m_\ell \ll m_y \approx m_s$ limit of
the {\em next-to-leading order} SU(3) SChPT result, 
provided that one allows the LECs to depend on $m_y$ and $m_s$ 
in an unknown (but analytic) way.
In addition, one can show that the size of the chiral
logarithms involving $m_x$ and $m_\ell$ relative to the LO
term is {\em not corrected by an unknown $m_s$ dependence}.
Thus the chiral logarithm remains a predicted correction,
as in SU(3) SChPT.

The validity of this ``recipe'' is demonstrated in Appendix~\ref{app:su2}.
The argument holds to all orders in an expansion in powers of 
$m_s/\Lambda_{\rm QCD}$.

Applying this recipe we find an important simplification:
the NLO terms in SU(3) SChPT that involve taste-breaking
LECs proportional to $a^2$ or
$\alpha_s^2$ get pushed to NNLO in SU(2) SChPT.
This drastically reduces the number of parameters required 
in a fit to the NLO expression.
How this happens is explained in detail
in Appendix~\ref{app:su2result}.

The final SU(2) SChPT result for $B_K$ is
\begin{equation}
B_K  = \sum_{i=1}^{4} d_i Q_4
\,,
\label{eq:su2-fit-func}
\end{equation}
where the functions that appear are
\begin{eqnarray}
Q_1 &=& 1 + \frac1{32\pi^2 f^2} \Bigg[
(L_\TI-X_\TI)\tilde\ell(X_\TI) 
\nonumber\\
&& \qquad\qquad +\ell(X_\TI)  - 2 \sum_\TB \tau^\TB \ell(X_\TB)\Bigg]\,,
\label{eq:q_1}
\\
Q_2 &=& \frac{X_P}{\Lambda_\chi^2}\,,
\label{eq:q_2}
\\
Q_3 &=& \Big( \frac{X_P}{\Lambda_\chi^2} \Big)^2\,,
\\
Q_4 &=& \frac{L_P}{\Lambda_\chi^2}\,.
\label{eq:q_4}
\end{eqnarray}
The chiral-logarithmic functions $\ell$ and $\tilde\ell$
are defined in Eqs.~(\ref{eq:app:l}) and (\ref{eq:app:tilde-l}),
and the factors $\tau^\TB$, which are the fractional multiplicities
of the different tastes, are given in Eq.~(\ref{eq:app:tauB}).
Note that we have kept a single analytic NNLO term, $Q_3$,
as for the SU(3) SChPT fits.
As in the SU(3) fits we fix $\Lambda_\chi=1\,$GeV,
although we could also use $\Lambda_\chi=m_K^{\rm phys}$.

The constants $d_i$ are arbitrary, unknown,
analytic functions of $m_s$ and $m_y$. In addition, at NLO
$d_1$ contains taste-conserving discretization and truncation errors
(which it inherits from $b_1$). Specifically, one has
\begin{equation}
d_1 = B_0^{SU(2)}(m_y,m_s) + O(a^2,\alpha^2) \,.
\label{eq:d_1}
\end{equation}
where the first term is the value of $B_K$ in the SU(2)
chiral limit ($m_x=m_\ell=0$) and with the given values of $m_y$
and $m_s$.
The $O(a^2,\alpha^2)$ terms also have an implicit dependence
on $m_y$ and $m_s$.

Clearly the SU(2) SChPT result is much simpler than that
in SU(3) SChPT---at NLO, it has only
3 unknown coefficients at a fixed lattice spacing,
to be compared to 16 for the SU(3) form. 
This gain is compensated in part by
the fact that it can be used only for a small subset of our data, and
because the fit parameters have an implicit dependence on $m_y$ and
$m_s$. Nevertheless, we find, as described below, 
that the SU(2) fitting is more
straightforward and leads to smaller final errors in $B_K$. 

\section{Data analysis \label{sec:data-anal}}
In this section we explain how we calculate $B_K$, how we convert
the numerical data into physical observables using matching factors, and
how we extrapolate to the physical kaon mass and
the continuum limit using S$\chi$PT.
%
\subsection{Computation of $B_K$ \label{ssec:bk-comp}}
%
\begin{table}[htb]
\caption{MILC lattices used for the numerical study.
  Here, ``ens'' represents the number of gauge configurations,
``meas'' is the number of measurements per configuration,
  and ID will be used later to identify the corresponding 
  MILC lattice. 
  \label{tab:milc-lat}}
\begin{ruledtabular}
\begin{tabular}{l  l  l  c  l }
$a$ (fm) & $am_l/am_s$ & \ \ size & ens $\times$ meas  & ID \\
\hline
0.12 & 0.03/0.05  & $20^3 \times 64$ & $564 \times 1$  & C1 \\
0.12 & 0.02/0.05  & $20^3 \times 64$ & $486 \times 1$  & C2 \\
0.12 & 0.01/0.05  & $20^3 \times 64$ & $671 \times 9$  & C3 \\
0.12 & 0.01/0.05  & $28^3 \times 64$ & $274 \times 8$  & C3-2 \\
0.12 & 0.007/0.05 & $20^3 \times 64$ & $651 \times 10$ & C4 \\
0.12 & 0.005/0.05 & $24^3 \times 64$ & $509 \times 1$  & C5 \\
\hline
0.09 & 0.0062/0.031 & $28^3 \times 96$  & $995 \times 1$ & F1 \\
0.09 & 0.0031/0.031 & $40^3 \times 96$  & $678 \times 1$ & F2 \\
\hline
0.06 & 0.0036/0.018 & $48^3 \times 144$ & $744 \times 2$ & S1
\end{tabular}
\end{ruledtabular}
\end{table}
We use the MILC ensembles listed in Table~\ref{tab:milc-lat}.
They are generated with $N_f=2+1$ flavors of improved (asqtad action)
staggered sea quarks, using the rooting trick to cancel the effects
of the extra tastes.

The values of the light sea-quark mass ($a m_l$) 
and strange sea-quark mass ($am_s$)
are given in Table~\ref{tab:milc-lat}.
Details of the lattice generation, decorrelation times, etc.
are given in Ref.~\cite{milc-04}.
 We use three lattice spacings---$a\approx 0.12$, $0.09$ and $0.06$
fm--- which are called coarse, fine and superfine, respectively.
These lattice spacings are the nominal values---in our analysis we
actually use the values obtained on each ensemble
from the value of $r_1/a$ obtained by the 
MILC collaboration~\cite{milc-rmp-09,milc-private}.
These are listed in Table~\ref{tab:r1/a}.
To obtain $a$ we set 
$r_1 =0.3108(15)({}^{+26}_{-79})\;$fm, 
as given in Ref.~\cite{milc-rmp-09} using $f_\pi$ to set the scale.

\begin{table}[htbp]
\caption{Values for $r_1/a$ and $1/a$. See text for discussion.
  \label{tab:r1/a}}
\begin{ruledtabular}
\begin{tabular}{ l l l }
ID   & $r_1/a$ &  $1/a$ (GeV) \\ 
\hline
C1   &  2.650(4) &  1.682(3) \\
C2   &  2.644(3) &  1.679(2) \\
C3   &  2.618(3) &  1.662(2) \\
C3-2 &  2.618(3) &  1.662(2) \\
C4   &  2.635(3) &  1.673(2) \\
C5   &  2.647(3) &  1.681(2) \\
\hline
F1   &  3.699(3) &  2.348(2) \\
F2   &  3.695(4) &  2.346(3) \\
\hline
S1   &  5.296(7) &  3.362(4) \\
\end{tabular}
\end{ruledtabular}
\end{table}
%

\begin{table}[htbp]
\caption{Valence quark masses (in lattice units).
  \label{tab:val-qmass}}
\begin{ruledtabular}
\begin{tabular}{ l  l  l}
$a$ (fm) & $a m_x$ and $a m_y$ & \\ 
\hline
0.12 &  $0.005 \times n$  & with $n=1,2,3,\ldots,10$ \\
0.09 &  $0.003 \times n$  & with $n=1,2,3,\ldots,10$ \\
0.06 &  $0.0018 \times n$ & with $n=1,2,3,\ldots,10$ \\
\end{tabular}
\end{ruledtabular}
\end{table}
The valence quarks are HYP-smeared staggered fermions, i.e. the
configurations are HYP-smeared~\cite{Hasenfratz:2001hp}
(using the coefficients of the HYP(II) choice 
in the convention of Refs.~\cite{wlee-03,wlee-02})
and then we use the unimproved staggered action. 
We have found this to be as effective at reducing taste-breaking
as the more elaborate HISQ action~\cite{wlee-08-2}. 
We use 10 different values of the valence quark masses (\textit{i.e.}
$m_x$ and $m_y$) running from $\sim m_s/10$ to $\sim m_s$.
They are given in Table \ref{tab:val-qmass}.
In total, we have 55 mass combinations for our valence PGBs:
10 degenerate ``pions'' ($m_x = m_y$) 
and 45 non-degenerate ``kaons'' ($m_x \ne m_y$).

To give a more precise indication of the size of our valence-quark
masses, we quote in Table~\ref{tab:physical-qmass} the values
of the valence ``physical'' down and strange quark masses.
These are obtained on each ensemble as follows.
The strange mass, $m_y$, is tuned so that $Y_P=(0.6858\ {\rm GeV})^2$, 
which is the continuum $s-\bar s$ mass keeping only
quark-connected contractions, as determined by Ref.~\cite{Lepagemss}.
The down mass, $m_x$, is then tuned so that
$G = m_{K_0,\rm phys}^2$.
The tunings require extrapolation, and we use the 
linear extrapolation in quark masses discussed below. 
For the coarse ensemble, we quote only results for the C3 ensemble,
since those on the other coarse ensembles are very similar.
Comparing to Table~\ref{tab:val-qmass},
we see that our heaviest quark mass is somewhat lighter than the physical
strange quark mass on all ensembles.

\begin{table}[htbp]
\caption{Physical down and strange valence-quark masses (in lattice units).
  \label{tab:physical-qmass}}
\begin{ruledtabular}
\begin{tabular}{ l  l  l}
Ensemble & $a m_d$ & $a m_s$  \\ 
\hline
C3 & 0.00212(2) & 0.05174(5) \\
\hline
F1   &0.00142(2) & 0.03523(5) \\
\hline
S1   & 0.00102(1) & 0.02362(3) \\
\end{tabular}
\end{ruledtabular}
\end{table}
%

To calculate the four-quark matrix elements needed for $B_K$
we follow the methodology developed in 
Refs.~\cite{Kilcup:1989fq,Kilcup:1997ye,wlee-05}.
We place wall-sources composed of U(1)-noise at timeslices 
$t_1$ and $t_2>t_1$, each
of which are used to produce both a valence quark and antiquark.
Boundary conditions are periodic in all directions.
The sources consist of random phases for
each color index on each site on the timeslice,
and have the property of selecting only the kaon having the
desired taste P (i.e. with quantum numbers $\gamma_5 \otimes \xi_5$)
and having zero spatial momentum. 
One source produces a valence kaon, the other an antikaon,
and these are contracted together in a gauge-invariant 
four-quark operator at an intermediate time $t$.
These operators involve quark and antiquark fields spread
out over $2^4$ hypercubes, and thus live on the two timeslices $t$
and $t+1$. 
They are made gauge invariant using products of HYP-smeared links
along the shortest paths joining quarks and antiquarks, averaged
over all such paths. The operators are then summed over spatial
directions so as to improve the signal. 

For completeness, we give expressions for the operators
that we use.
The four-fermion operator is, at tree-level,
\begin{eqnarray}
{\cal O}_K &=& \sum_{j=1,4}  O_j^{\rm latt}
\\
O_1^{\rm latt}(t) &=& \sum_{\vec y} \sum_\mu
[V_\mu\times P][V_\mu\times P]_I(\vec y,t)
\label{eq:O1latt}\\
O_2^{\rm latt}(t) &=& \sum_{\vec y} \sum_\mu
[V_\mu\times P][V_\mu\times P]_{II}(\vec y,t)
\label{eq:O2latt}\\
O_3^{\rm latt}(t) &=& \sum_{\vec y} \sum_\mu
[A_\mu\times P][A_\mu\times P]_I(\vec y,t)
\label{eq:O3latt}\\
O_4^{\rm latt}(t) &=& \sum_{\vec y} \sum_\mu
[A_\mu\times P][A_\mu\times P]_{II}(\vec y,t)\,,
\label{eq:O4latt}\\
\end{eqnarray}
where $\vec y$ is a vector labeling spatial hypercubes
(so that its components are all even).
The operators are composed of hypercube-based bilinears, 
each denoted by $[S\times F]$, where $S$ is the spin and $F$ the taste.
The subscripts $I$ and $II$ distinguish the two ways in which the
color indices in the bilinears are connected by products of links.
A complete definition of these operators
is given in Ref.~\cite{wlee-03}.
At one-loop order, the coefficient multiplying each 
operator differs from unity, as shown in Eq.~(\ref{eq:pm}) below.

The four-quark matrix element should be independent of $t$ when
$t_1 < t < t_2$, as long as the operator is far enough
from each source to remove contamination from excited states.
This is because we always use a kaon and antikaon with the same mass.
To obtain $B_K$ we then divide this matrix element by that
obtained via the ``vacuum saturation'' approximation,
as shown in Eq.~\ref{eq:bk-def}. 
Explicitly, the ratio of correlation functions we use is
\begin{equation}
B_K(t) = \frac{2 \langle W(t_1) {\cal O}_K(t) W(t_2)\rangle}
        {(8/3) \langle W(t_1) [A_0\times P](t)\rangle
               \langle [A_0 \times P](t) W(t_2)\rangle}\,,
\end{equation}
where $W$ represents a wall source, and the factor of
2 in the numerator accounts contractions
present in the continuum which are absent for our lattice
operator~\cite{Kilcup:1997ye}.
This ratio
should be also be independent of $t$ away from the sources, and
cancels the coupling of the wall sources to the (anti)kaons.
Thus our aim is to choose the sources so that there
is a clear plateau, and then fit $B_K$ to a constant in this region.

\begin{figure}[htbp]
\includegraphics[width=20pc]{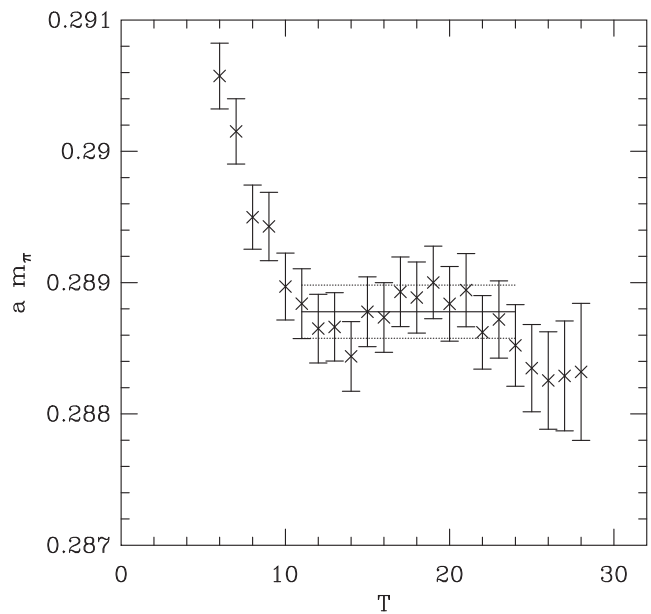}
\caption{Pion effective mass versus time from wall-source.
  Parameters are $am_x = am_y = 0.025$ on the C3 coarse ensemble.
  The fit is described in the text.
  \label{fig:a4-1}}
\end{figure}

To gauge how far apart to place the sources, we look at the
two-point correlator from the wall-source (say at $t=0$)
to the taste-P axial current 
(i.e. with quantum numbers $\gamma_4\gamma_5\otimes \xi_5$)
summed over the spatial time-slices at times $t$ and $t+1$. 
This is one of the two factors in the denominator of $B_K$
(and also shows how we obtain the taste-P kaon mass).
We show an example of the resulting effective mass
in Fig.~\ref{fig:a4-1}. This is defined by equating the
correlator at times $t$ and $t+1$ to the form
\begin{equation}
f(t) = Z [ \exp(-m_\pi (t+0.5)) - \exp(-m_\pi(T-t-0.5))] \,,
\label{eq:singleexp}
\end{equation}
which includes both a direct and a ``wrap-around'' exponential
(with $T$ the lattice extent in the time direction).
The plot is for a pion on one of the coarse ensembles
with a mass close to that of the physical kaon.
We observe a noticeable contamination from the
excited states up to $t \approx 10$,
and that the signal degrades for $26 \lesssim t$.
Also shown is a fit of the correlator
to the form (\ref{eq:singleexp}) in the plateau region.

From this and similar plots,
we know that on the coarse lattices
we need to place our four-fermion operator at least 10 timeslices 
from both sources, i.e. that $\Delta t=t_2-t_1 > 20$.
Here we assume that the coupling of the four-quark operator
to excited states is comparable to that of the axial current---a
reasonable assumption given that the vacuum saturation approximation
turns out to be good to within a factor of two.
The source separation cannot be too large, however, 
because the signal will then be contaminated by 
wrap-around contributions arising from our use of a
periodic lattice.
Thus we have chosen $\Delta t$ to be somewhat larger than the minimum,
but not too much larger. Our specific choices are given in
Table~\ref{tab:bk-para}.
We have held the physical distance between the sources,
$\Delta t/a$, roughly constant as the lattice spacing is reduced.
The table also gives the parameters that determine
the fitting range: we fit from $t=t_1+t_L$ to $t=t_1+t_R=t_2-t_L-1$.
Note that our fit ranges are symmetrical with respect to the
two wall sources since the four-quark operator resides on timeslices
$t$ and $t+1$.

\begin{table}[htbp]
\caption{Choices for the wall-source separation, $\Delta t$,
and its ratio to the temporal length of the lattices, $T$,
as well as the parameters determining the fitting range.
  \label{tab:bk-para}}
\begin{ruledtabular}
\begin{tabular}{ l  l  l l l}
$a$ (fm) & $\Delta t$ & $\Delta t / T$ & $t_L$ & $t_R$\\ 
\hline
0.12 &  26  & 0.41  & 10 & 15\\
0.09 &  40  & 0.42  & 13 & 26\\
0.06 &  60  & 0.42  & 22 & 37\\
\end{tabular}
\end{ruledtabular}
\end{table}

We can estimate the systematic error
due to wrap-around contributions as follows. 
These contributions have the kaon from one source propagating
the ``wrong'' way around the lattice, so that the matrix element
which occurs is $\langle K_0 K_0|{\cal O}|0\rangle$ or its conjugate.
At leading order in continuum ChPT, this matrix element has the
same magnitude (and opposite sign) as the desired matrix element
$\langle K_0|{\cal O}|\bar K_0\rangle$. 
Since the wrap-around effect is small, when estimating its size we
ignore the differences between
these two matrix elements due to higher order SChPT effects.
Then the contamination is simply given by the ratio of the kaon
propagators in the two cases. Adding the two contributions
involving a single (anti)kaon wrapping around the lattice,
we find that the ratio of contamination to signal is
\begin{eqnarray}
R_K(t) &=& 2 \exp\{-a m_K (T-\Delta t)\} 
\nonumber\\
    &&\ \times \cosh\{2 a m_K [(t - t_1) - (\Delta t - 1)/2]\}
\,. \label{eq:RKdef}
\end{eqnarray}
This is suppressed by the overall exponential factor,
but grows on either side of its minimum (which occurs midway
between the sources) as a cosh with an exponent containing
$2 a m_K$. To give an idea of the size of the effect
we give two examples of the value of $R_K$ at the edge of
the fitting range (where it is maximal). 
On the coarse lattices, with $T=64$, $\Delta t=26$
and $t-t_1=t_L-t_1=10$ we find $R_K(t)=0.015$ for the lightest
pion ($a m_{xy:\rm P}=0.1342$ for $am_x=am_y=0.005$),
while $R_K(t)=1.2 \times 10^{-4}$ for the lightest kaon used
in the SU(2) chiral fit ($a m_{xy:\rm P}=0.2749$ for
$am_x=0.005$ and $am_y=0.04$).
The latter result is most relevant for an estimate of the size of the
induced error in our final result for the physical kaon (which has
a mass $a m_K^{\rm phys}\approx 0.3$ on the coarse lattices).
This is because, whether we use the SU(3) or SU(2) fits,
the most important points are those closest to the place
to which we are extrapolating.
Thus we conclude that the wrap-around systematic is negligible
for $B_K$, although it would be significant (at the few percent
level) were we to quote a result for $B_0$
($B_K$ in the chiral and continuum limits).

Based on the autocorrelation lengths for light hadron properties,
the lattices in the MILC ensembles are somewhat correlated~\cite{milc-04}.
Because of this, we attempt to reduce correlations between
our measurements of $B_K$ by randomly choosing the overall
source position, $t_1$, on each configuration (with 
$t_2$ fixed to $t_1+\Delta t$).
For three ensembles (C3, C3-2 and C4) we have increased 
our statistics by doing multiple measurements
on each lattice (as shown in Table \ref{tab:milc-lat}).
We chose the $t_1$ values for these measurements randomly
and also set the random number seeds for the U(1) noise sources 
randomly.

Using these enlarged data-sets, we have done a preliminary study
of autocorrelations by seeing how the error changes when we bin
configurations together. 
This has been done for a pion two-point correlator, for which
the errors are smaller than those for $B_K$.
The results for ensemble C3 are presented
in Ref.~\cite{wlee-09-3}. We find that the errors increase somewhat
as the bin size increases, 
reaching a plateau for bins of size 3 or above,
in which the errors are about 20\%
larger than those obtained treating each configuration
as independent. This increase is only visible after we use
the larger sample (9 measurements per configuration).

We take this result as indicating that autocorrelations are
a small effect. In particular, because our statistical
errors turn out to be smaller than the systematic errors, 
a 20\% increase in the former would have little impact on
our final result. Thus we
have ignored the possible impact of autocorrelations in this paper.
%

%
\begin{figure}[htbp]
\includegraphics[width=20pc]{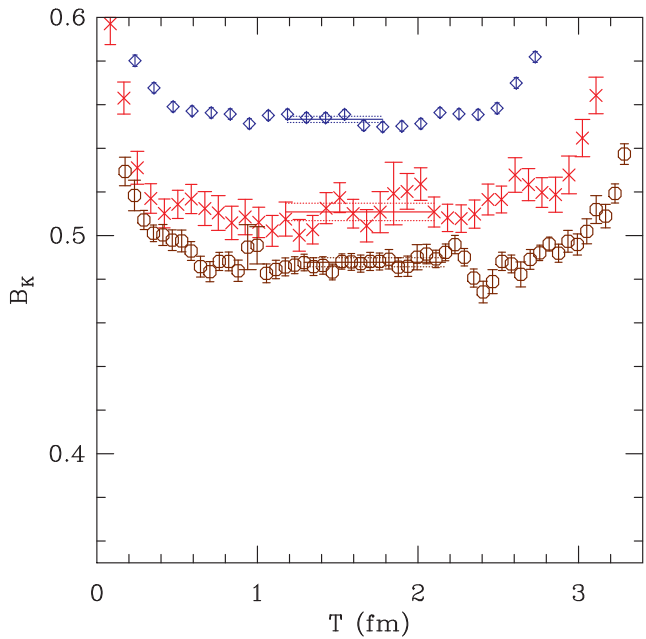}
\caption{$B_K(\mu=1/a)$ as a function of $T=t-t_1$.
(Blue) diamonds are from the
coarse ensemble C3, with $am_x = am_y = 0.025$;
(red) crosses are from the fine ensemble F1, 
with $am_x = am_y = 0.015$; and
(brown) octagons are from the superfine ensemble S1, 
with $am_x = am_y = 0.009$.
The result of a fit to a constant over the range given in
Table~\protect\ref{tab:milc-lat} is shown by the horizontal lines.
Note that the $T$ is given in fermis.
  \label{fig:bk-fit-b5}}
\end{figure}

Typical results for one-loop renormalized $B_K$, along with the
fitting range and resulting value and error, are shown in
Fig.~\ref{fig:bk-fit-b5}.  This figure shows the
dependence on the lattice spacing when we use a kaon
composed of degenerate quarks but with a mass close to that of the
physical kaon.  The data have a similar profile for all three lattice
spacings, showing contamination from excited states near the boundary
but a reasonable plateau that extends further out than our fit region.
The smaller size of the errors on the coarse lattice reflects the
larger sample (9 measurements on each lattice).

\begin{figure}[htbp]
\includegraphics[width=20pc]{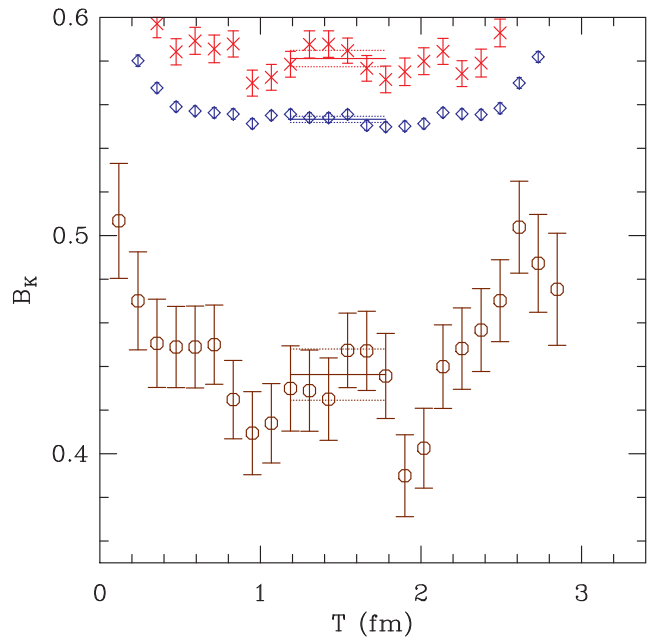}
\caption{As for Fig.~\protect\ref{fig:bk-fit-b5}, except that
all data is from the C3 ensemble.
(Blue) diamonds show results for $am_x = 0.025$ and $am_y = 0.025$,
(red) crosses for $am_x = 0.005$ and $am_y = 0.045$, and
(brown) octagons for $am_x = 0.005$ and $am_y = 0.005$.
  \label{fig:bk-fit-b17}}
\end{figure}

The effect of varying the quark masses (at fixed lattice spacing)
is illustrated by Fig.~\ref{fig:bk-fit-b17}. 
The (blue) diamonds show the same data as in Fig.~\ref{fig:bk-fit-b5},
with degenerate quarks and an approximately physical kaon mass.
The (red) crosses show the change when $m_y-m_x$ is maximized (in our
data set) with $m_x+m_y$ fixed (so that the kaon mass is almost the same).
One sees that the central value increases a little,
and that the errors are significantly larger.
The (brown) octagons shows what happens if both $m_x$ and $m_y$
are reduced to our smallest value, 
so that the resulting ``kaon'' is closer to a pion
(having a mass $\approx 220\,$MeV).
Here the central value is significantly reduced,
and the errors are much larger.
Fortunately, our results from light ``kaons'' play little role
in determining the extrapolation to the physical kaon.

In these figures, and in all subsequent analysis, errors are obtained
using the jackknife procedure.  This is true both for the errors on
individual points and for the errors in fit parameters.  When doing
fits to multiple timeslices, and when fitting to multiple quark masses
(as required for the chiral fits discussed below) we do not use the
(inverse of the) full covariance matrix, which he have found to lead
to instabilities, but rather keep only the diagonal elements. This
means that we cannot quantitatively judge the goodness of fit, since
the $\chi^2$ is generically underestimated.  Assuming, however,
that the fits are reasonable, the jackknife errors should be reliable. 

We close this subsection by describing how we determine the
values of $m_{xy:\TB}$ needed as inputs to the SChPT fits to $B_K$.
For the taste-P meson, we measure the mass directly using our
wall sources. We have four sets of two-point correlation functions:
\begin{subequations}
\begin{eqnarray}
C_1(t) &=& \langle A_4(t) P(t_1) \rangle \,, \\
C_2(t) &=& \langle A_4(t) P(t_2) \rangle \,, \\
C_3(t) &=& \langle P(t) P(t_1) \rangle \,, \\  
C_4(t) &=& \langle P(t) P(t_2) \rangle \,.  
\end{eqnarray}
\end{subequations}
We determine the Goldstone pion masses by fitting these four sets
of data individually and averaging the results.

\begin{figure}[htbp]
\includegraphics[width=20pc]{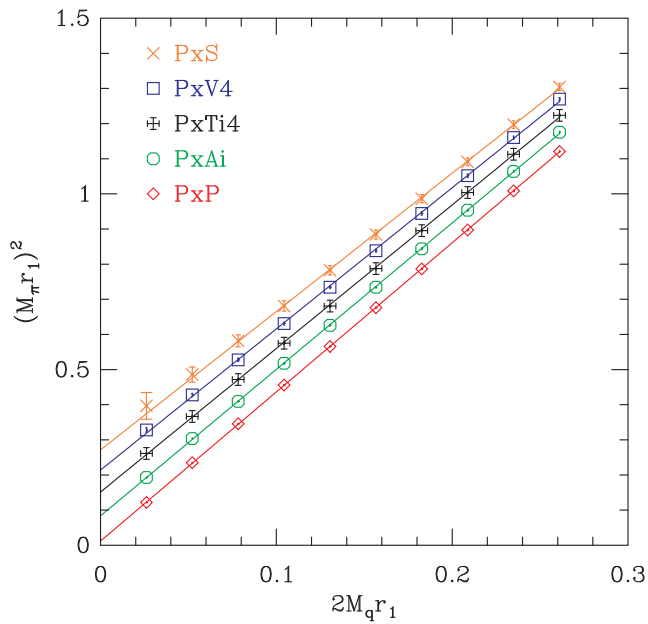}
\caption{$m_{xy}^2 r_1^2 $ versus $(m_x+m_y)r_1$, for states with
  $m_x=m_y$, and tastes (from bottom to top) $\xi_5$, $\xi_i\xi_5$,
  $\xi_i\xi_4$, $\xi_4$ and $1$. $r_1$ is the modified
  Sommer-parameter, whose value is $2.619a$.  The fits are discussed in
  the text.  Results are for the coarse ensemble C3.
  \label{fig:pion-mass-1}}
\end{figure}

\begin{figure}[htbp]
\includegraphics[width=20pc]{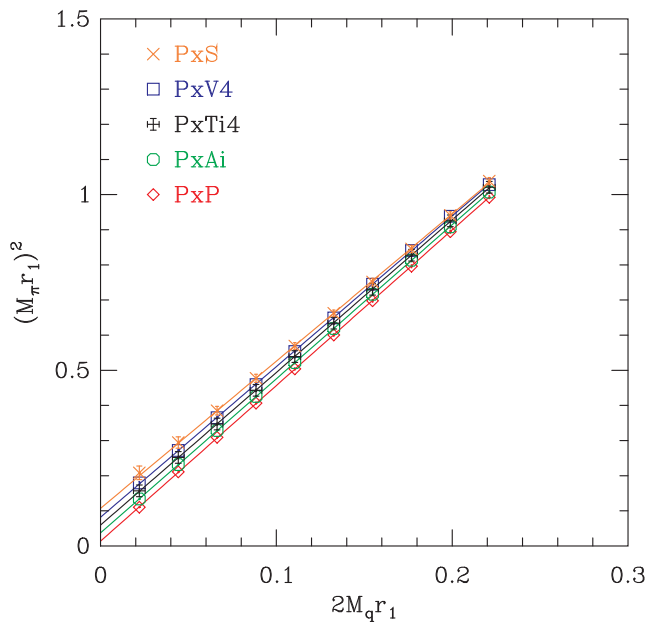}
\caption{As in Fig.~\protect\ref{fig:pion-mass-1}, but for
the fine ensemble F1. Here $r_1=3.701a$.
  \label{fig:pion-mass-2}}
\end{figure}

We have calculated the masses of PGBs having all other tastes
on most of the ensembles.
The methodology is explained in Refs.~\cite{wlee-08-1,wlee-08-2}.
Specifically, we use 8 ``cubic wall sources'' per configuration,
and ``Golterman-style'' sink operators.
Results for the best-measured tastes are shown in
Figs.~\ref{fig:pion-mass-1} and \ref{fig:pion-mass-2} for a coarse and
the fine ensemble, respectively.  The nondegenerate points are also
consistent within errors with the observed linear behavior, though we do
not show them for the sake of clarity.  The results for the other
tastes have masses that are completely consistent with the $SO(4)$
symmetry expected at LO.  Given this, we choose to assume exact
$SO(4)$ symmetry for the masses used in the ChPT expressions, taking
the values from the best measured tastes (such as those in the
Figures).


The LO prediction of Eq.~(\ref{eq:pion-mass-sq-1}) has the
slopes being the same for all tastes.
NLO effects will, however, lead to a difference between the slopes. 
We find that the slopes do have statistically significant differences, 
but that these differences are very small, 
a few percent or less~\cite{wlee-08-1}.
In practice we do linear fits allowing the slopes to differ
(i.e. allowing $b_1$ to depend on taste B), 
and from these determine the intercepts $\Delta_B$.
We then reconstruct the PGB masses that we use in chiral loops using
\begin{subequations}
\label{eq:pion-mass-sq-2}
\begin{eqnarray}
K_\TB &=& G + \Delta_\TB - \Delta_\TP\,,\\
X_\TB &=& X_\TP + \Delta_\TB - \Delta_\TP \,,\\
Y_\TB &=& Y_\TP + \Delta_\TB - \Delta_\TP \,.
\end{eqnarray}
\end{subequations}
This choice effectively assumes that the slopes are equal to
that of the taste P mesons.
This approach reproduces the actual masses to within a few percent or better,
and is, in any case, theoretically consistent
as the LO masses can be used in loops.
Put differently, the inclusion of the small differences between slopes 
in the chiral logarithms would lead to an analytic
contribution of NNLO, i.e. of higher order than our expressions.

\begin{figure}[htbp]
\includegraphics[width=20pc]{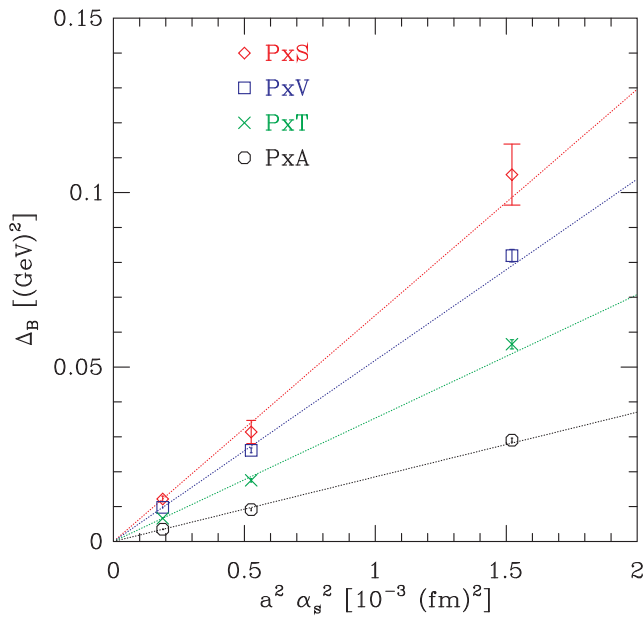}
\caption{$\Delta_\TB$ from ensembles C3 and F1 (the rightmost
two sets of points). These are extrapolated, as described
in the text, to obtain the results for the S1 ensemble
(leftmost set of points). Numerical values are given in
Table~\protect\ref{tab:Delta_B}.
  \label{fig:delta-scale}}
\end{figure}

We proceed in the manner just described on all ensembles except
the large-volume coarse ensemble (C3-2), the fine ensemble (F2), and the
superfine ensemble (S1). On these three ensembles we have not measured
the full taste spectrum, due to the high computational cost.
We estimate the taste splittings on these ensembles as follows.
For (C3-2) we simply use the taste splittings
from the smaller lattices with the same quark masses [ensemble (C3)].
For (F2) we assume the same taste splittings as for the other fine
ensemble, (F1), based on the fact that taste splittings on the coarse
lattices are almost independent of the light sea-quark mass.
Finally, for the superfine ensemble, we obtain the $\Delta_\TB$
by extrapolating from the coarse and fine lattices.
These quantities are expected to vanish in the continuum limit
as $a^2 \alpha^2$. We thus fit our results from the
C3 and F1 ensembles to
\begin{equation}
\Delta_\TB(a) = b_1 \ a^2 \alpha_s^2(\mu=1/a) 
\label{eq:Delta_B_extrap}
\end{equation}
with $\alpha_s$ determined using 4-loop evolution from the
Particle Data Group value at $\mu=M_Z$.
The result is satisfactory, as shown in Fig.~\ref{fig:delta-scale},
and leads to the values collected in Table~\ref{tab:Delta_B}.
Since the taste splittings on the superfine lattice are so small,
the uncertainty introduced by the need to do this extrapolation 
has a negligible impact on our chiral fitting.

\begin{table}[htbp]
\caption{Results for valence taste-splittings,
$\Delta_\TB$ (GeV${}^2$), on ensembles C3, F1 and S1.
Those on C3 and F1 are obtained by calculating PGB masses
using cubic wall sources (see Ref.~\cite{wlee-08-1}).
Those on S1 are obtained by extrapolating using 
Eq.~(\protect\ref{eq:Delta_B_extrap}).
  \label{tab:Delta_B}}
\begin{ruledtabular}
\begin{tabular}{ c | l  l  l}
$a$ (fm) & 0.12 (C3) &  0.09 (F1) & 0.06 (S1) \\ 
\hline
$\Delta_A$ &  0.02905(59)  & 0.00912(29)  & 0.003487(60) \\
$\Delta_T$ &  0.0565(13)   & 0.01753(49)  & 0.00666(12) \\
$\Delta_V$ &  0.0819(18)   & 0.02612(68)  & 0.00977(16) \\
$\Delta_S$ &  0.1052(88)   & 0.0314(33)   & 0.01219(80)
\end{tabular}
\end{ruledtabular}
\end{table}

\subsection{Matching factors \label{ssec:matching}}
To convert the results of our lattice calculation into physical
observables, we need to match lattice and continuum operators.
The latter we define in the conventional way using 
naive dimensional regularization with $\overline{\rm MS}$ subtraction.
For the matching, we use one-loop perturbation theory, which leads to
the general form 
\begin{eqnarray}
O^\text{cont}_i(\mu) = \sum_j \bigg[ \delta_{ij}
\!-\! \frac{\alpha(q^*)}{4\pi} ( \gamma_{ij} \log(a \mu)\!+\! c_{ij} ) \bigg]
O^\text{latt}_j(a)\,.
\label{eq:pm}
\end{eqnarray}
Here superscripts ``cont'' and ``latt'' indicate continuum and lattice
operators, respectively,
$\gamma_{ij}$ is the anomalous dimension matrix,
the $c_{ij}$ are finite constants,
and $\mu$ is the renormalization scale for the continuum operator.
We choose the coupling constant here and below to be that defined in the
continuum $\overline{MS}$ scheme (which we evaluate numerically
using four-loop running from $m_Z$).
The scale at which it is evaluated, $q^*$, is undetermined by
one-loop matching.

In our calculation we are interested in a single continuum operator, so
$i$ takes one value,
but the matching involves many lattice operators, labeled by $j$. 
The operators which appear are listed in Ref.~\cite{wlee-03}.
The values of $c_{ij}$ for our HYP-smeared operators, 
using the appropriate Symanzik-improved gauge action,
have been calculated by three of us and
will be presented in Ref.~\cite{wlee-10-2}.

When using this matching formula we do not, in fact, include
all the lattice operators which appear at one-loop, but instead
keep only those with external tastes $P$. There are
four of these, and they are listed above in 
Eqs.~(\ref{eq:O1latt}-\ref{eq:O4latt}).
Dropping the other operators leads to
the $O(\alpha/\pi)$ truncation errors discussed earlier.

Having matched to the continuum operator at an intermediate scale
$\mu$, we then evolve using the continuum renormalization group (RG)
down to a canonical scale $p=2\;$GeV. We do so using the two-loop
formula
\begin{eqnarray}
B_K(\text{NDR},p) &=&
\frac{ [ 1 - \dfrac{\alpha(\mu)}{4\pi } Z ] }
{ [ 1 - \dfrac{\alpha(p)}{4\pi} Z ] }
\bigg( \frac{\alpha(p)}{\alpha(\mu)} \bigg)^{ d^{(0)} }
B_K(\text{NDR},\mu)
\nonumber \\
\label{eq:bk-rg-evol}
\\
Z &=& \frac{ \gamma^{(1)} }{2 \beta_0} -
d^{(0)} \frac{ \beta_1 }{ \beta_0 }  
\nonumber \\
d^{(0)} &=& \frac{ \gamma^{(0)} }{ 2 \beta_0 }\,.
\nonumber
\end{eqnarray}
Expressions for the $\beta_i$ and $\gamma^{(i)}$ are given,
for example, in Ref.~\cite{buras-90-1}.
In this way we obtain $B_K({\rm NDR}, p = 2\text{ GeV})$.

In order to implement this matching and running
we have to choose values for $\mu$ and $q^*$.
The result depends on these choices because we have truncated the
matching at one-loop and the running at two-loops. 
In both cases the error we make is of $O(\alpha^2)$.
Based on the arguments of Ref.~\cite{Gupta:1996yt}, we use
``horizontal matching'' in which $q^*=\mu$.
The choice of $q^*$ should be made so as to minimize higher-order
terms in perturbation theory, which requires that $q^*$ be a typical
scale of momenta flowing in loops. We simply take 
$q^*=1/a$. 
It is possible to make a better-informed estimate using the
one-loop integrands themselves, 
but we have not yet attempted this.

Finally, we recall that one can also
define the RG invariant quantity $\widehat{B}_K$,
\begin{eqnarray}
\widehat{B}_K &=& [ 1 - \frac{\alpha(\mu)}{4\pi } Z ]
\bigg( \frac{1}{\alpha(\mu)} \bigg)^{ d^{(0)} }
B_K^\text{NDR}(\mu)\,,
\label{eq:bk-rgi}
\end{eqnarray}
which is often used in the continuum literature.
This quantity has the apparent advantage that all truncation errors
are manifestly proportional to $\alpha(\mu)^2$, with the
scale $\mu$ being naturally chosen of $O(1/a)$ (and in our case,
chosen to be exactly $1/a$). Thus these errors vanish
(albeit logarithmically) in the continuum limit.
By comparison, using $B_K({\rm NDR},2\;{\rm GeV})$ apparently leads
to truncation errors which have a nominal size of $\alpha(2\;{\rm GeV})^2$,
which does not vanish in the continuum limit.
This is, however, misleading, in that if different lattice calculations
use the same truncated formula to run to $2\;$GeV, they are in
effect comparing $\widehat{B}_K$, so the real truncation errors in the
comparison are $\sim \alpha(1/a)^2$.
%

\subsection{SU(3) fitting: Strategy \label{ssec:fit}}
%
%
We would like to fit our results for $B_K$ to the
functional form of Eq.~(\ref{eq:bk-su3-ma}). 
For a given ensemble (fixed $a$ and sea-quark masses) the fit form
has 15 parameters.\footnote{%
There are 15 rather than 16 parameters
because $H_5$ is a constant for
fixed sea-quark masses, and so can be absorbed into $b_1$
(the coefficient of $H_1$). Doing so leads to additional
chiral logarithms multiplied by quark masses, but these
are of NNLO, and thus irrelevant to our NLO expression.
Of course, having absorbed $b_5$ into $b_1$, we
must subsequently take into account the
resulting sea-quark mass dependence of $b_1$.
This we do in Sec.~\ref{ssec:sea-quark}.}
Despite the fact that we have
10 degenerate and 45 non-degenerate data points on each
ensemble, we find
a direct fit to Eq.~(\ref{eq:bk-su3-ma}) to be very difficult.
Here we explain why and describe our strategies for dealing with
this problem.

We begin by sorting the 15 terms into 4 categories, according to
the following criteria:
\begin{itemize}
\item 
Does the term survive in the continuum limit, i.e.
does it correspond to a term in continuum ChPT? 
Contributions from discretization errors (taste-breaking or
otherwise) and from truncation errors do not have such a
correspondence.
\item 
Does the term contribute to $B_K$ in the degenerate case of $m_x = m_y$,
or does it vanish in that limit?
\end{itemize}
The result of this categorization is shown in Table~\ref{tab:sort-H}. 

\begin{table}[htbp]
\caption{ Classification of contributions, $H_i$,
to the NLO SU(3) SChPT expression for $B_K$, 
Eq.~(\ref{eq:bk-su3-ma}),
when fitting at a fixed $a$ and sea-quark masses.
Terms in the ``cont'' column have a corresponding continuum form,
while those labeled ``non-cont'' do not.
Terms in the row labeled ``deg'' contribute both for $m_x=m_y$
and $m_x\ne m_y$, while those in the
``non-deg'' row contribute only for $m_x\ne m_y$.
Numbers in the table give the subscript, $i$, of the fit form $H_i$.
Numbers \underline{underlined} indicate
those $H_i$ which are kept in our final SU(3) fits.
  \label{tab:sort-H}}
\begin{ruledtabular}
\begin{tabular}{ l l l }
  &  cont & non-cont \\ 
\hline
deg     & \underline{1}, \underline{2}, \underline{3}  
& 6, 7, 8, 9, \underline{10} \\
non-deg & \underline{4}        
& 11, \underline{12}, 13, \underline{14}, 15, 16
\end{tabular}
\end{ruledtabular}
\end{table}
%

From Table \ref{tab:sort-H}, we observe that the fitting functional
is somewhat simplified if we fit only to degenerate data points.
There are then three ``continuum-like'' terms, $H_{1-3}$,
and five terms which are lattice artifacts, $H_{6-10}$.
Since we have 10 degenerate data points, we can, in principle,
fit them to a functional form composed of 8 terms.
In practice, this turns out to be very difficult,
primarily because the functional forms of the five lattice-artifact terms 
are very similar for our range of PGB masses.
This is not unexpected: the five functions have the same form
[that of $F^{(4)}$ defined in Eq.~(\ref{eq:app:F4def})],
but differ in the taste of the PGB whose mass appears.
Thus one expects the functions to be similar except at the
lightest quark masses where the taste-splittings are comparable
to the masses themselves (see Fig.~\ref{fig:pion-mass-1}).
This is borne out by direct numerical evaluation.
The similarity of the functions becomes more pronounced
as one approaches the continuum limit, because of the
reduction in taste splittings. Indeed, on the superfine lattices,
one can argue that the $\CO(a^2)$ effects should be treated as
of NLO, and thus that the difference between these five functions
is really an effect of NNLO. 

In light of these considerations, we approximate all five functions
by $F_\TT^{(4)}$, i.e. the function in which the mass
is that of the taste-T PGB. We choose the tensor taste as its mass
is roughly the average of those of all tastes.
We stress that this is not a systematic procedure, since if
$\CO(m)\sim\CO(a^2)$, as is the case for the smallest
quark masses on the coarse lattices, 
it is {\em not} legitimate to expand $\log(X_\TB)$ about $X_\TB=X_\TT$
and treat the difference as of higher order.
Our procedure
is simply a phenomenological way of resolving the presence of
nearly flat directions in the fit function.

With this approximation, 
the fitting function for the degenerate data-set collapses to
\begin{equation}
f_\textrm{th}^{\rm deg}  = \sum_{i=1}^{4} c_i F_i
\label{eq:su3-fit-deg-form}
\end{equation}
where $F_1=H_1$, $F_2=H_2$, $F_3=H_3$ 
[with the $H_i$ as defined in Eqs.~(\ref{eq:H1def}-\ref{eq:H3def})]
and $F_4 = H_{10}=F^{(4)}_{\TT}$.
Fits to this 4-parameter form are stable,
and we use them as the first stage of most of
our fitting strategies.

The fit to (\ref{eq:su3-fit-deg-form}) can also be constrained
given our prior knowledge of the size of the coefficients.
We expect from continuum ChPT 
power-counting that $c_1\sim c_2\sim c_3\sim \CO(1)$.
For $c_4=b_{10}$ we have two possible estimates,
given in Eq.~(\ref{eq:b6-12est}),
depending on whether discretization or truncation errors dominate.
One can systematically include this prior information in the
fitting using the Bayesian method~\cite{lepage-02},
as discussed in the next subsection.
%

We now turn to the remaining 45 non-degenerate data points.
As can be seen from Table~\ref{tab:sort-H},
including these points in the fit requires the introduction
of one additional term with a non-vanishing continuum limit,
and six terms arising from lattice artifacts. 
Once again, however, directly fitting to all these terms
is difficult because some of the functions are similar.
Thus we choose again to approximate the fitting function.

We first observe that $H_{11}=F^{(5)}$
[given in Eq.~(\ref{eq:app:F5def})] is
numerically close to twice $H_{12}=F^{(6)}$
[given in Eq.~(\ref{eq:app:F6def})].
This proportionality is exact if taste-splittings are
ignored. Thus we choose to drop $H_{11}$ and keep
$H_{12}$ as a surrogate for both these functions.

Next we note that $H_{13}=F^{(1)}_\TV$ 
is very similar to $H_{14}=F^{(1)}_\TA$---differing
only in the taste of the PGBs whose masses appear in the function.
Thus we choose to keep only $H_{14}$ from this pair.

Finally, we come to $H_{15}=F^{(2)}_\TV$ and $H_{16}=F^{(2)}_\TA$.
These are also numerically similar, but, more importantly, 
their coefficients are proportional to
$\delta^\textrm{MA2}_\TA$ and $\delta^\textrm{MA2}_\TV$, 
quantities that we have argued are substantially suppressed
(see Sec.~\ref{ssec:mixed}).
Thus we consider these contributions to be of NNLO and simply drop them.

The outcome of these considerations is that the functional
form we use when fitting our entire data-set contains 7 terms:
\begin{equation}
f_\text{th}  = \sum_{i=1}^{7} c_i F_i
\label{eq:su3-fit-non-deg-form}
\end{equation}
where the first 4 terms are the same as in
Eq.~(\ref{eq:su3-fit-deg-form}) 
and the rest are $F_5=H_4$ (the analytic term), 
$F_6=H_{12}=F^{(6)}$ and $F_7 = H_{14}=F^{(1)}_\TA$.
We expect $c_5=\CO(1)$ and $c_6\sim c_4$, while
$c_7=b_{14}$ is given by the estimates in Eq.~(\ref{eq:b13-16est}).
In some of our fits we impose the latter two estimates using
the Bayesian method.

\subsection{SU(3) fitting: Implementation \label{ssec:fit-su3}}
%
In this section, we use our fits to the C3 and S1 ensembles as
examples to explain our detailed strategy and the results.

We first fit the degenerate data to the 4-parameter form of
Eq.~(\ref{eq:su3-fit-deg-form}) with no constraints.
We label this fit ``D-U'', with D for degenerate and
U for unconstrained.\footnote{This fit is called ``D-T3''
in our previous publications.}
\begin{figure}[htbp]
\includegraphics[width=20pc]{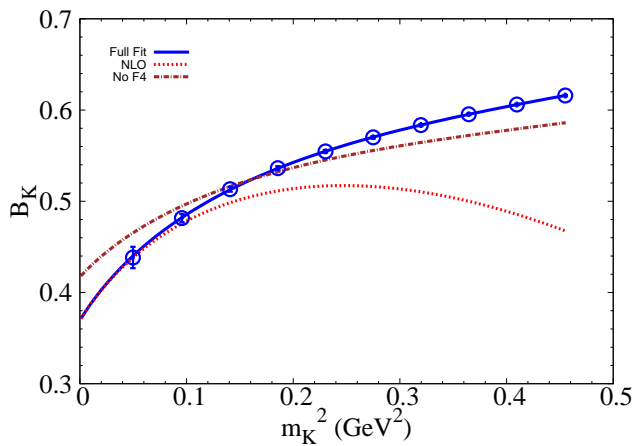}
\caption{$B_K({\rm NDR},\mu=1/a)$ vs. $m_K^2$
for degenerate kaons ($m_x=m_y$), on the C3 ensemble. 
A fit of type D-U 
is shown by the solid (blue) curve.
The dotted (red) curve labeled ``NLO''
shows the impact of dropping the NNLO term.
The dashed (brown) curve labeled ``No F4'' 
shows the impact of dropping the $F_4$ term.
  \label{fig:fit-D-T3:C3}}
\end{figure}
\begin{figure}[htbp]
\includegraphics[width=20pc]{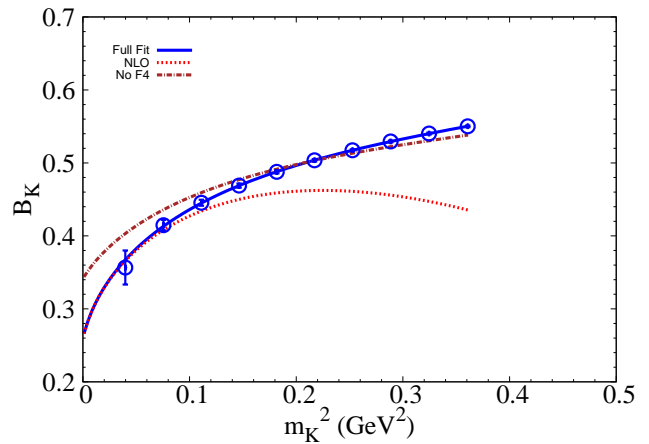}
\caption{$B_K({\rm NDR},\mu=1/a)$ vs. $m_K^2$
for degenerate kaons on the S1 ensemble showing a D-U fit.
Notation as in Fig.~\protect\ref{fig:fit-D-T3:C3}
  \label{fig:fit-D-T3:S1}}
\end{figure}
We show the result of this fit on the C3 and S1 ensembles in
Figs.~\ref{fig:fit-D-T3:C3} and \ref{fig:fit-D-T3:S1}, respectively.
These ensembles have, to good approximation, the same sea-quark masses
and thus differ mainly in the value of $a$.
The fits give a good representation of the data,
as is the case also on the other ensembles (not shown). 
The parameters from the D-U fits
are collected in Table~\ref{tab:fit-D-T3}.\footnote{%
Ensemble C3-2 is only used for an estimate of finite-volume errors,
as described in Sec.~\ref{ssec:err-bud}.}
We stress again that the $\chi^2/{\rm dof}$ uses only
diagonal elements of the covariance matrix and thus is expected
to be much less than unity. 
In fact, a value approaching unity indicates a poor fit.

\begin{table}[htbp]
\caption{Parameters of D-U fits. 
The $c_i$ are defined in
  Eq.~(\ref{eq:su3-fit-deg-form}). $\chi^2/\text{dof}$ represents
  $\chi^2$ per degree of freedom calculated using only diagonal
  elements of the covariance matrix.
  \label{tab:fit-D-T3}}
\begin{ruledtabular}
\begin{tabular}{ l | l l l l l }
 ID &  $c_1$ & $c_2$ & $c_3$ & $c_4$(GeV${}^2$) & $\chi^2/\text{dof}$ \\ 
\hline
C1 &  0.283(41) & 0.88(40) &    0.24(41) &    0.0018(19) & 0.004(20) \\
C2 &  0.351(42) & 0.18(41) &    0.85(41) & $-$0.0022(19) & 0.012(17) \\
C3 &  0.339(13) & 0.37(13) &    0.72(13) & $-$0.0008(6)  & 0.041(54) \\
C4 &  0.327(13) & 0.47(13) &    0.60(13) & $-$0.0005(6)  & 0.051(59)  \\
C5 &  0.286(32) & 0.84(30) &    0.22(30) &    0.0013(14) & 0.08(11) \\
\hline
F1 &  0.349(39) & 0.05(33) &   1.06(40) & $-$0.0012(11) & 0.10(12) \\
F2 &  0.356(32) & 0.01(27) &   1.13(33) & $-$0.0013(9)  & 0.11(13) \\
\hline
S1 &  0.323(25) & 0.18(21) &   0.88(29) & $-$0.0005(6)  & 0.05(12) \\
\end{tabular}
\end{ruledtabular}
\end{table}

We find that $c_{1-3}$ are of $\CO(1)$, indicating
that the chiral expansion is behaving as expected.
The convergence of the ChPT series is illustrated in the figures.
At LO, $B_K$ is a constant.
The NLO form is shown by the (brown) dashed curves,
and has curvature due to the chiral logarithms in $F_1$ and $F_4$. 
Our phenomenological NNLO term
makes up the difference between the full fit and the NLO form.
The convergence is reasonable for $m_K\lesssim m_K^{\rm phys}$,
(e.g. on C3 we find LO:NLO:NNLO $\approx 1:0.5:0.12$ for 
$m_K\approx m_K^{\rm phys}$,), 
but clearly poor for our heaviest kaon.
We stress, however, that, because we do not have the complete NNLO
expression, this analysis of convergence is only approximate.

Since $c_{1-3}$ are continuum parameters, and are dimensionless,
we would expect
approximate consistency between their values in all the fits, up to
a modest dependence on lattice spacing and on sea-quark masses.
This expectation holds within errors---although we note that
the errors in $c_2$ and $c_3$ are quite large.

The coefficient $c_4$ multiplies the sole
NLO discretization/truncation term in our fit function.
We find that our fits do not require such a term---it is consistent 
with zero on all ensembles.
The range of values of $c_4$ allowed by the fits 
(as indicated by the errors)
are such, however, that this term can make a noticeable contribution.
This is illustrated in the figures, where the contribution of
the $c_4$ term is given by the
difference between the full fits and the ``No F4'' curves.
It is seen to be relatively small ($< 10\%$) 
over the entire mass range on the C3 ensemble,
and in particular is considerably smaller than continuum-like NLO
terms (and comparable to the NNLO terms) at $m_K^{\rm phys}$. 
As expected, its contribution on the superfine ensemble is generally
much smaller than on the coarse ensemble.
These results imply that we have been conservative in
treating this term as of NLO in our power-counting.

Another way of seeing this
is to compare the values of $c_4$ 
(and their errors) to the magnitudes expected
from power counting. As noted above, these range from
\begin{eqnarray}
|c_4| &=& \CO(a^2) \approx \Lambda_\textrm{QCD}^2 
(a\Lambda_\textrm{QCD}^2)^2 
\nonumber \\
&\approx& \left\{ 
\begin{array}{l l} 
0.0031 \GeV^2 & \text{on the coarse lattices,} \\
0.0015 \GeV^2 & \text{on the fine lattice,} \\
0.00075 \GeV^2 & \text{on the superfine lattice.}
\end{array} \right.
\,,
\label{eq:c4_DB1}
\end{eqnarray}
(where we have used $\Lambda_{\rm QCD}=0.3\;$GeV)
to
\begin{eqnarray}
|c_4| &=& \CO(\alpha^2) \approx \Lambda_\textrm{QCD}^2 
\alpha^2 
\nonumber \\
&\approx& \left\{ 
\begin{array}{l l} 
0.010  \GeV^2 & \text{on the coarse lattice} \\
0.0069 \GeV^2 & \text{on the fine lattice} \\
0.0050 \GeV^2 & \text{on the superfine lattice}
\end{array} \right.
\label{eq:c4_DB2}
\end{eqnarray}
(where we have evaluated $\alpha$ at the scale $\mu=1/a$).
Our results for $|c_4\pm \delta c_4|$ are at or
below the {\em smaller} of these estimates on all ensembles.

We have also considered the effect of constraining $c_4$ using
Bayesian fits. Specifically, following Ref.~\cite{lepage-02},
we augment the usual $\chi^2$ with an additional term:
\begin{eqnarray}
\chi^2_\textrm{aug} &=& \chi^2 + \chi^2_\textrm{prior}
\\
\chi^2_\textrm{prior} &=& \frac{( c_4 - a_4)^2}{\tilde{\sigma}^2_4}
\,.
\end{eqnarray}
Since we have no prior information on the sign of $c_4$ we set
$a_4=0$. We do have prior information on
the expected size of $c_4=b_{10}$, given
in Eq.~(\ref{eq:b6-12est}), and
we choose $\tilde{\sigma}_4$ accordingly.
In particular we consider two Bayesian fits,
one assuming that discretization errors dominate
[$\tilde{\sigma}_4 \approx \Lambda_\textrm{QCD}^2 
(a\Lambda_\textrm{QCD}^2)^2$],
the other assuming that truncation errors dominate
[$\tilde{\sigma}_4 \approx \Lambda_\textrm{QCD}^2 \alpha^2$].
We label these ``D-B1'' and ``D-B2'', respectively.\footnote{%
These correspond to D-BT7 and D-BT7-2 in our previous publications.}
The numerical values of $\tilde{\sigma}_4$ are given
in Eqs.~(\ref{eq:c4_DB1}) and (\ref{eq:c4_DB2}) above.

We expect that adding these constraints will have little impact
on the fits, given that the unconstrained fits yield values of
$c_4$ and $\delta c_4$ that are mostly consistent even with the 
stronger of the constraints. This is borne out by the results
of the fits. We show the results for the D-B1 case in 
Table~\ref{tab:fit-D-BT7}. There is no significant change in
$c_{1-3}$, but the central values, and the errors, in $c_4$ have
been reduced on some ensembles (with a corresponding increase
in the augmented $\chi^2$).
As expected, the less constrained
D-B2 fits are almost identical to the D-U fits, 
so we do not show them.

\begin{table}[htbp]
\caption{Parameters of D-B1 fits. 
  \label{tab:fit-D-BT7}}
\begin{ruledtabular}
\begin{tabular}{ l | l l l l l }
ID &  $c_1$ & $c_2$ & $c_3$ & $c_4$(GeV${}^2$) & $\chi^2_\textrm{aug}/\text{dof}$ \\ 
\hline
C1 &  0.300(25) & 0.71(23) & 0.42(23) &    0.0011(12) & 0.044(72) \\
C2 &  0.328(23) & 0.39(22) & 0.61(21) & $-$0.0013(12) & 0.068(99) \\
C3 &  0.338(12) & 0.38(12) & 0.70(12) & $-$0.0008(6)  & 0.053(70) \\
C4 &  0.326(12) & 0.48(12) & 0.59(12) & $-$0.0005(6)  & 0.057(54) \\
C5 &  0.295(23) & 0.76(21) & 0.31(21) &    0.0009(10) & 0.11(15) \\
\hline
F1 &  0.327(21) & 0.24(17) & 0.81(20) & $-$0.0007(6)  & 0.17(22) \\
F2 &  0.338(21) & 0.15(17) & 0.93(20) & $-$0.0009(6)  & 0.20(22) \\
\hline
S1 &  0.311(13) & 0.28(11) & 0.73(15) & $-$0.0003(3)  & 0.11(21)
\end{tabular}
\end{ruledtabular}
\end{table}

\begin{figure}[htbp]
  \includegraphics[width=20pc]
                  {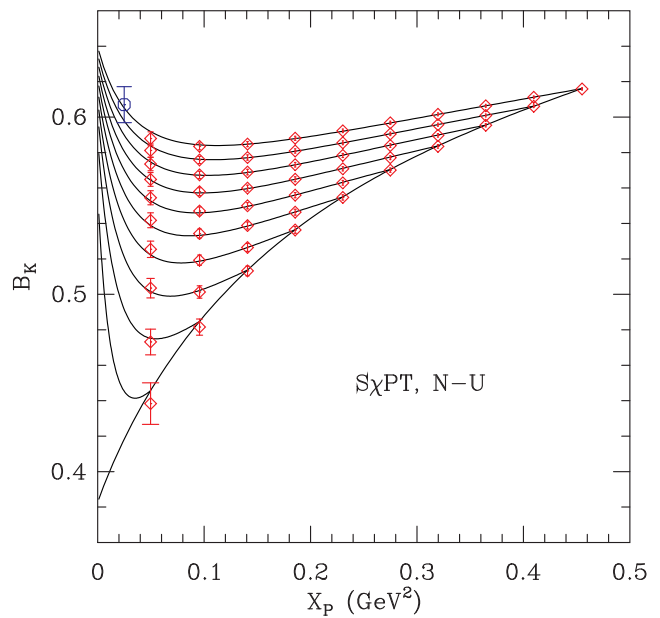}
\caption{$B_K({\rm NDR},\mu=1/a)$ as a function of $X_P$ on the C3
  ensemble. For each value of $X_P\propto m_x$, 
  the tower of points corresponds to the different values of $m_y$,
  with $m_y$ increasing from $m_x$ at the bottom to $m_y^{\rm max}=0.05/a$
  at the top. Curves show the N-U fit, both for fixed values
  of $m_y$ and for $m_x=m_y$. Error bands are not included
  for the sake of clarity. The (blue) octagon
  gives $B_K(1/a)$ evaluated at the physical kaon mass {\em with 
  lattice artifacts removed} (as described in the text).
  \label{fig:fit-N-T2}}
\end{figure}

\bigskip
%
\begin{table*}[htbp]
\caption{Parameters from N-U fits.
 The $c_i$ are defined in
  Eq.~(\ref{eq:su3-fit-non-deg-form}), and $\chi^2/\text{dof}$
  is the uncorrelated $\chi^2$ per degree of freedom. 
  The last column gives the value for $B_K(2\text{GeV})$ that results when we set
  quark masses to their physical values and remove discretization errors
  (as described in the text) and then run the resulting $B_K(1/a)$ down to
  2 GeV.
  \label{tab:fit-N-T2}}
\begin{ruledtabular}
\begin{tabular}{ l | l l l l l l l l l }
ID &  $c_1$ & $c_2$ & $c_3$ & $c_4$(GeV${}^2$) & $c_5$ & $c_6$ (GeV${}^2$)
& $c_7$(GeV${}^4$) & 
$\chi^2/\text{dof}$ & $B_K(\mu=2\text{GeV})$ \\
\hline
C1 &  0.275(38) & 0.96(37) & 0.15(37) & 0.0021(17) & 0.050(34)
   &  $-$0.0054(37) & 0.0012(5) & 0.0063(80) & 0.616(30) \\ 
C2 &  0.351(38) & 0.17(37) & 0.86(37) & $-$0.0022(18) & 0.103(36)
   &  $-$0.0051(42) & 0.0015(6) & 0.0039(38) & 0.602(33) \\ 
C3 &  0.331(12) & 0.46(12) & 0.63(12) & $-$0.0004(6) & 0.057(11) 
   &  $-$0.0025(12) & 0.0008(2) & 0.135(41) & 0.596(10) \\ 
C4 &  0.319(12) & 0.56(12) & 0.51(12) & $-$0.0001(6) & 0.062(11)
   &  $-$0.0034(12) & 0.0008(2) & 0.196(47) & 0.603(10) \\ 
C5 &  0.285(29) & 0.85(28) & 0.21(28) & 0.0014(13) & 0.069(24)
   &  $-$0.0052(27) & 0.0008(3) & 0.040(30) & 0.621(22) \\ 
\hline
F1 &  0.330(37) & 0.21(31) & 0.87(38) & $-$0.0007(10) & 0.053(25)
   &  $-$0.0017(23) & 0.0006(3) & 0.038(33) & 0.564(20) \\ 
F2 &  0.343(30) & 0.10(25) & 1.00(31) & $-$0.0010(9) & 0.039(18)
   &     0.0000(18) & 0.0003(2) & 0.034(30) & 0.550(16) \\ 
\hline
S1 &  0.316(23) & 0.23(20) & 0.80(26) & $-$0.0004(5) & 0.072(15)
   &  $-$0.0031(12) & 0.0005(2) & 0.018(22) & 0.581(12) 
\end{tabular}
\end{ruledtabular}
\end{table*}
We now turn to fits to our full data set, i.e.
to all 55 degenerate and non-degenerate points.
We first fit to the functional form (\ref{eq:su3-fit-non-deg-form})
without constraints, labeling the fits ``N-U''
(with N for non-degenerate, U for unconstrained).\footnote{%
This corresponds to N-T2 in our previous publications.} 
The parameters of the N-U fit for all ensembles are given in 
Table~\ref{tab:fit-N-T2}, and the quality of the fit is
illustrated by Figs.~\ref{fig:fit-N-T2} and ~\ref{fig:fit-N-T2:S1}.
{%
\begin{figure}[htbp]
  \includegraphics[width=20pc]
                  {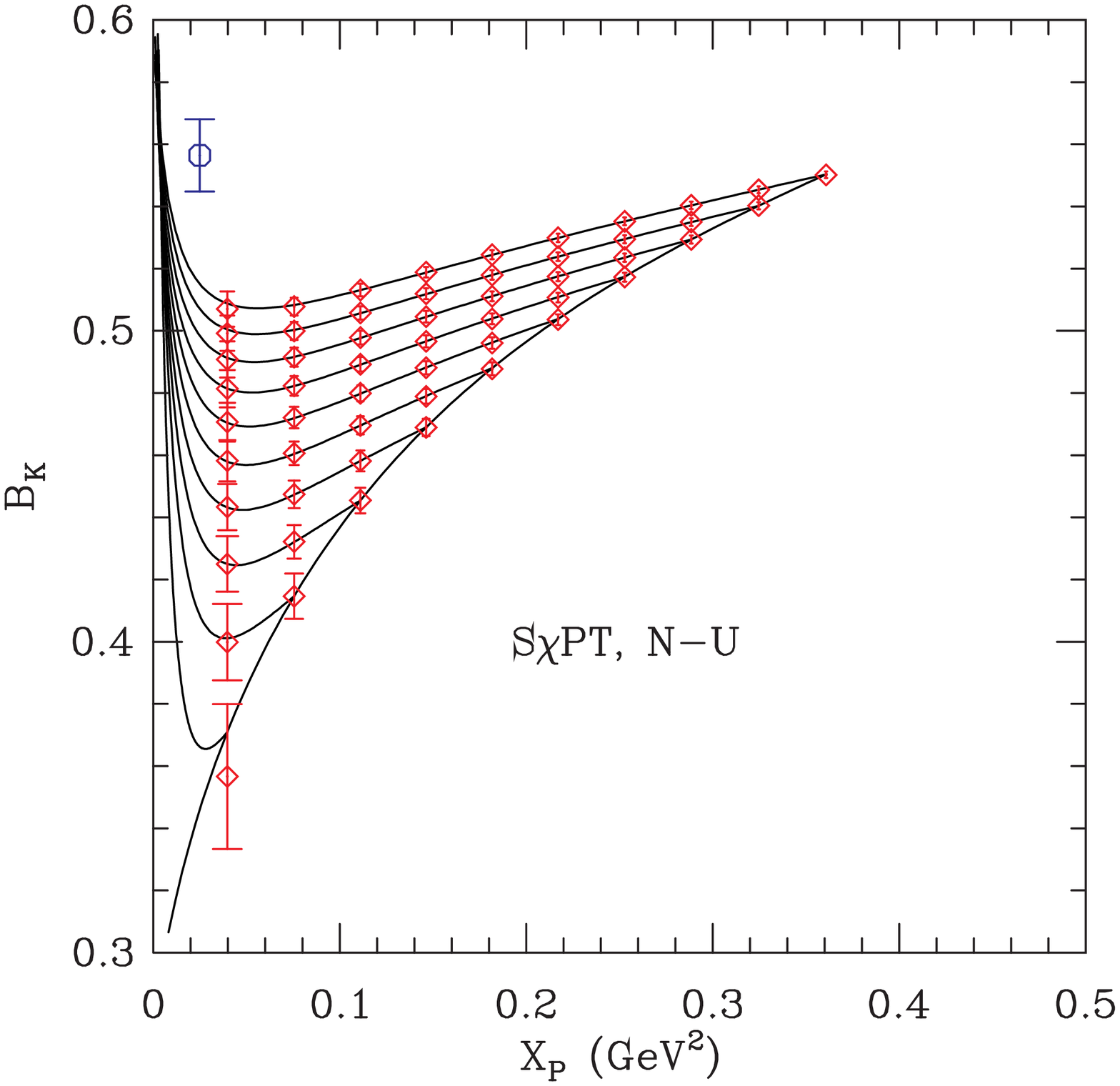}
\caption{$B_K({\rm NDR},\mu=1/a)$ as a function of $X_P$ on the S1
  ensemble. Notation as in Fig.~\ref{fig:fit-N-T2}.
  \label{fig:fit-N-T2:S1}}
\end{figure}
To better display the fit, we plot $B_K$ versus
$X_P$ rather than $G=m_K^2$. This has several advantages:
it spreads the points out, giving a sense of the quality of the entire fit;
it displays the dependence on the light valence quark mass $m_x\propto X_P$,
which is the critical parameter which must be extrapolated in order to
reach the physical point;
and it allows a more direct comparison with the SU(2) ``X-fits'' presented
below.
The fits are reasonable---as is the case on all other ensembles.
The C3 and C4 ensembles have larger $\chi^2/$dof because
the statistical errors are substantially smaller, which
brings to light the
inadequacies in our partial-NNLO fit form.\footnote{%
As noted by the MILC collaboration, more precise lattice
data requires the use of ChPT expressions of higher order,
with the order chosen so the error from the
truncation of the series is reduced to the
level of the statistical errors.}

We also include in the figure the value of $B_K(NDR,\mu=1/a)$ extrapolated
to the physical quark masses
using the fit function, {\em with all 
taste-breaking discretization and truncation errors removed}.\footnote{%
In detail, we remove the terms induced by
lattice artifacts by setting $c_4=c_6=c_7=0$,
remove all taste splittings in sea and valence mesons
($\Delta_\TB=0$),
and choose physical valence-quark masses by setting 
$Y_\TB= m_{s\bar s,\rm phys}^2=(0.6858\,{\rm GeV})^2$~\cite{Lepagemss},
$G=K_\TB=m_{K_0,\rm phys}^2$, 
and 
$X_\TB = 2 m_{K_0,\rm phys}^2- m_{s\bar s,\rm phys}^2$
(the latter assuming the observed linear dependence on quark masses,
Eqs.~(\ref{eq:pion-mass-sq-1})).
In addition we choose physical sea-quark masses by setting
$L=m_{\pi_0,\rm phys}^2$ and $S=m_{K_0,\rm phys}^2$.
Here we are assuming Dashen's theorem, namely that there are
no electromagnetic corrections to the neutral pion and kaon masses.
Other choices for tuning the quark masses lead to results that
differ by much less than the statistical errors.}
This ``physical'' $B_K$ is plotted at the physical value of $X_P$.
We also include physical $B_K$ values in Table~\ref{tab:fit-N-T2},
and subsequent tables for other fits,
except that in the tables we run them to a common scale, $\mu=2\;$GeV, 
so as to allow comparison between ensembles.

We stress that we are only able to remove taste-breaking discretization
errors because we have used a {\em staggered} ChPT fit in which such breaking
is explicit. We also note that taste-conserving discretization and
truncation errors are not removed by this procedure,
and must be dealt with separately.

The values of the parameters $c_1-c_4$ given in Table~\ref{tab:fit-N-T2}
are consistent with those from the fits to the degenerate data alone.
This is a non-trivial result, suggesting that our truncation of the full
NLO SU(3) expression down to 7 parameters can adequately describe the data.
The ``continuum non-degenerate'' parameter $c_5$ appears to be non-zero
(this is most significant on the high-statistics ensembles C3 and C4),
and is consistent across all ensembles.
The ``lattice non-degenerate'' parameters $c_6$ and $c_7$ appear also
to be non-zero on the high-statistics ensembles C3 and C4,
although they are consistent with zero on other ensembles.
Our errors are too large to tell whether $c_6$ and $c_7$ show
the expected decrease as the continuum limit is approached.

The most striking result, however, is that $c_5$ is very small.
It is expected to be of order unity, but turns out to be more
than an order of magnitude smaller.
We do not know of an explanation for this smallness.
Indeed, its value is not of direct physical significance,
since it depends on the scale chosen in the chiral logarithms 
(which we have taken to be $\mu_\textrm{DR}=0.77\;$GeV).
Nevertheless, there is an important practical consequence of 
our finding that $c_5$ is small. If $c_5$ vanished, 
then fits to the degenerate data alone would allow a determination
of the {\em physical} $B_K$, since $F_5$ is
the only {\em continuum} term in our fit function that vanishes
for degenerate quark masses. Since we find $c_5$ to be very small,
the degenerate data must play an important role in 
determining the extrapolated, physical $B_K$.

The magnitude of the parameter $c_6$ is expected to lie in the same
range as $c_4$, namely from
$\Lambda_\textrm{QCD}^2 (a \Lambda_\textrm{QCD})^2$
(discretization errors)
to $\Lambda_\textrm{QCD}^2 \alpha^2$ (truncation errors).
Numerical values for these ranges are given in Eqs.~(\ref{eq:c4_DB1}) 
and (\ref{eq:c4_DB2}) above. We see from Table~\ref{tab:fit-N-T2}
that some of the N-U fit values of $c_6$ exceed the smaller of
these expectations, though not the larger.
As for $c_7$, if discretization errors dominate we expect
\begin{eqnarray}
c_7 &\approx&
\Lambda_\textrm{QCD}^4 (a \Lambda_\textrm{QCD})^2 
\nonumber \\
\label{eq:B1constraint}
&\approx& \left\{ 
\begin{array}{l l} 
0.00027 \GeV^4 & \text{(coarse)} \\
0.00014 \GeV^4 & \text{(fine)} \\
0.000068 \GeV^4 & \text{ (superfine)}
\end{array} \right.
\end{eqnarray}
while if truncation errors dominate we expect
\begin{eqnarray}
c_7 &\approx& \Lambda_\textrm{QCD}^4 \alpha_s^2 
\nonumber \\
&\approx& \left\{ 
\begin{array}{l l} 
0.00090 \GeV^4 & \text{(coarse)} \\
0.00062 \GeV^4 & \text{(fine)} \\
0.00045 \GeV^4 & \text{(superfine)}
\end{array} \right.
\end{eqnarray}
In this case, some of the fit values of $c_7$ are larger than
both of these expectations.

We have thus carried out Bayesian fits simultaneously
enforcing the expected
sizes of $c_4$, $c_6$ and $c_7$.
As in the degenerate case, 
$\chi^2$ is augmented by
\begin{equation}
\chi^2_\textrm{prior} = \sum_{i=4,6,7} 
\frac{( c_i - a_i)^2}{\tilde{\sigma}^2_i}\,.
\end{equation}
We set $a_{4,6,7}=0$, and make two choices for the $\tilde{\sigma}_i$.
The first assumes that discretization errors dominate,
\begin{equation}
\tilde{\sigma}_4 = \tilde{\sigma}_6 = 
\Lambda_\textrm{QCD}^2 (a \Lambda_\textrm{QCD})^2 \,,\quad
\tilde{\sigma}_7 = 
\Lambda_\textrm{QCD}^4 (a \Lambda_\textrm{QCD})^2 \,,
\label{eq:B1constraint-1}
\end{equation}
while the second assumes that truncation errors dominate,
and leads to the weaker constraints:
\begin{equation}
\tilde{\sigma}_4 =\tilde{\sigma}_6 
= \Lambda_\textrm{QCD}^2 \alpha^2 \,,\quad
\tilde{\sigma}_7 =
\Lambda_\textrm{QCD}^4 \alpha^2\,.
\label{eq:B2constraint}
\end{equation}
We label these fits N-B1 and N-B2, respectively.\footnote{%
They correspond respectively to N-BT8 and N-BT8-2 in our 
previous publications.}
\begin{figure}[htbp]
\includegraphics[width=20pc]
                {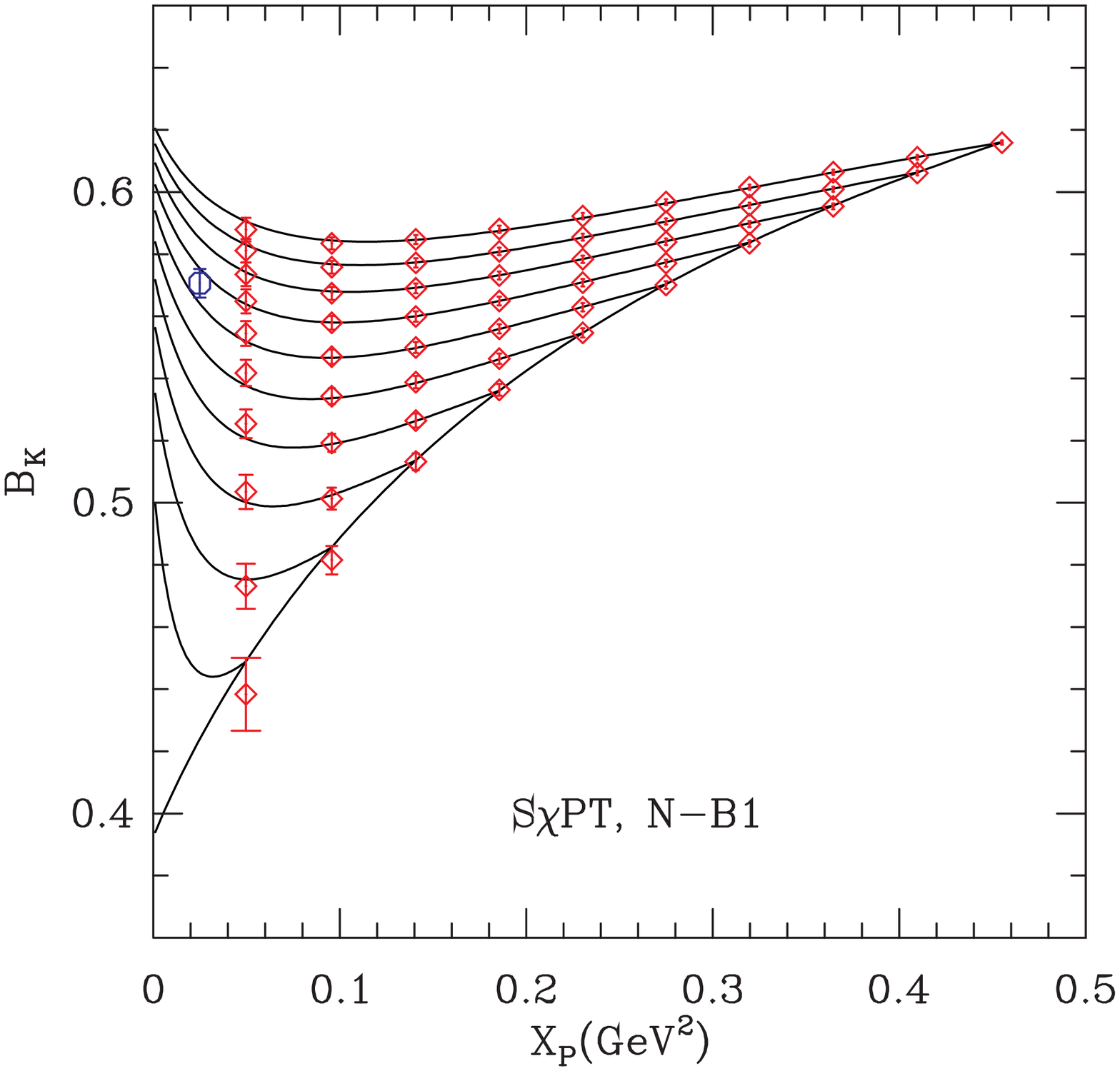}
\caption{
$B_K(1/a)$ vs. $X_P$ on the
 C3 ensemble, showing the N-B1 fit. 
Notation as in Fig.~\protect\ref{fig:fit-N-T2}.
  \label{fig:fit-N-BT8}}
\end{figure}
\begin{table*}[htbp]
\caption{Parameters of N-B1 fits. Notation as in
Table~\protect\ref{tab:fit-N-T2}.
  \label{tab:fit-N-BT8}}
\begin{ruledtabular}
\begin{tabular}{ l | l l l l l l l l l }
ID &  $c_1$ & $c_2$ & $c_3$ & $c_4$ & $c_5$ & $c_6$ & $c_7$ & 
$\chi^2_\textrm{aug}/\text{dof}$ & $B_K(\mu=2\,\text{GeV})$ \\
\hline
C1 &  0.265(30) & 1.05(29) & 0.06(29) & 0.0026(14) & $-$0.033(23)
   &  0.0021(9) & 0.0002(1) & 0.067(36) & 0.557(12) \\ 
C2 &  0.312(31) & 0.56(30) & 0.46(30) & $-$0.0005(15) & $-$0.003(27)
   &  0.0035(9) & 0.0003(1) & 0.106(53) & 0.540(13) \\ 
C3 &  0.321(12) & 0.55(12) & 0.53(12) & 0.0000(6) & 0.005(7) 
   &  0.0017(3) & 0.0004(1) & 0.22(40) & 0.560(5) \\ 
C4 &  0.309(12) & 0.65(12) & 0.40(12) & 0.0003(6) & 0.004(7)
   &  0.0013(3) & 0.0004(1) & 0.283(41) & 0.562(4) \\ 
C5 &  0.275(27) & 0.94(25) & 0.11(26) & 0.0017(12) & $-$0.006(19)
   &  0.0009(7) & 0.0002(1) & 0.072(39) & 0.567(10) \\ 
\hline
F1 &  0.306(29) & 0.42(24) & 0.61(29) & $-$0.0001(8) & 0.008(19)
   &  0.0014(5) & 0.0001(0) & 0.103(57) & 0.533(10) \\ 
F2 &  0.327(26) & 0.24(22) & 0.82(26) & $-$0.0006(8) & 0.020(17)
   &  0.0012(4) & 0.0001(0) & 0.080(51) & 0.539(9) \\ 
\hline
S1 &  0.296(18) & 0.41(16) & 0.55(21) & 0.0000(4) & 0.011(12)
   &  0.0005(2) & 0.0001(0) & 0.094(56) & 0.533(7) \\ 
\end{tabular}
\end{ruledtabular}
\end{table*}
\begin{table*}[htbp]
\caption{Parameters of N-B2 fits. Notation as in
Table~\protect\ref{tab:fit-N-T2}. 
  \label{tab:fit-N-BT8-2}}
\begin{ruledtabular}
\begin{tabular}{ l | l l l l l l l l l }
ID &  $c_1$ & $c_2$ & $c_3$ & $c_4$ & $c_5$ & $c_6$ & $c_7$ & 
$\chi^2_\textrm{aug}/\text{dof}$ & $B_K(\mu=2\text{GeV})$ \\
\hline
C1 &  0.263(38) & 1.07(37) & 0.03(37) & 0.0026(17) & $-$0.027(20)
   &  0.0010(11) & 0.0004(2) & 0.021(12) & 0.562(14) \\ 
C2 &  0.333(39) & 0.35(38) & 0.67(38) & $-$0.0014(18) & 0.003(23)
   &  0.0030(11) & 0.0006(2) & 0.028(15) & 0.533(14) \\ 
C3 &  0.328(12) & 0.48(12) & 0.60(12) & $-$0.0003(6) & 0.041(8) 
   &  $-$0.0012(9) & 0.0007(1) & 0.151(39) & 0.585(8) \\ 
C4 &  0.316(12) & 0.58(12) & 0.48(12) & 0.0000(6) & 0.045(8)
   &  $-$0.0020(9) & 0.0007(1) & 0.214(44) & 0.591(8) \\ 
C5 &  0.277(30) & 0.93(29) & 0.13(29) & 0.0016(13) & 0.019(13)
   &  $-$0.0013(11) & 0.0004(2) & 0.050(31) & 0.585(11) \\ 
\hline
F1 &  0.321(38) & 0.28(32) & 0.77(39) & $-$0.0005(11) & 0.011(15)
   &  0.0010(7) & 0.0003(1) & 0.047(33) & 0.532(8) \\ 
F2 &  0.340(31) & 0.13(26) & 0.96(31) & $-$0.0009(9) & 0.024(11)
   &  0.0010(7) & 0.0002(1) & 0.038(32) & 0.539(6) \\ 
\hline
S1 &  0.309(23) & 0.30(20) & 0.71(27) & $-$0.0003(5) & 0.038(8)
   &  $-$0.0012(6) & 0.0003(1) & 0.036(23) & 0.554(5) \\ 
\end{tabular}
\end{ruledtabular}
\end{table*}

The N-B1 fit on the C3 ensemble is shown in Fig.~\ref{fig:fit-N-BT8}.
The fit appears to be of comparable quality to that 
in Fig.~\ref{fig:fit-N-T2}---the
somewhat larger $\chi^2$ indicates, however, 
that the constraints are having a non-trivial impact.
The most notable changes compared to Fig.~\ref{fig:fit-N-T2} are
that the curvature at small $X_P$ (and fixed $m_y$) is smaller,
and that the value of $B_K(1/a)$ is reduced. 
Similar changes are seen on all ensembles.

The parameters from the N-B1 and N-B2 fits are given in Tables
\ref{tab:fit-N-BT8} and \ref{tab:fit-N-BT8-2}. The values of
$c_1-c_4$ from the two fits are consistent with each other and with those from
the N-U fits. For several ensembles, however,
$c_5$ is not consistent between the fits, most notably on the
C3 and C4 ensembles. This suggests that the errors are underestimated
and/or that the truncated fit function does not fully represent our
data. We conclude only that $|c_5|\ll 1$.

Our tentative conclusion of a non-zero $c_6$ from the N-U fit
is not confirmed by the two new fits. Results of both signs are
found. On most ensembles the results from the three fits are consistent
with each other, and with a vanishing result. On C3 and C4, there are
some marginal inconsistencies. We conclude that we cannot reliably
determine $c_6$.
The situation for $c_7$ is slightly better, since all three fits
are consistent within 2-$\sigma$, and all have the same sign.

The results for $B_K(2\;{\rm GeV})$ from the three fits
(N-U, N-B1 and N-B2) are not consistent. The differences
are most significant on the C3 and C4 ensembles, where they
exceed 3-$\sigma$. This is caused by the poor determination
of some of the fit parameters, particularly
$c_5$ and $c_6$. Although all the fits give a reasonable
representation of the data, the results obtained after
extrapolation to physical quark masses and removal of taste-breaking 
differ. This is because $c_5$ and $c_6$ terms have significantly different
dependence on quark masses and taste breaking.

To do better, we recall that, because of the smallness of $c_5$,
we can almost determine the {\em physical} $B_K$ using the
degenerate data alone. Furthermore, the D-U, D-B1 and D-B2 fits
give consistent results for $c_{1-4}$, which we take to indicate
that these results are reliable.
Thus, in our last stage of SU(3) fitting, we use the results
from the degenerate fits as constraints on the parameters
$c_{1-4}$ when doing fits to the full data set.
We dub these ``double Bayesian'' fits, and think that they
provide the most reliable way of determining the parameters
$c_5$ and $c_6$.

In more detail,
we first do a D-B1 or D-B2 fit to degenerate points as described above.
The central values and errors from these fits are then
used as constraints on $c_{1-4}$ in a second fit to the full data-set,
along with the constraints on $c_{6,7}$ just described.
In other words, if the first fit yields
$a_i \pm \tilde\sigma_i$ for $i=1,4$, then the second fit augments the
$\chi^2$ by
\begin{subequations}
\begin{align}
\chi^2_\textrm{prior} & = 
\chi^2_\textrm{prior (1)} + \chi^2_\textrm{prior (2)}
\\
\chi^2_\textrm{prior (1)} & = \sum_{i=1}^{4} 
\frac{( c_i - a_i)^2}{\tilde{\sigma}^2_i}
\\
\chi^2_\textrm{prior (2)} & = \sum_{i=6,7} 
\frac{( c_i )^2}{\tilde{\sigma}^2_i}
\end{align}
\end{subequations}
with $\tilde{\sigma}_{6,7}$ given either by
Eq.~(\ref{eq:B1constraint-1}), if the first fit
was D-B1, or by (\ref{eq:B2constraint}) otherwise.
We label these fits N-BB1 and N-BB2, respectively.\footnote{%
These correspond respectively to the N-BT7  and N-BT7-2 fits
in our previous publications.}

Examples of the resulting fits are shown in Figs.~\ref{fig:fit-N-BT7}
and \ref{fig:fit-N-BT7-2}, and fit parameters from all ensembles
are collected in Tables~\ref{tab:fit-N-BT7} and \ref{tab:fit-N-BT7-2}.
\begin{figure}[htbp]
  \includegraphics[width=20pc]
                  {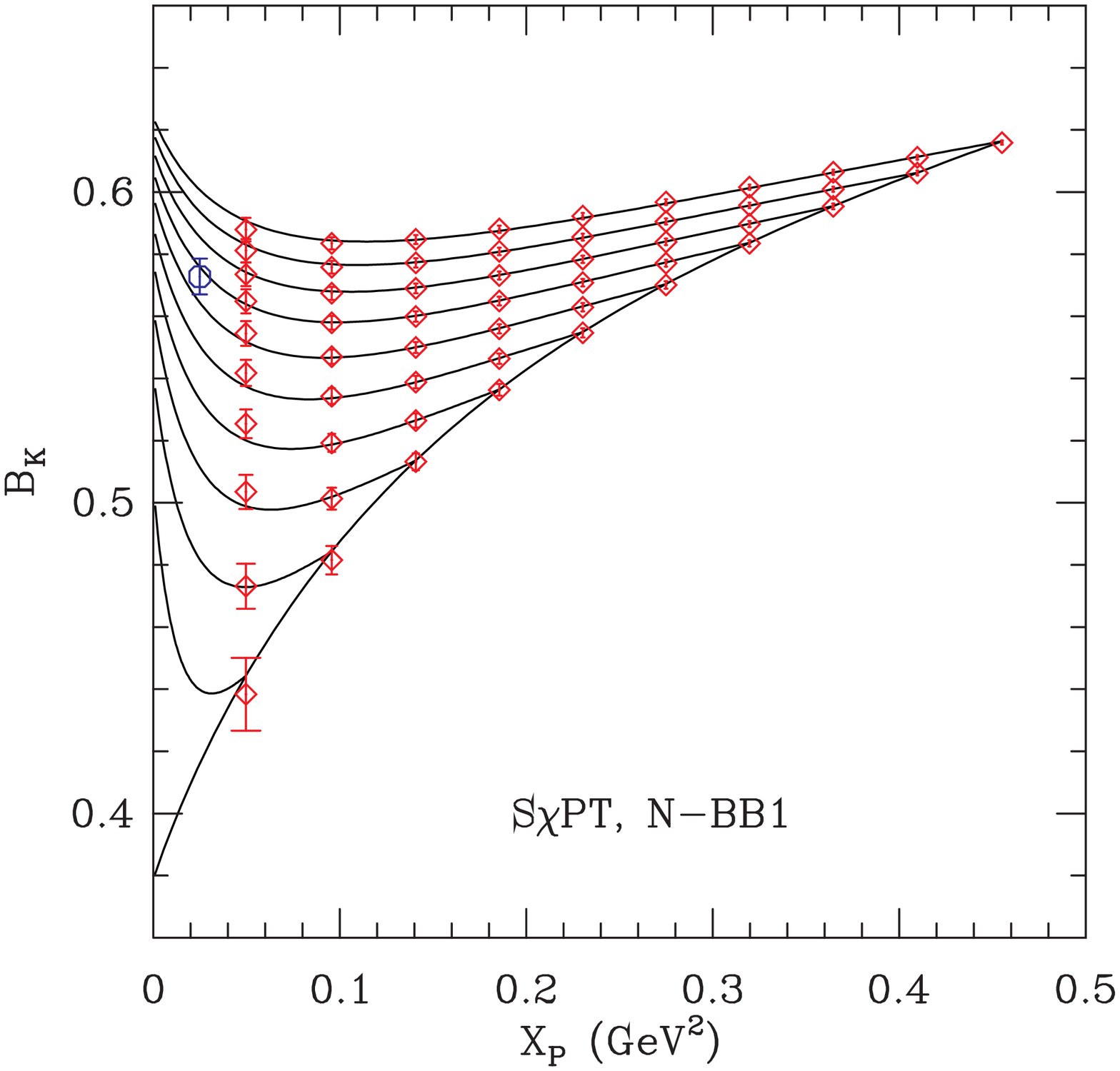}
  \caption{
$B_K(1/a)$ vs. $X_P$ on the
 C3 ensemble, showing the N-BB1 fit. 
Notation as in Fig.~\protect\ref{fig:fit-N-T2}.
   \label{fig:fit-N-BT7}}
\end{figure}
\begin{table*}[htbp]
\caption{Fit parameters for N-BB1 fits. Notation is as in
Table~\protect\ref{tab:fit-N-T2}. 
  \label{tab:fit-N-BT7}}
\begin{ruledtabular}
\begin{tabular}{ l | l l l l l l l l l }
ID &  $c_1$ & $c_2$ & $c_3$ & $c_4$ & $c_5$ & $c_6$ & $c_7$ & 
$\chi^2_\textrm{aug}/\text{dof}$ & $B_K(\mu=2\text{GeV})$ \\
\hline
C1 &  0.2944(54) & 0.767(45) & 0.386(36) & 0.0015(5) & $-$0.017(15)
   &  0.0023(9) & 0.0002(1) & 0.084(61) & 0.555(12) \\ 
C2 &  0.3269(51) & 0.409(41) & 0.623(34) & $-$0.0011(6) & 0.004(16)
   &  0.0036(8) & 0.0003(1) & 0.108(54) & 0.538(12) \\ 
C3 &  0.3345(28) & 0.418(27) & 0.677(20) & $-$0.0006(2) & 0.017(5) 
   &  0.0013(5) & 0.0004(1) & 0.250(55) & 0.562(6) \\ 
C4 &  0.3229(27) & 0.514(26) & 0.560(19) & $-$0.0003(2) & 0.016(5)
   &  0.0009(5) & 0.0004(1) & 0.320(72) & 0.564(5) \\ 
C5 &  0.2913(51) & 0.790(43) & 0.285(32) & 0.0011(5) & 0.003(12)
   &  0.0009(6) & 0.0002(1) & 0.078(61) & 0.567(10) \\ 
\hline
F1 &  0.3243(47) & 0.253(28) & 0.825(35) & $-$0.0005(3) & 0.015(12)
   &  0.0015(5) & 0.0001(0) & 0.111(55) & 0.535(9) \\ 
F2 &  0.3367(49) & 0.155(32) & 0.939(30) & $-$0.0008(3) & 0.024(10)
   &  0.0013(4) & 0.0001(0) & 0.078(45) & 0.540(8) \\ 
\hline
S1 &  0.3088(33) & 0.293(19) & 0.727(28) & $-$0.0002(2) & 0.016(7)
   &  0.0006(2) & 0.0001(0) & 0.105(63) & 0.535(6) \\
\end{tabular}
\end{ruledtabular}
\end{table*}
\begin{figure}[htbp]
\includegraphics[width=20pc]
                {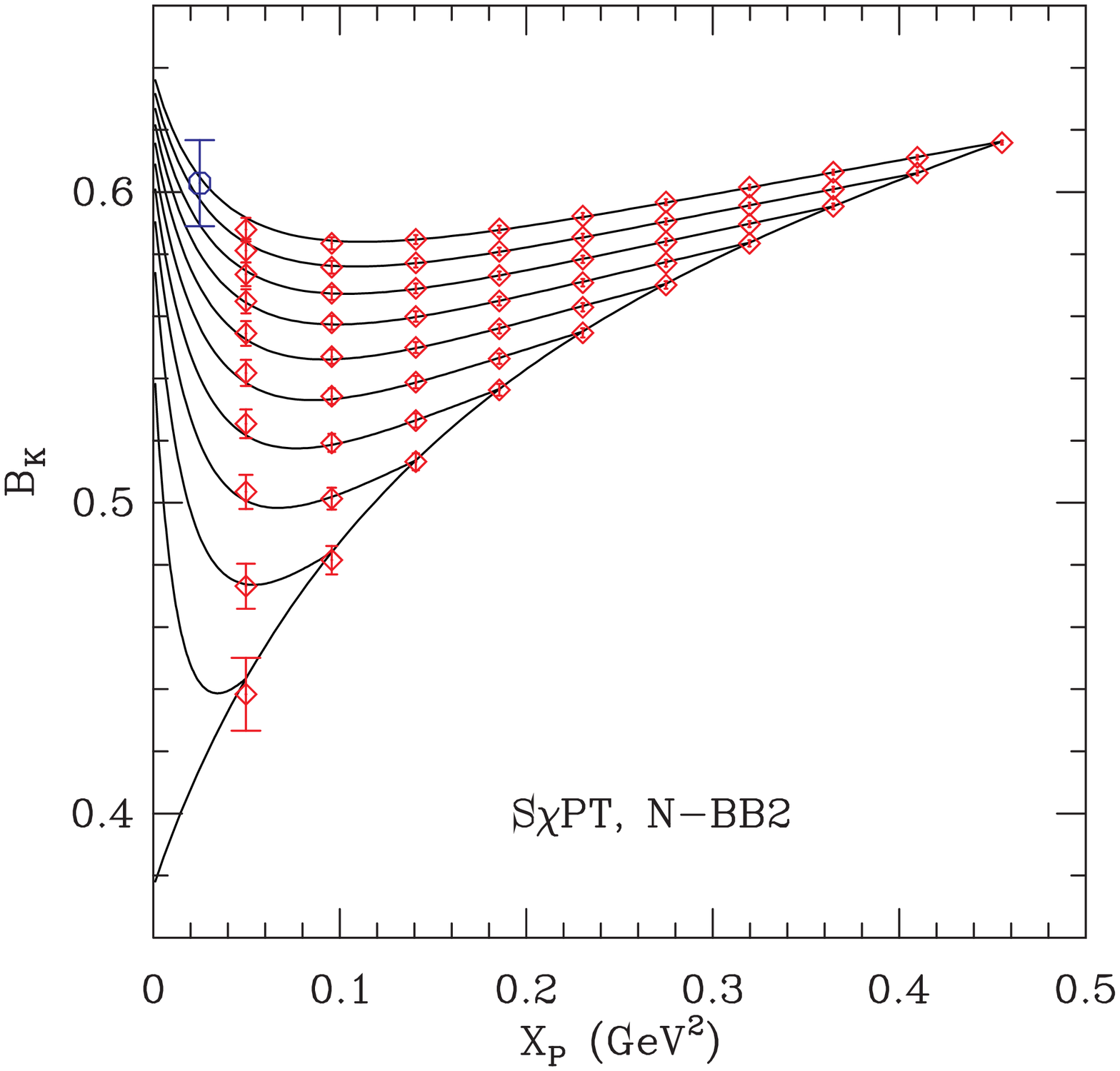}
\caption{
$B_K(1/a)$ vs. $X_P$ on the
 C3 ensemble, showing the N-BB2 fit. 
Notation as in Fig.~\protect\ref{fig:fit-N-T2}.
  \label{fig:fit-N-BT7-2}}
\end{figure}
\begin{table*}[htbp]
\caption{Fit parameters for N-BB2 fits. Notation is as in
Table~\protect\ref{tab:fit-N-T2}. 
  \label{tab:fit-N-BT7-2}}
\begin{ruledtabular}
\begin{tabular}{ l | l l l l l l l l l }
ID &  $c_1$ & $c_2$ & $c_3$ & $c_4$ & $c_5$ & $c_6$ & $c_7$ & 
$\chi^2_\textrm{aug}/\text{dof}$ & $B_K(\mu=2\text{GeV})$ \\
\hline
C1 &  0.2801(92) & 0.905(86) & 0.217(67) & 0.0019(7) & $-$0.013(15)
   &  0.0006(17) & 0.0005(2) & 0.026(26) & 0.564(17) \\ 
C2 &  0.3444(87) & 0.235(82) & 0.795(59) & $-$0.0019(7) & 0.013(15)
   &  0.0027(17) & 0.0007(2) & 0.029(20) & 0.535(17) \\ 
C3 &  0.3369(27) & 0.395(26) & 0.700(19) & $-$0.0007(2) & 0.056(15) 
   &  $-$0.0020(15) & 0.0008(2) & 0.164(42) & 0.592(14) \\ 
C4 &  0.3251(26) & 0.493(25) & 0.580(18) & $-$0.0004(2) & 0.060(14)
   &  $-$0.0029(15) & 0.0008(2) & 0.227(56) & 0.598(13) \\ 
C5 &  0.2851(68) & 0.850(61) & 0.213(50) & 0.0013(5) & 0.027(17)
   &  $-$0.0017(20) & 0.0004(2) & 0.052(40) & 0.588(19) \\ 
\hline
F1 &  0.3409(92) & 0.112(69) & 0.996(65) & $-$0.0010(4) & 0.025(12)
   &  0.0006(12) & 0.0004(2) & 0.055(31) & 0.539(12) \\ 
F2 &  0.3511(76) & 0.030(56) & 1.095(50) & $-$0.0012(4) & 0.034(13)
   &  0.0006(14) & 0.0003(2) & 0.042(28) & 0.545(13) \\ 
\hline
S1 &  0.3189(63) & 0.207(46) & 0.843(52) & $-$0.0005(2) & 0.048(11)
   &  $-$0.0015(10) & 0.0003(1) & 0.042(25) & 0.560(11) \\ 
\end{tabular}
\end{ruledtabular}
\end{table*}

The values of the $c_i$, the augmented $\chi^2$, and $B_K$, are
consistent between fits N-B1 and N-BB1, and between N-B2 and N-BB2.
The major changes induced by using two levels of constraints are that
the errors in $c_1-c_3$ are reduced, while those in $B_K$ are increased.
The former is expected, since we are placing constraints on $c_1-c_3$
in the second stage of fitting. 
The latter is indicative of the uncertainties in the fit
parameters induced by discretization errors, particularly in $c_6$.
With $c_1-c_3$ pinned down, a greater range of $c_6$ values are explored,
leading to more variation in the artifact subtraction needed
to obtain the physical $B_K$.
Since this variation represents a genuine systematic error in
our procedure, and is best represented by the N-BB fits,
we choose the N-BB fits for our central values. 
Specifically, we choose fit N-BB1, using N-BB2
and N-B1 to estimate different fitting systematics.

\begin{table*}[htbp]
\caption{Breakdown of contributions to $B_K$ 
in the SChPT N-BB1 and N-BB2 fits on the C3 and S1 ensembles.
Valence quark masses are set to their physical values.
  \label{tab:BKchpt-C3}}
\begin{ruledtabular}
\begin{tabular}{l l | l | l l l l l}
Ens. & Fit & $B_K(1/a)$ & LO & NLO-$c_{1-2}$ 
                  &  NLO-$c_5$ & NLO-$c_4,c_{6,7}$& NNLO-$c_3$  \\
\hline
C3 & N-BB1 & 0.598 & 0.335& 0.157& 0.013& 0.052 & 0.042\\
C3 & N-BB2 & 0.609 & 0.337& 0.151& 0.045& 0.033 & 0.043\\
\hline
S1 & N-BB1 & 0.530 & 0.309& 0.144& 0.012& 0.020 & 0.045\\
S1 & N-BB2 & 0.529 & 0.319& 0.125& 0.038& -0.006 & 0.052\\
\end{tabular}
\end{ruledtabular}
\end{table*}

We see from tables that
the difference between the physical $B_K$ obtained from the N-BB1 and N-BB2
fits can be as large as 6\% on the coarse and superfine lattices. 
To understand this difference in more detail,
Table~\ref{tab:BKchpt-C3} gives  
a breakdown of the contributions to $B_K$ on both the C3 
and S1 ensembles.
Here, valence quark masses have been set to their physical values,
but taste-splittings have not been removed.
In other words, this is the result of using the SChPT fit form to
extrapolate the lattice data on this ensemble to the physical 
valence-quark masses.\footnote{%
We stress that the values for $B_K(1/a)$ differ from those
in Figs.~\ref{fig:fit-N-BT7} and \ref{fig:fit-N-BT7-2} because
the latter have taste splittings set to zero.
}
The chiral series shows poor convergence, with NLO:LO $\approx 2/3$,
although the NNLO term is of reasonable size
(NNLO:LO $\approx 0.13$).
We see also that the lattice artifacts (given in the
NLO-$c_{4,6,7}$ column) amount to no more than a quarter
of the total NLO contribution on the coarse ensemble,
and significantly less on the superfine ensemble.
The major difference between the two N-BB1 and N-BB2 fits is in the relative
size of these lattice artifacts and the continuum $c_5$ contribution.
Since the artifacts are removed when we quote a physical value,
it is this difference which is mainly responsible for
the differences one sees in the physical $B_K$ between
the N-BB1 and N-BB2 fits. Given our data, it appears
difficult to resolve this uncertainty.

We close this subsection by further illustrating the impact of
lattice artifacts.
Figures~\ref{fig:NBB1} and ~\ref{fig:NBB1S1}
show the dependence of $B_K$ on $X_P \propto m_x$
for the N-BB1 fit on the C3 and S1 ensembles,
with $m_y$ set to the physical strange-quark mass.
The three curves on each plot
show the fit function itself (solid), the
fit function with lattice-artifact terms removed (dashed),
and the fit function with 
all taste-violating lattice artifacts removed, 
and sea-quark masses set to their physical values (dotted).
The points with errors indicate the resulting ``physical'' $B_K(1/a)$.
Only the ``SU(2) regime'' is shown, for which $m_x/m_s^{\rm phys}<1/2$.
We observe that the size
of the total taste-violating discretization errors 
(the difference between the solid and [red] dotted curves) 
is moderate on the course lattices 
(6\% or less for $m_x\ge m_x^{\rm phys}$), 
and somewhat reduced on the superfine lattices. 

We also observe that the curvature in the fit functions
is almost entirely due to the lattice-artifact terms. Removing these
gives the (blue) dashed curves, which show much less curvature.
The chiral logarithms that remain if lattice-artifact terms are removed
are proportional to $\log(X_I)$,
and thus do not diverge in the chiral limit because $X_I$ is 
non-vanishing in this limit due to taste splitting.
After removing taste splitting (giving the [red] dotted curves), 
there is a logarithmically divergent
term (an SU(2) partially quenched chiral logarithm), but it has a
small coefficient and is barely visible.
\begin{figure}[htbp]
\includegraphics[width=20pc]
                {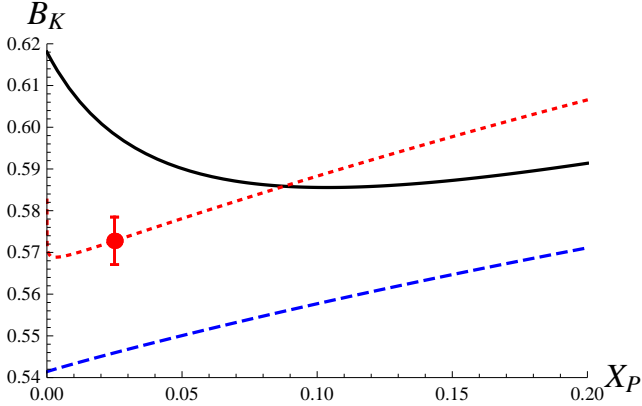}
\caption{
$B_K(1/a)$ vs. $X_P$ (in GeV${}^2$) on the
C3 ensemble from the N-BB1 fit. The solid (black) curve shows
the fit function with $m_y=m_s^{\rm phys}$, the dashed (blue) curve
shows the result of removing the lattice-artifact terms
(i.e. those proportional to $c_4$, $c_6$ and $c_7$),
and the dotted (red) curve shows the ``physical'' $B_K$, 
i.e. with lattice artifacts and taste-splittings removed, and
with the sea-quarks set to their physical values. The data-point
shows the result for $m_x=m_d^{\rm phys}$.
  \label{fig:NBB1}}
\end{figure}
\begin{figure}[htbp]
\includegraphics[width=20pc]
                {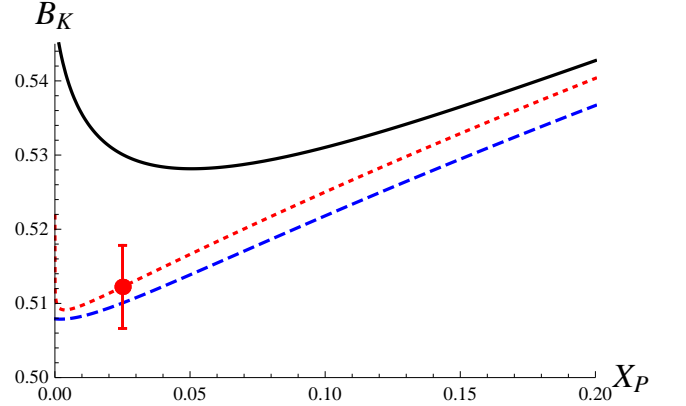}
\caption{
As in Fig.~\protect\ref{fig:NBB1} but for the S1 ensemble.
  \label{fig:NBB1S1}}
\end{figure}

\bigskip
We summarize our experience with SU(3) fitting as follows.
The fits are reasonably successful at reproducing the data,
but have two major drawbacks. 
First, the convergence of the chiral series is suspected
for physical quark masses, which undermines our use of SChPT to
remove unphysical lattice artifacts.
Second, our fits do not determine the size of the lattice-artifact 
terms, leading to significant differences between the
``physical'' results from different fits.
To some extent these two drawbacks ``cancel'', and we can hope
that the systematic error we estimate by comparing fits takes
into account the uncertainties in removing the artifacts.
Nevertheless, the reliance on SU(3) SChPT is unsatisfactory.

\subsection{SU(2) fitting \label{ssec:fit-su2}}
%
We now turn to our second fitting strategy, in which we
fix $m_y$ close to $m_s^{\rm phys}$, extrapolate $m_x$ to the
physical light-quark mass, and then extrapolate $m_y$ to the
physical strange quark mass. At all stages we need 
$m_x \ll m_y\sim m_s^{\rm phys}$
in order that SU(2) S$\chi$PT can be used.
The fitting function is given in Eqs.~(\ref{eq:su2-fit-func}-\ref{eq:q_4}), 
and has the key feature that there are no terms arising solely
from discretization or truncation errors.

On each ensemble, we  extrapolate to $m_s^{\rm phys}$
using our three heaviest $y$ quarks (e.g. 
$a m_y = \{0.04, 0.045, 0.05\}$ on the coarse lattices).
We label this choice ``3Y''.
The extent of the extrapolation varies substantially between the
ensembles, as can be seen by comparing
Tables~\ref{tab:val-qmass} and ~\ref{tab:physical-qmass}.
Our largest values of $a m_y$ lie 
3\%, 15\% and 24\% below the physical strange quark mass, 
on coarse, fine and superfine lattices, respectively.

For the chiral extrapolation we use either our lightest 4 or 5
values of $m_x$, choices we label 4X and 5X, respectively.
For the ``4X3Y'' choice, which leads to our preferred fit, the
SU(2) expansion parameter satisfies $m_x/m_y \le 1/2$.

For fixed sea-quark masses, the fitting function
(\ref{eq:su2-fit-func}) has three terms.
The first two are present at NLO, while the third is of NNLO. 
We have thus tried two types of ``X-fits'': NLO (two parameter)
and NNLO (three parameter).

We begin by considering the 4X3Y-NLO fits.
The first stage is the ``X-fit'', which we do separately
for the 3 largest values of $a m_y$. The fits for the
largest $a m_y$ on ensembles C3 and S1 are shown, respectively,
in Figs.~\ref{fig:su2-fit-4X3Y-NLO-Xfit:C3} and 
\ref{fig:su2-fit-4X3Y-NLO-Xfit:S1}.
\begin{figure}[htbp]
\includegraphics[width=20pc]
                {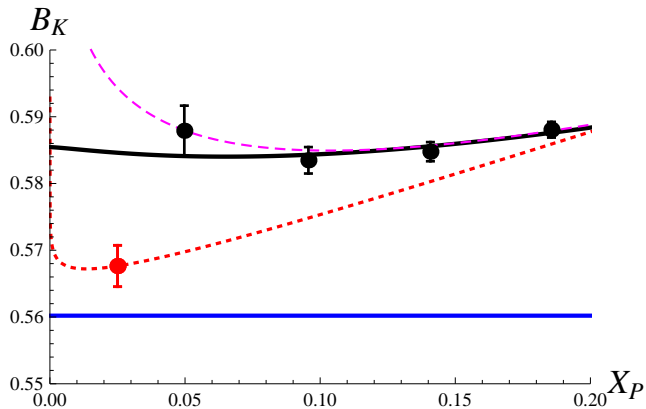}
\caption{$B_K(1/a)$ versus $X_P$ on coarse ensemble C3, for $a m_y=0.05$.
  The 4X NLO fit is shown by the solid (black) line.
  The horizontal (blue) line gives the contribution of the LO term. 
  The (magenta) dashed curve has taste-splittings removed, while the (red) 
  dotted curve shows the result of removing taste-splittings and setting the
  the light sea-quark to its physical value. The (red) point on this line
  is at $X_P = m_{\pi_0}^2$ and shows the errors from the fit. 
  \label{fig:su2-fit-4X3Y-NLO-Xfit:C3}}
\end{figure}
\begin{figure}[htbp]
\includegraphics[width=20pc]
                {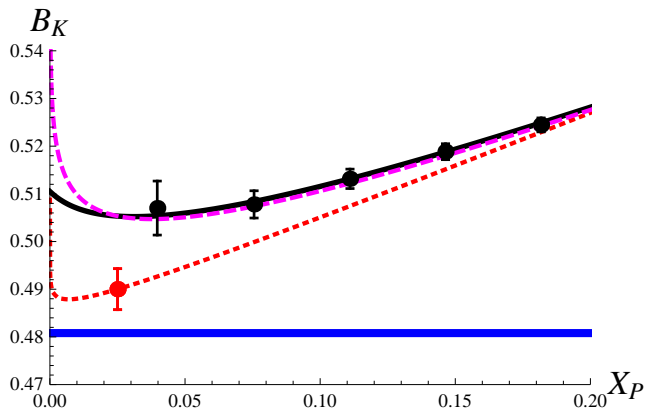}
\caption{$B_K(1/a)$ versus $X_P$ on the superfine ensemble S1, 
  for $a m_y=0.018$.  The data point at the largest $X_P$ is not included
  in the fit.
  Notation as in Fig.~\ref{fig:su2-fit-4X3Y-NLO-Xfit:C3}.
  \label{fig:su2-fit-4X3Y-NLO-Xfit:S1}}
\end{figure}
The (mild) curvature in the fit function
is due to the chiral logarithms. As in the SU(3) fit,
these do not diverge in the chiral limit because of taste-breaking.
The figures also show the relative size of the LO and NLO contributions.
The NLO contribution (difference between the solid [black]
and horizontal [blue] lines) is small, 
not exceeding $10\%$ over the range of the fits.
The (magenta) dashed curves show the impact of removing taste-splittings
from the fit function, so that the chiral logarithms do diverge.
The impact is seen to be small in the region of our data,
most notably on the superfine lattice.
The (red) dotted curves show the form in the continuum limit,
with physical light sea-quarks, as predicted by the fit.
The single point on these curves
shows the result (including errors from the fit) for
$m_x=m_d^{\rm phys}$.
The difference between the fit and these ``continuum'' curves
is seen mainly to be the result of setting the light sea-quark
mass to its physical value, which alters the size of the chiral
logarithm. This correction for unphysical effects is clearly
dependent on our use of SChPT.

The solid [black] and dotted [red] curves in these figures can be
compared, approximately,
to the corresponding curves from the SU(3) fits shown 
in Figs.~\ref{fig:NBB1} and \ref{fig:NBB1S1}.
The comparison is not exact because the values of $m_y$ differ
(most notably for the S1 ensemble).
For the C3 ensemble, the
curvature in the SU(2) fit is much smaller than in the SU(3) fit.
This is because the curvature in the SU(3) fits is
dominated by the lattice-artifact terms, which are absent in the SU(2)
fits, and very poorly determined in the SU(3) fits. 
The SU(2) fit is
superior in that it incorporates the expectation that these artifacts
are small in the SU(2) regime. 
The continuum curves also differ between SU(2) and SU(3) fits,
with the chiral log being weaker in the latter.
This is possible because of $O(m_x/m_y)$ corrections to its coefficient
that are of NNLO in SU(2) SChPT and thus dropped in the SU(2) fit.
Again, the SU(2) fit is superior, since the size of these
corrections is not known in the SU(2) limit.

The parameters of the NLO X-fits, and the resulting 
``physical'' $B_K$, are given (for the largest values
of $a m_y$ only) in Table \ref{tab:fit-4X3Y-NLO-Xfit}
for all ensembles.
\begin{table}[htbp]
\caption{Fit parameters for the 4X-NLO fits.
  The valence strange quark mass is fixed at our 
  heaviest value, i.e. $a m_y = 0.05$, $0.03$, and $0.018$
  on coarse, fine and superfine lattices, respectively.
  The result for $B_K(1/a)$ is obtained from the fit by
  setting taste-splittings to zero, $L=m_{\pi_0}^2$ and
  $m_x=m_d^{\rm phys}$.
  \label{tab:fit-4X3Y-NLO-Xfit}}
\begin{ruledtabular}
\begin{tabular}{ l | l l l l }
ID &  $d_1$ & $d_2$ & $\chi^2/\text{dof}$ & $B_K(\mu=1/a)$ \\
\hline
C1 &  0.548(11) & 0.107(49) & 0.19(26) & 0.5568(96) \\ 
C2 &  0.556(12) & 0.060(54) & 0.36(37) & 0.564(11) \\ 
C3 &  0.5602(34) & 0.035(16) & 0.83(54) & 0.5676(31) \\ 
C4 &  0.5581(35) & 0.039(16) & 0.57(45) & 0.5656(31) \\ 
C5 &  0.5501(85) & 0.066(40) & 0.08(16) & 0.5582(77) \\ 
\hline
F1 &  0.5107(86) & 0.089(45) & 0.06(14) & 0.5189(76) \\ 
F2 &  0.5146(75) & 0.066(40) & 0.03(11) & 0.5222(66) \\ 
\hline
S1 &  0.4808(50) & 0.143(30) & 0.06(14) & 0.4900(43) \\ 
\end{tabular}
\end{ruledtabular}
\end{table}
Some fits have a relatively high (uncorrelated) $\chi^2/\text{dof}$,
indicating that the NLO fit function may be insufficient.
The fit parameters themselves are consistent on all the coarse
lattices, with evidence for monotonic variation with $a^2$.

After repeating this procedure for the three heaviest values of $m_y$,
we then proceed to the ``Y-fit''. The results for ``continuum,
physical'' $B_K$ from the X-fits are expected to be a smooth, analytic
function of $Y_P \propto m_y$. Examples of the data are shown in 
Figs.~\ref{fig:su2-fit-4X3Y-NLO-Yfit} and
\ref{fig:su2-fit-4X3Y-NLO-Yfit:S1}.
We extrapolate to the ``physical'' value of the $\bar ss$ meson,
$Y_P =(0.6858\ {\rm GeV})^2$~\cite{Lepagemss}.
In all cases, our results are consistent with a linear dependence,
and so we use a fit to
\begin{equation}
f_\textrm{ph} = b_1 + b_2 Y_P 
\label{eq:yfit-1}
\end{equation}
to obtain our central value.
We also do a quadratic ``fit'' in order to estimate the
systematic error arising from this extrapolation.
\begin{figure}[htbp]
\includegraphics[width=20pc]
                {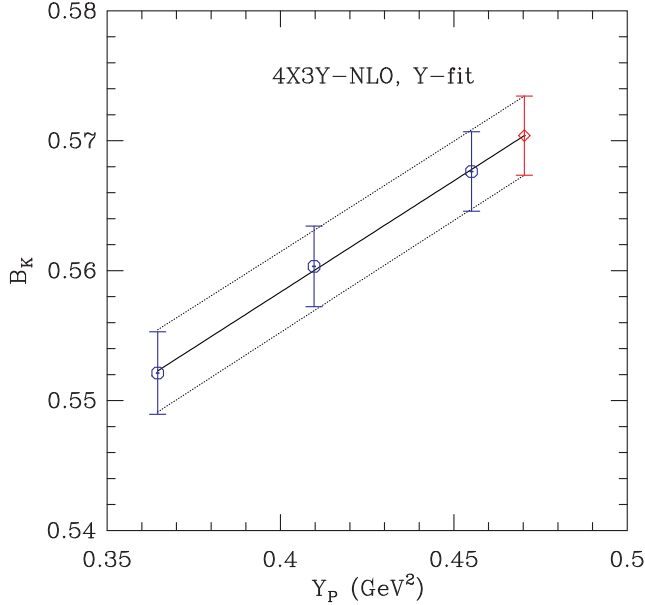}
\caption{$B_K(1/a)$ from the 4X-NLO fit vs. $Y_P$, on the coarse ensemble C3.
  A linear extrapolation to the physical strange quark mass is shown.
 \label{fig:su2-fit-4X3Y-NLO-Yfit}}
\end{figure}
\begin{figure}[htbp]
\includegraphics[width=20pc]
                {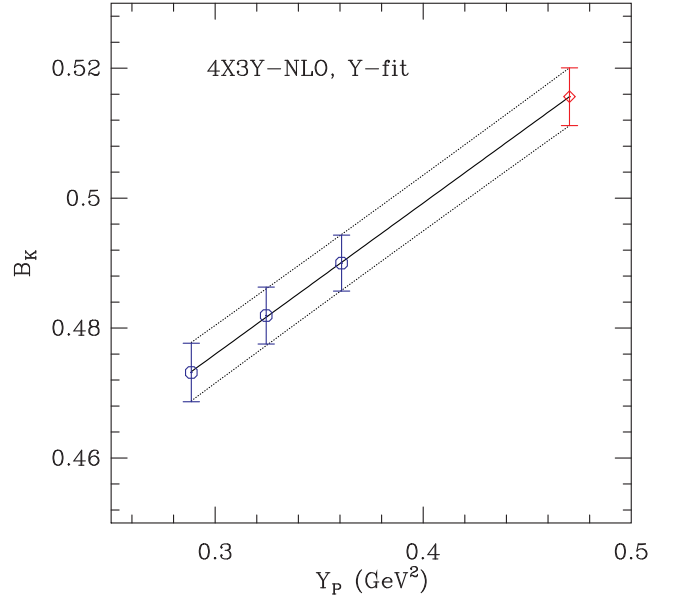}
\caption{$B_K(1/a)$ from the 4X-NLO fit, versus $Y_P$, on the 
superfine ensemble S1.
  A linear extrapolation to the physical strange quark mass is shown.
 \label{fig:su2-fit-4X3Y-NLO-Yfit:S1}}
\end{figure}
The parameters from the linear 3Y fits
are collected in Table~\ref{tab:fit-4X3Y-NLO}.
The fits are good, and the parameters are consistent
across all ensembles.
\begin{table}[htbp]
\caption{
Parameters of 3Y fits (using input from
4X-NLO fits) and the resulting $B_K(\mu=2\text{GeV})$.
  \label{tab:fit-4X3Y-NLO}}
\begin{ruledtabular}
\begin{tabular}{ l | l l l l }
ID &  $b_1$ & $b_2$ & $\chi^2/\text{dof}$ & $B_K(\mu=2\text{GeV})$ \\
\hline
C1 &  0.482(15)  & 0.162(22)  & 0.0009(6)  & 0.5488(95) \\ 
C2 &  0.466(16)  & 0.214(27)  & 0.0026(10) & 0.556(10) \\ 
C3 &  0.4899(47) & 0.1712(73) & 0.0154(26) & 0.5599(30) \\ 
C4 &  0.4866(50) & 0.1725(75) & 0.0135(28) & 0.5577(31) \\ 
C5 &  0.476(12)  & 0.179(19)  & 0.0015(8)  & 0.5503(75) \\ 
\hline
F1 &  0.433(13) & 0.215(23) & 0.0017(10) & 0.5416(77) \\ 
F2 &  0.437(11) & 0.214(19) & 0.0037(15) & 0.5454(66) \\ 
\hline
S1 &  0.4061(69) & 0.233(14) & 0.0049(17) & 0.5385(46) \\ 
\end{tabular}
\end{ruledtabular}
\end{table}
Also quoted are the values of $B_K$ obtained by extrapolation
in $Y_P$ and subsequent running to the
common scale $2\;$GeV.

%
%
\begin{figure}[htbp]
\includegraphics[width=20pc]
                {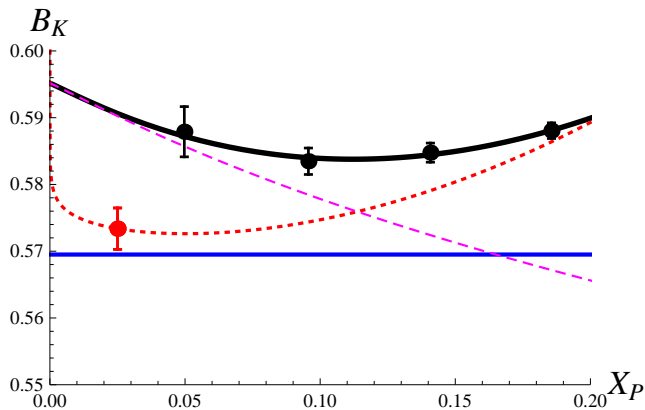}
\caption{$B_K(1/a)$ versus $X_P$ on coarse ensemble C3, for $a m_y=0.05$.
  The 4X-NNLO fit is shown by the solid (black) line.
  The horizontal (blue) line gives the contribution of the LO term. 
  The (magenta) dashed curve shows the fit with NNLO term removed.
  The (red) dotted curve has taste-splittings removed
  and the light sea-quark mass set to its physical value. 
  The (red) point on this line
  is at $X_P = m_{\pi_0}^2$ and shows the errors from the fit. 
  \label{fig:su2-fit-4X3Y-NNLO-Xfit:C3}}
\end{figure}
\begin{figure}[htbp]
\includegraphics[width=20pc]
                {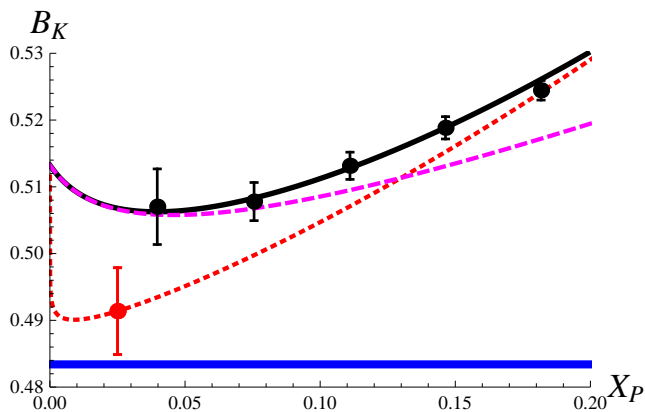}
\caption{$B_K(1/a)$ versus $X_P$ on the superfine ensemble S1, 
  for $a m_y=0.018$. Notation is as in
  Fig.~\protect\ref{fig:su2-fit-4X3Y-NNLO-Xfit:C3}.  The
data-point with the largest value of $X_P$ is {\em not} included
in the fit.
  \label{fig:su2-fit-4X3Y-NNLO-Xfit:S1}}
\end{figure}
We have repeated this analysis using NNLO X-fits (i.e. keeping
the $Q_3$ term). The X-fits 
are shown in Figs.~\ref{fig:su2-fit-4X3Y-NNLO-Xfit:C3}
and \ref{fig:su2-fit-4X3Y-NNLO-Xfit:S1}
for the C3 and S1 ensembles, respectively.
The resulting fit parameters are
given in Tables \ref{tab:fit-4X3Y-NNLO-Xfit}
and \ref{tab:fit-4X3Y-NNLO}.
\begin{table}[htbp]
\caption{Fit parameters for the 4X-NNLO fits. Notation as
in Table~\protect\ref{tab:fit-4X3Y-NLO-Xfit}.
\label{tab:fit-4X3Y-NNLO-Xfit}}
\begin{ruledtabular}
\begin{tabular}{ l | l l l l l }
ID &  $d_1$ & $d_2$ & $d_3$ & $\chi^2/\text{dof}$ & $B_K(\mu=1/a)$ \\
\hline
C1 &  0.561(19) & $-$0.13(20) & 0.88(60) & 0.031(51) & 0.565(14) \\ 
C2 &  0.577(21) & $-$0.29(23) & 1.31(69) & 0.090(93) & 0.577(16) \\ 
C3 &  0.5695(60) & $-$0.127(67) & 0.61(20) & 0.16(12) & 0.573(5) \\ 
C4 &  0.5660(62) & $-$0.097(68) & 0.50(20) & 0.12(10) & 0.571(5) \\ 
C5 &  0.557(15) & $-$0.05(16) & 0.42(48) & 0.037(56) & 0.562(11) \\ 
\hline
F1 &  0.517(15) & $-$0.04(19) & 0.54(66) & 0.007(25) & 0.523(11) \\ 
F2 &  0.519(13) & $-$0.02(17) & 0.35(57) & 0.002(14) & 0.525(10) \\ 
\hline
S1 &  0.4834(93) & $+$0.09(13) & 0.27(51) & 0.075(94) & 0.491(7) \\ 
\end{tabular}
\end{ruledtabular}
\end{table}
\begin{table}[htbp]
\caption{ 
Parameters of 3Y fits (using input from
4X-NNLO fits), and the resulting $B_K(\mu=2\text{GeV})$.
  \label{tab:fit-4X3Y-NNLO}}
\begin{ruledtabular}
\begin{tabular}{ l | l l l l }
ID &  $b_1$ & $b_2$ & $\chi^2/\text{dof}$ & $B_K(\mu=2\text{GeV})$ \\
\hline
C1 &  0.501(19) & 0.139(31) & 0.0002(3) & 0.557(14) \\ 
C2 &  0.478(22) & 0.215(39) & 0.0012(6) & 0.569(16) \\ 
C3 &  0.5081(64) & 0.144(11) & 0.0053(18) & 0.5651(46)  \\ 
C4 &  0.5080(69) & 0.137(11) & 0.0038(17) & 0.5621(47) \\ 
C5 &  0.493(17) & 0.150(27) & 0.0003(4) & 0.554(11) \\ 
\hline
F1 &  0.443(18) & 0.199(34) & 0.0005(6) & 0.544(11) \\ 
F2 &  0.444(15) & 0.203(27) & 0.0016(11) & 0.547(10) \\ 
\hline
S1 &  0.4121(98) & 0.220(20) & 0.0020(12) & 0.5385(72) \\ 
\end{tabular}
\end{ruledtabular}
\end{table}

The NNLO X-fits are better able to
capture the curvature seen in the data
(particularly on the coarse ensembles),
and have correspondingly reduced $\chi^2/{\rm dof}$.
The convergence of the chiral series begins to break
down on the C3 ensemble at the largest fit values of $X_P$,
in the sense that $NNLO:NLO\approx 1$. The convergence on
the S1 ensembles is satisfactory for all four fit points.
The linear Y-fits remain very good on all ensembles
(and are not shown). 

Comparing Tables~\ref{tab:fit-4X3Y-NLO} and \ref{tab:fit-4X3Y-NNLO},
we see that NLO and NNLO fits yield ``physical'' $B_K$ values 
that are consistent within errors.
We have also done NNLO 5X3Y fits, 
and simultaneous X and Y extrapolations (rather than sequential),
both of which give consistent results.

Clearly, SU(2) SChPT fitting is much more straightforward than
that to the SU(3) form. No fitting parameters need be dropped,
and no Bayesian constraints are required. While we use only
a fraction of our data (roughly a quarter), the points we
keep are those which lie closest to the physical kaon and thus
should dominate the extrapolation to the physical point.
As for the SU(3) fits, the use of SChPT (rather than, say,
an analytic extrapolation) is crucial,
as is allows us to remove the effect of taste-breaking,
and to account for the shift in the chiral logarithm
due to the use of an unphysical light sea-quark mass.
We cannot, however, claim to have provided a 
test of the predicted chiral logarithms; instead, we
have shown consistency with our data.

We compare our SU(2) results with those from SU(3) fitting when
we discuss error budgets in Sec.~\ref{ssec:err-bud} below.

\subsection{Dependence of $B_K$ on sea quark masses \label{ssec:sea-quark}}

The fits presented above are on individual ensembles,
and thus at fixed values of lattice spacing and sea-quark masses.
Both SU(2) and SU(3) SChPT predict, at NLO,
linear dependence of $B_K$ on the light sea-quark mass 
(once chiral logarithms have been accounted for, as has been done above).
This enters through the $H_5$ term in SU(3) SChPT and
the $Q_4$ term in SU(2) SChPT.
At NLO, this dependence can either be treated as that of
the leading order coefficients ($b_1$ and $d_1$) or of
$B_K$ itself.
We take the latter approach, and investigate this dependence 
on the coarse and fine lattices, where we have results with
more than one light sea-quark masses. We also discuss the
impact of the mismatch of the strange sea-quark mass with its
physical value.

We plot $B_K(\mu=2\text{GeV})$ 
on the coarse lattices from the N-BB1 SU(3) fits  
against $a m_\ell$ in Fig.~\ref{fig:su3-fit-N-BT7-sea:coarse}.
The data show only weak dependence on $a m_\ell$.
We perform both constant and linear extrapolations to the physical
value of $a m_\ell$, and find that they lead to consistent results.
\begin{figure}[htbp]
\includegraphics[width=20pc]{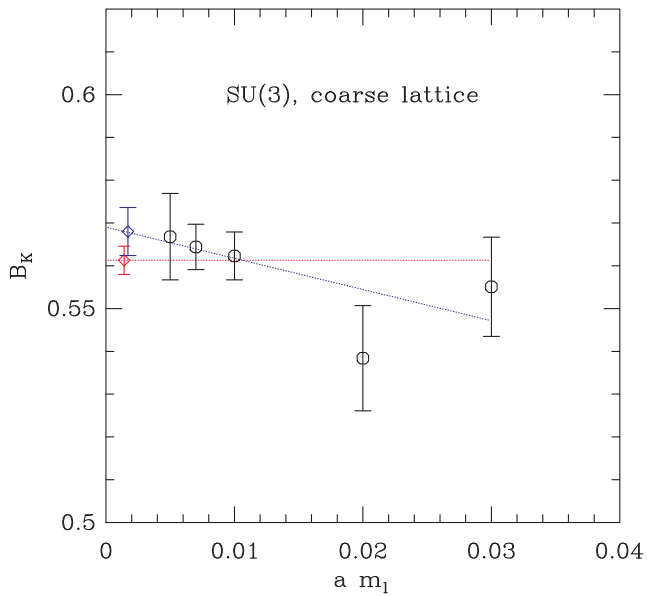}
\caption{$B_K(\mu=2\text{GeV})$ from N-BB1 SU(3) fits versus $am_l$
  (light sea-quark mass) on MILC coarse ensembles.  The data come from
  Table~\ref{tab:fit-N-BT7}.  Linear and constant fits are shown,
  along with the resulting value after extrapolation to the physical
  value of $a m_l$.  The errors on the point at $a m_\ell=0.01$ and
  0.007 are smaller due to the larger number of measurements made on
  the C3 and C4 ensembles.
  \label{fig:su3-fit-N-BT7-sea:coarse}}
\end{figure}
The corresponding results from the 4X3Y-NNLO SU(2) fit are
shown in Fig.~\ref{fig:su2-fit-4X3Y-NNLO-no-sea:coarse}.
Here the linear and constant fits are almost indistinguishable.
\begin{figure}[htbp]
\includegraphics[width=20pc]{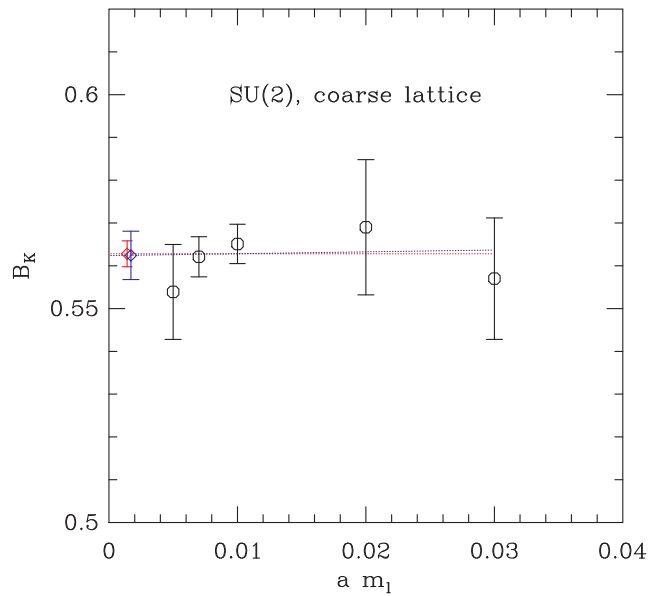}
\caption{$B_K(\mu=2\text{GeV})$ from 4X3Y-NNLO SU(2) fits
versus $am_l$ (light sea-quark mass) on MILC coarse ensembles.
The data come from Table~\ref{tab:fit-4X3Y-NNLO}.
Details are as in Fig.~\protect\ref{fig:su3-fit-N-BT7-sea:coarse}.
  \label{fig:su2-fit-4X3Y-NNLO-no-sea:coarse}}
\end{figure}
We collect the results of these extrapolations
in Table~\ref{tab:bk-extrap-aml}.
\begin{table}[htbp]
\caption{ $B_K(\mu=2\text{GeV})$ 
on the coarse lattices after extrapolation 
to the physical light sea-quark mass,
for various choices of fit. 
  \label{tab:bk-extrap-aml}}
\begin{ruledtabular}
\begin{tabular}{ l l l l }
S$\chi$PT & valence fit &  $a m_\ell$ fit  & $B_K$ \\
\hline
SU(3) & N-BB1     & constant  & 0.5613(33) \\
SU(3) & N-BB1     & linear    & 0.5680(56) \\
\hline
SU(2) & 4X3Y-NNLO & constant  & 0.5628(30) \\
SU(2) & 4X3Y-NNLO & linear    & 0.5624(57) \\
\end{tabular}
\end{ruledtabular}
\end{table}

For the fine lattices, we have two light sea-quark masses,
corresponding approximately to $am_\ell = 0.005$ (C5) and
$am_\ell = 0.01$ (C3) on the coarse lattices.
As on the coarse lattices, the data have very weak
dependence on $am_\ell$, as shown in
Figs.~\ref{fig:su3-fit-N-BT7-sea:fine} and
\ref{fig:su2-fit-4X3Y-NNLO-no-sea:fine}.
\begin{figure}[htbp]
\includegraphics[width=20pc]{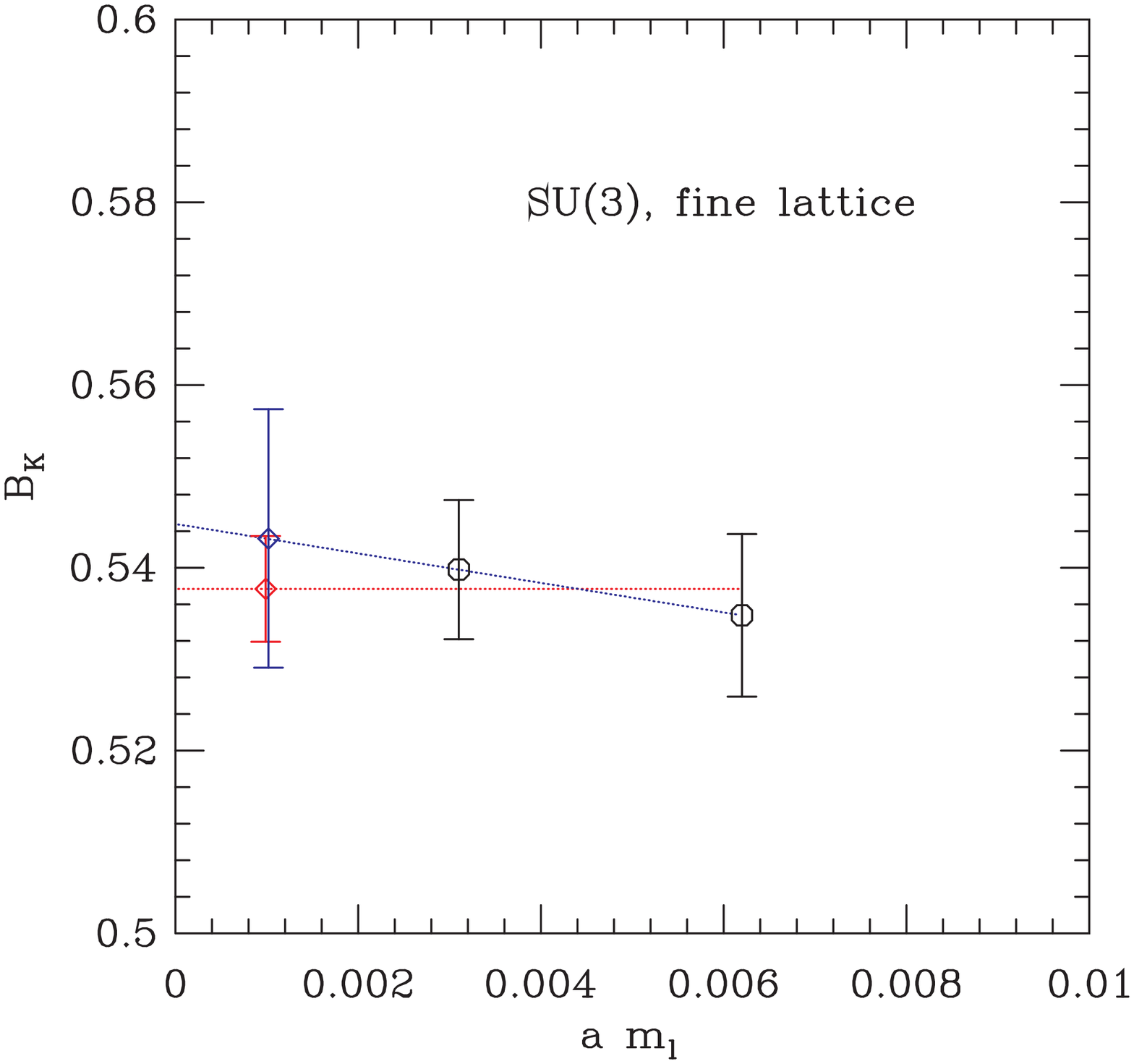}
\caption{ As in Fig.~\ref{fig:su3-fit-N-BT7-sea:coarse} but for the
  MILC fine ensembles. 
  \label{fig:su3-fit-N-BT7-sea:fine}}
\end{figure}
\begin{figure}[htbp]
\includegraphics[width=20pc]{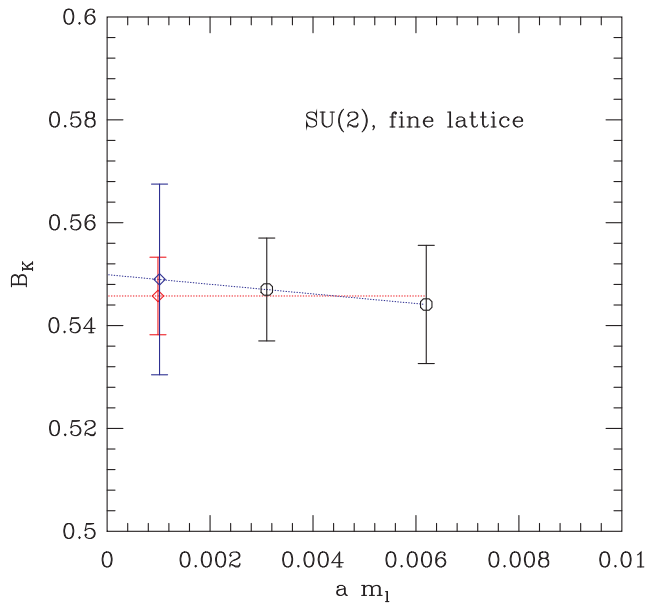}
\caption{As in Fig.~\ref{fig:su2-fit-4X3Y-NNLO-no-sea:coarse}
  but for the MILC fine ensembles. 
  \label{fig:su2-fit-4X3Y-NNLO-no-sea:fine}}
\end{figure}

To quantify the size of the mass dependence, we use the
form expected from either SU(2) or SU(3) SChPT
\begin{equation}
B_K(a m_\ell) = B_K(a m_\ell=0) \left[1 - L_P/\Lambda^2\right]
\,.
\label{eq:BKvsLP}
\end{equation}
The linear fit shown in Fig.~\ref{fig:su3-fit-N-BT7-sea:coarse} 
has $\Lambda\approx 2.8\;$GeV.
This is much smaller than the canonical size 
expected in ChPT, namely $\Lambda\approx 1\;$GeV,
indicating that the sea-quark mass dependence is indeed weak.
We note that mass dependence with this value of $\Lambda$
is also consistent with the fine lattice data.

For the superfine lattices, we have only a single
light sea-quark mass, and so
cannot perform an extrapolation to $a m_\ell^{\rm phys}$.
Instead, we assume (based on the fits above) that,
to first approximation, there is no dependence on $a m_\ell$.
We do so consistently for all three lattice spacings,
taking ensembles C3, F1 and S1 for subsequent continuum
extrapolation. We use these three ensembles since they all
have $m_\ell/ m_s = 1/5$, with $m_s\approx m_s^{\rm phys}$,
and thus have closely matched sea-quark masses.

Clearly this procedure leads to a systematic error if
there is, in fact, a dependence on $a m_\ell$. We estimate
this error assuming that the dependence is that given by
the linear fit of
Fig.~\ref{fig:su3-fit-N-BT7-sea:coarse}.
Specifically, we take the difference between the result on
ensemble C3 with that after linear extrapolation to $a m_\ell^{\rm phys}$
(as given in Table~\ref{tab:bk-extrap-aml}),
divided by the C3 value, as our estimate of the fractional error
from this source. The resulting error is $1.2\%$ for the SU(3) fits,
and $0.06\%$ for the SU(2) fits.
This is quite likely an underestimate in the SU(2) case, but since
other sources of error are significantly larger, we have not
attempted to improve on this estimate.

%
\begin{table}[htbp]
\caption{ Nominal ($am_s$) and actual
($am_s^\textrm{phys}$) strange sea-quark masses~\cite{milc-rmp-09}.
  \label{tab:am_s}}
\begin{ruledtabular}
\begin{tabular}{ l l l }
MILC &  $am_s$  & $am_s^\textrm{phys}$ \\
\hline
coarse & 0.05   &  0.0350(7) \\
fine   & 0.031  &  0.0261(5) \\
superfine & 0.018 & 0.0186(4)
\end{tabular}
\end{ruledtabular}
\end{table}
Finally, we should account for the 
difference between the values of $a m_s$ used in the MILC ensembles
and $m_s^{\rm phys}$. These values are given in Table~\ref{tab:am_s}. 
The most reliable approach would be to have results with
more than one value of $a m_s$, and extrapolate/interpolate to the
physical value. Absent this possibility, we can use SU(3) ChPT to
estimate the resulting error. In SU(3) ChPT, the
dependence on sea-quark masses is obtained from
Eq.~(\ref{eq:BKvsLP}) by the substitution $L_P \to L_P+ S_P/2$.
Assuming this, and using the masses in Table~\ref{tab:am_s}
as well as the value of $\Lambda$ from the linear fit in
Fig.~\ref{fig:su3-fit-N-BT7-sea:coarse},
we find that the correction to $B_K(2\;{\rm GeV})$ is
$+0.0075$, $+0.003$, and $-0.0006$ for the coarse, fine and
superfine ensembles, respectively.
Extrapolating to the continuum linearly in $a^2$ leads to
a final correction of $-0.0028$. We take the magnitude of
this correction as an estimate of the $m_s$ mismatch error
for {\em both} SU(3) and SU(2) fits.

The analysis described in this subsection does not make optimal
use of all our data. For example, ensemble F2 is only used to
check the (lack of) $a m_\ell$ dependence. By doing a
combined continuum and $a m_\ell$ fit we could likely do better.
We have not attempted such a fit, however, since this source
of error is subdominant.
Concerning the $a m_s$ mismatch, we plan in the near future to do an
exact interpolation to the physical strange sea quark mass using the
reweighting technique \cite{chulwoo-2009-1}.

\subsection{Continuum extrapolation \label{ssec:scaling}}

At this stage of the analysis we have results
for $B_K$ on three lattice spacings. As noted above,
we take these to be the results from the C3, F1 and S1 ensembles.
Running to the common scale $\mu=2\;$GeV yields, 
for various SU(3) and SU(2) fits,
the results in Tables~\ref{tab:bk-disc-su3}
and \ref{tab:bk-disc-su2}, respectively. Most of these
results have appeared in earlier Tables, but we also include
results for the SU(2) 5X fits for completeness.
The dominant errors remaining at this stage are those due to taste-conserving
discretization and truncation errors.
These depend on the lattice spacing as $a^2 \alpha^n$,
where $n=0,1, \dots$.
(n=0 is allowed since we do not use Symanzik-improved operators),
and as $\alpha^2$. Since we cannot disentangle these effects
with a fit to three points, we choose to fit to a linear function
of $a^2$, and estimate the error we make by dropping other
dependences. The resulting fits are shown in
Figs.~\ref{fig:su3-fit-sv} and \ref{fig:su2-fit-sv}.

%
\begin{table}[htbp]
\caption{$B_K(\mu=2\text{GeV})$ from various SU(3) fits. 
  \label{tab:bk-disc-su3}}
\begin{ruledtabular}
\begin{tabular}{ l | l l l l l}
$a$ (fm) & N-U & N-B1 & N-B2 & N-BB1 & N-BB2 \\
\hline
0.12 & 0.596(10) & 0.560(5) & 0.585(8) & 0.562(6) & 0.592(14) \\
0.09 & 0.564(20) & 0.533(10)& 0.532(8) & 0.535(9) & 0.539(12) \\
0.06 & 0.581(12) & 0.533(7) & 0.554(5) & 0.535(6) & 0.560(11)
\end{tabular}
\end{ruledtabular}
\end{table}
\begin{table}[htbp]
\caption{$B_K(2\,\text{GeV})$ from various SU(2) fits. 
  \label{tab:bk-disc-su2}}
\begin{ruledtabular}
\begin{tabular}{ l | l l l l}
$a$ (fm) & 4X3Y-NLO & 4X3Y-NNLO & 5X3Y-NLO & 5X3Y-NNLO\\
\hline
0.12 & 0.560(3) & 0.565(5) & 0.558(3) & 0.563(4)\\
0.09 & 0.542(8) & 0.544(11)& 0.541(7) & 0.543(10)\\
0.06 & 0.538(5) & 0.538(7) & 0.539(4) & 0.538(6) \\
\end{tabular}
\end{ruledtabular}
\end{table}
\begin{figure}[htbp]
\includegraphics[width=20pc]{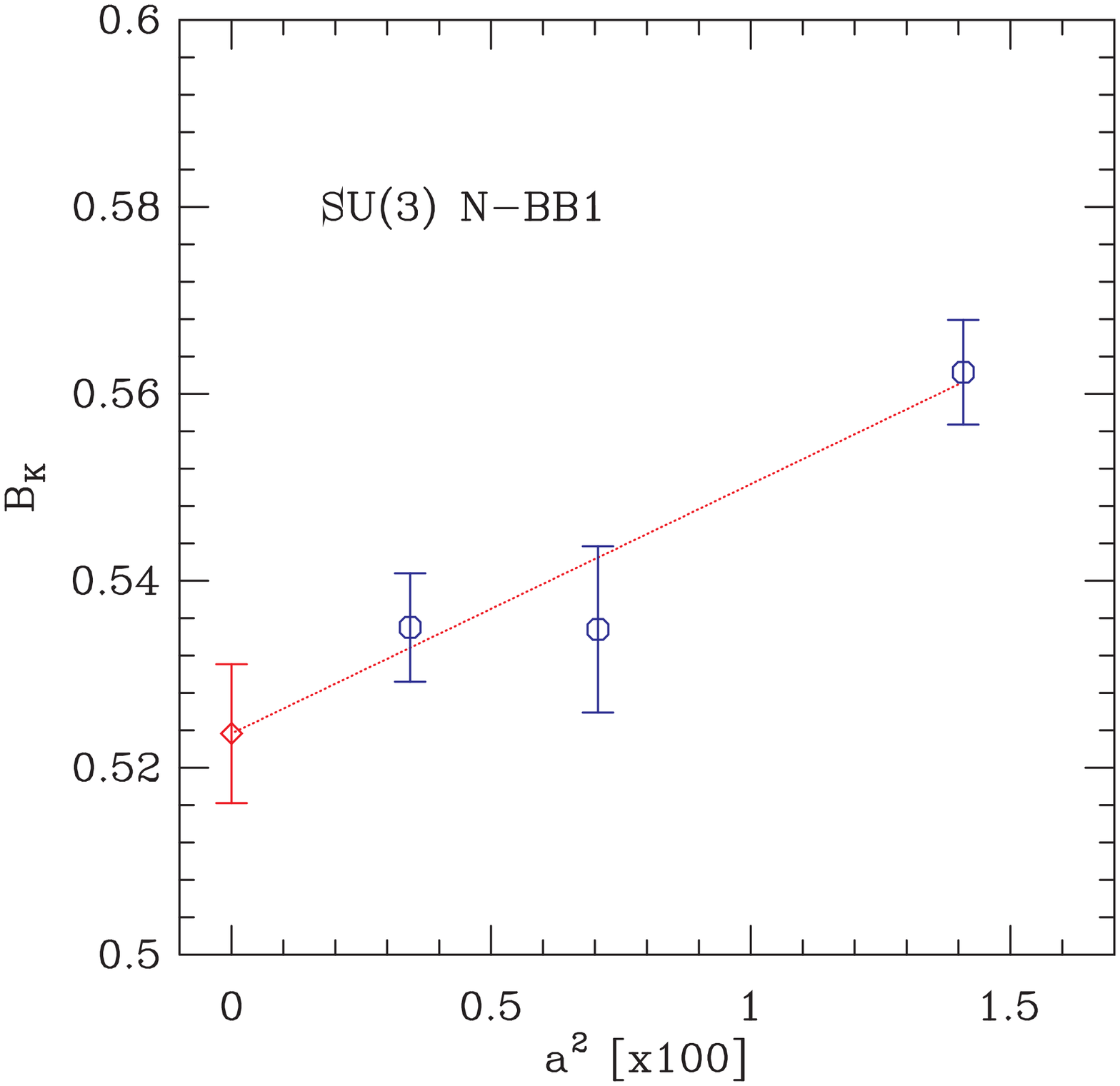}
\caption{$B_K(2\,{\rm GeV})$ versus $a^2$, together with
a linear extrapolation to the continuum limit. 
  The data is obtained using SU(3) N-BB1 fits.
  \label{fig:su3-fit-sv}}
\end{figure}
\begin{figure}[htbp]
\includegraphics[width=20pc]{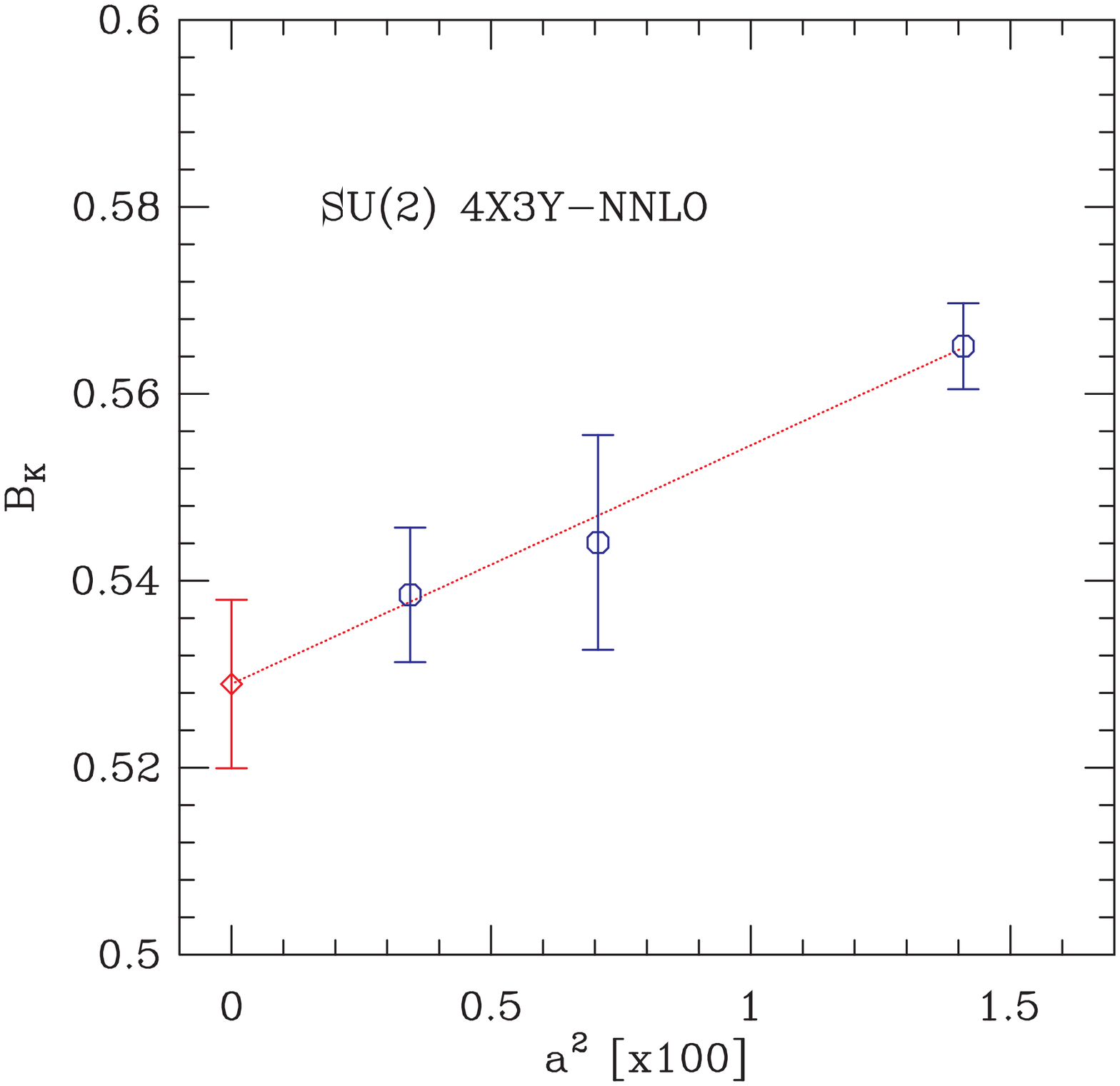}
\caption{$B_K(2\,{\rm GeV})$ versus $a^2$, together with
a linear extrapolation to the continuum limit. 
  The data is obtained using SU(2) 4X3Y-NNLO fits.
  \label{fig:su2-fit-sv}}
\end{figure}
\begin{table}[htbp]
\caption{$B_K(\mu=2\text{GeV}, a=0)$ from various fits. 
  \label{tab:bk-a=0}}
\begin{ruledtabular}
\begin{tabular}{ c c l }
Group & fit & $B_K(\mu=2\text{GeV},a=0)$ \\
\hline
SU(3) & N-BB1 &  0.524(7) \\
SU(3) & N-BB2 &  0.539(14) \\
\hline
SU(2) & 4X3Y-NLO  & 0.531(6) \\ 
SU(2) & 4X3Y-NNLO & 0.529(9) \\ 
\end{tabular}
\end{ruledtabular}
\end{table}

For the SU(3) analysis, we take the continuum value obtained
with N-BB1 fits as our central value. The difference between
the N-BB1 and N-BB2 fits will be used as a fitting systematic,
as discussed in the next subsection.
To account for the fact that the discretization errors can
depend on $a^2$ times $\alpha$ to some power, rather than just $a^2$,
we take the difference between the results on the S1 ensemble and
in the continuum as the systematic error in the extrapolation.
This is reasonable because including powers of $\alpha$ moves
the S1 ensemble closer to the continuum limit relative to the
other two ensembles. It is conservative because our data strongly
indicates that the actual continuum value lies below that obtained
on ensemble S1.

We follow the same strategy for the SU(2) analysis, using
the 4X3Y-NNLO fits for our central value.

Corrections proportional to $\alpha^2$, arising from
truncation of the matching factors, are estimated separately,
as we now discuss.

\subsection{Error budget \label{ssec:err-bud}}

In this subsection we estimate the remaining systematic errors
and combine them with those obtained above to give the complete
error budget.

%
%
%
We begin by estimating the impact of truncating matching factors.
We assume that the dominant missing term
in the perturbative expansion is of magnitude $1\times \alpha_s(1/a)^2$,
and thus take as our error estimate
\begin{eqnarray}
\Delta B_K^{(2)} \approx B_K^{(1)}(\mu=1/a) \times 
\Big[ \alpha_s(\mu=1/a) \Big]^2
\,.
\end{eqnarray}
Here $B_K^{(1)}$ is the one-loop corrected result, and
$\alpha_s$ is the $\overline{\rm MS}$ coupling constant.
The resulting error estimates for the SU(3) analysis are
given in Table~\ref{tab:bk-2loop-su3}.
\begin{table}[htbp]
  \caption{Estimates of the truncation error in the SU(3) analysis.
  We use the results of the N-BB1 fits, and ensembles C3, F1 and S1.
  \label{tab:bk-2loop-su3}}
\begin{ruledtabular}
\begin{tabular}{ l l l l l l}
$a$ (fm) & $B^{(1)}_K$ & $\alpha_s$ & $\Delta B_K^{(2)}$ \\
\hline
0.12 & 0.5728(57) & 0.3286 & 0.062\\
0.09 & 0.5271(88) & 0.2730 & 0.039\\
0.06 & 0.5123(56) & 0.2337 & 0.028\\
\end{tabular}
\end{ruledtabular}
\end{table}
The results for the SU(2) analysis are almost identical.

The continuum extrapolation discussed above will 
extrapolate away some of this error, so that the truncation
error in the continuum result will be smaller than that
on the superfine ensemble. To be conservative, however, we use
the size of $\Delta B_K^{(2)}$ on the superfine ensemble as
an estimate of this error.
It is appropriate to be particularly conservative for the
truncation error, since the choice of unity multiplying
$\alpha^2$ is a guess.
We note, however, that the same approach applied to the tree-level
data (i.e. using $\Delta B_K^{(1)}=B_K^{(0)} \alpha$)
leads to an overestimate of the shift between $B_K^{(0)}$
and $B_K^{(1)}$~\cite{wlee-09-4}.

A further source of error is due to our use of a finite volume (FV).
This error can be approximately accounted for by fitting to
the finite volume form of the NLO SChPT result
(discussed in Appendix~\ref{app:defs}).
We have not yet implemented this approach, however,
as it is very computationally demanding. 
We have used SChPT to make preliminary
estimates of the size of the corrections 
that might be induced~\cite{wlee-09-3}.
Considering the C3 ensemble, we found that the FV error 
predicted by SU(3) SChPT was $\sim 1\%$ after extrapolation
to physical masses. For SU(2) SChPT the error was substantially smaller.
Since these two results disagree, while the actual FV error has a
definite (though unknown) size, we concluded that SChPT at NLO is not
a reliable quantitative guide to the size of FV effects.

Here we adopt a more direct approach.
We determine the impact of changing
the volume by comparing the results on ensembles C3 and C3-2.
These differ only in their spatial volumes,
which are $20^3$ and $28^3$, respectively.
The statistical weights of the two lattices are almost equal
because 
\begin{equation}
R_\text{stat} = \frac{20^3 \times 9 \times 671}
{28^3 \times 8 \times 274} = 1.004\ldots\,.
\end{equation}
A comparison of results from these two lattices 
(after extrapolation to physical quark masses) 
is given in Table~\ref{tab:bk-fv}.
We take the difference in $B_K$ between C3 and C3-2 ensembles
as our estimate of the finite volume error.
By doing so, we are effectively assuming that the $28^3$ lattices
have negligible FV effects, which is what
is found using NLO SChPT.
\begin{table}[htbp]
  \caption{Volume dependence of $B_K$.
   We show $B_K(\mu=2\text{ GeV})$ from N-BB1 fits for SU(3),
    and from 4X3Y-NNLO fits for SU(2).
    \label{tab:bk-fv}}
\begin{ruledtabular}
\begin{tabular}{ c c c}
analysis & $20^3$ (C3) & $28^3$ (C3-2) \\
\hline
SU(2)    & 0.5651(46)  & 0.5699(45) \\
SU(3)    & 0.5623(56)  & 0.5754(56) \\
\end{tabular}
\end{ruledtabular}
\end{table}

The final source of significant error is our choice for
the decay constant $f_\pi$ that appears (as $1/f_\pi^2$)
in the coefficient of the chiral logarithms.
As should be clear from the description of our fits, 
our data are not sensitive enough to the contributions of
chiral logarithms to allow a fit to $f_\pi$.
Thus we must fix it. We choose $f_\pi=132\;$MeV for our
central value, which is a somewhat outdated approximation
to the physical value, $f_\pi=130.4\;$MeV~\cite{PDG10}.
(We have checked that results for $B_K$ are indeed insensitive
to the small difference between these two values.)
It is reasonable to use the physical pion decay constant since
the loops which give the dominant curvature are those involving pions.
It is, however, equally valid at NLO to use the decay constant in
the chiral limit, which is $\approx 108\;$MeV for SU(3) ChPT
and $\approx 124.2\;$MeV for SU(2) ChPT~\cite{milc-rmp-09}
In the SU(3) fits, it is also equally valid at NLO to use
the physical value of the kaon decay constant,
 $f_K\approx 156.5\;$MeV~\cite{milc-rmp-09}.

It is straightforward to determine an ``$f_\pi$ error''
for our SU(2) fits---we quote the difference between the
results obtained with $132\;$MeV and $124.2\;$MeV.
For the SU(3) fits, we have concluded after some experiments
that the $f_\pi$ error is already reasonably accounted for
by the error we include to account for the dependence on
the Bayesian fitting scheme [``fitting (2)'' below].
Since we do not use the SU(3) fits for our central
value, we have not pursued this further.

\bigskip
We now collect all sources of error and present the error budget.
We do this separately for the SU(3) and SU(2) analyses.
The SU(3) budget is given in Table~\ref{tab:su3-fit-err-budget}.
Most of these errors have been discussed either in this subsection or
earlier ones. The exceptions are the two ``fitting'' errors and 
the $r_1$ error.
The first fitting error estimates the uncertainty due to the
possibility of implementing the Bayesian method in different ways.
The N-B1 fit uses one level of Bayesian constraints
while the N-BB1 fit uses two levels. We prefer the latter
method, for reasons explained earlier, but to be conservative
we take the difference between the results of the two fits
as a systematic error.
The second fitting error estimates the
impact of making different assumptions about the
prior information used in the Bayesian fits.
The N-BB1 fit assumes that the dominant lattice artifacts are of
${\cal O}(a^2)$ while the N-BB2 fit assumes that they are of
${\cal O}(\alpha_s^2)$. We take the difference between these fits
as an estimate of this error.
The resulting rather large error roughly encompasses the range
of results obtained with different SU(3) fits, as can be seen
from Table~\ref{tab:bk-disc-su3}. Thus it can be interpreted
more generally as an estimate of the error due to uncertainties in
the parameters of the fit function corresponding to lattice artifacts.
%
\begin{table}[htbp]
\caption{ Error budget (in percent) for $B_K$ obtained
using SU(3) fitting.
  \label{tab:su3-fit-err-budget}}
\begin{ruledtabular}
\begin{tabular}{ l | l l }
cause & error (\%) & memo \\
\hline
statistics       & 1.4   & N-BB1 fit \\
matching factor  & 5.5   & $\Delta B_K^{(2)}$ (S1) \\
discretization   & 2.2   & diff.~of (S1) and (a=0) \\
fitting (1)      & 0.36  & diff.~of N-BB1 and N-B1 (C3) \\
fitting (2)      & 5.3   & diff.~of N-BB1 and N-BB2 (C3) \\
$a m_l$ extrap   & 1.0   & diff.~of (C3) and linear extrap \\
$a m_s$ extrap   & 0.5   & constant vs. linear extrap \\
finite volume    & 2.3   & diff.~of $20^3$ (C3) and $28^3$ (C3-2)\\
scale $r_1$      & 0.12  & uncertainty in $r_1$
\end{tabular}
\end{ruledtabular}
\end{table}

Finally, we have propagated the uncertainty in $r_1$, which sets the
scale, through the analysis. We do so by repeating the analysis
using $r_1 \pm \sigma_{r_1}$ to determine the lattice spacing.

Our final result from the SU(3) analysis is
\begin{equation}
B_K(\text{NDR}, \mu=2\text{ GeV}) = 0.5237 \pm 0.0074 \pm 0.0438
\,,\label{eq:bksu3final}
\end{equation}
where the first error is statistical while the second is the
sum of the systematic errors in quadrature.
The former error has magnitude 1.4\% and the latter 8.4\%.
The error budget for the SU(2) analysis is presented in
Table \ref{tab:su2-fit-err-budget}.
The errors are as for the SU(3) analysis except for the two
fitting errors and the $f_\pi$ error.
The first of these estimates the impact on the
fit of the addition of higher-order terms in the SChPT fit function
used for the X-fits. Specifically, we use the
difference between the NLO and NNLO fits on the 4X3Y data set.
The second fitting error results from
the uncertainty in the Y-fit: we
quote the difference between linear and quadratic
extrapolations to the physical point $Y_P = 2 m_K^2 - m_\pi^2$.
%
\begin{table}[htbp]
\caption{ Error budget of $B_K$ in the SU(2) fitting.
  \label{tab:su2-fit-err-budget}}
\begin{ruledtabular}
\begin{tabular}{ l | l l }
cause & error (\%) & memo \\
\hline
statistics      & 1.7    & 4X3Y NNLO fit \\
matching factor & 5.5    & $\Delta B_K^{(2)}$ (S1) \\
discretization  & 1.8    & diff.~of (S1) and (a=0) \\
fitting (1)     & 0.92   & X-fit (C3) \\
fitting (2)     & 0.08   & Y-fit (C3) \\
$a m_l$ extrap  & 0.48   & diff.~of (C3) and linear extrap \\
$a m_s$ extrap  & 0.5    & constant vs. linear extrap \\
finite volume   & 0.85   & diff.~of $20^3$ (C3) and $28^3$ (C3-2)\\
$r_1$           & 0.14   & $r_1$ error propagation \\
$f_\pi$         & 0.38   & $132\;$MeV vs. $124.4\;$MeV
\end{tabular}
\end{ruledtabular}
\end{table}

Our final result of $B_K$ from the SU(2) analysis is
\begin{equation}
B_K(\text{NDR}, \mu=2\text{ GeV}) = 0.5290 \pm 0.0090 \pm 0.0316
\,,\label{eq:bksu2final}
\end{equation}
with the statistical error having magnitude 1.7\%
and the total systematic error 5.9\%.
Although the statistical error in this result is somewhat
larger than that from the SU(3) analysis---a reflection of
the smaller data set used in the chiral extrapolation---the 
systematic error is considerably smaller, as is the total error.
This improvement results from the simplifications
which occur in SU(2) SChPT, which allow more straightforward fitting.
Because of this we use the SU(2) analysis for our final result,
while the SU(3) result provides a consistency check.

\section{Conclusion \label{sec:conclude}}
We have presented a calculation of $B_K$ using improved
staggered fermions with $2+1$ flavors of dynamical fermions. 
We have used three different lattice
spacings and multiple choices for the valence and sea quark masses
in order to perform the extrapolations to the continuum limit
and the physical quark masses.
A key ingredient in our analyses are the fitting forms
predicted by staggered chiral
perturbation theory, for they allows us to determine, and remove,
the impact of taste-breaking discretization errors.
To carry out our analysis,
we needed to generalize previous 
SU(3) SChPT results for $B_K$ by including
the effects of a mixed action (HYP-smeared valence quarks on
asqtad sea) and by extending the results to SU(2) SChPT.

A striking feature of the SU(2) SChPT result is that taste-violating
discretization and truncation errors do not enter until NNLO.
This means that fitting to the NLO form (as we do) is as simple
as for fermions with exact chiral symmetry.
Although we can use only a quarter of our complete data set when
doing the SU(2) fits, this is more than compensated for by the
simplicity of fitting.

Using these ChPT results, we have performed two 
independent analyses based, respectively,
on SU(3) and SU(2) SChPT at NLO.
The former requires some ad-hoc simplifications of the fitting
expression and the use of Bayesian constraints to obtain stable fits.
We find significant sensitivity to the prior information used in
the Bayesian procedure, as well as to the precise implementation.
The SU(2) fitting does not require the use of Bayesian priors, and
leads to significantly smaller overall errors.
Hence, we use the SU(2) analysis for our final result 
(\ref{eq:bksu2final}), while the SU(3) result, (\ref{eq:bksu3final}), 
provides a consistency check. 
Converting to the renormalization group invariant $B_K$, we find
\begin{equation}                                                               
\widehat{B}_K 
= 0.7243 \pm 0.0123 \pm 0.0433
= 0.724 \pm 0.045 \,.
\label{eq:bkrgifinal}
\end{equation}

Our result is in good agreement with the two
existing results which use 2+1 flavors of dynamical fermions
and control all sources of error.
Aubin, Laiho and Van de Water use a mixed action
with valence domain wall fermions 
on the MILC coarse and fine lattices and find~\cite{ALV-09}
\begin{equation}
\widehat{B}_K =0.724 \pm 0.008 \pm 0.028
= 0.724 \pm 0.029\,.
\,,
\end{equation}
%
The RBC-UKQCD collaboration use both domain wall valence and
sea quarks, finding~\cite{rbc-uk-08-1}
\begin{equation}
\widehat{B}_K =0.720 \pm 0.013 \pm 0.037
= 0.724 \pm 0.039
\end{equation}
on a single lattice spacing of $1.16\;$fm, 
with a preliminary result (without an estimate of finite volume effects)
\begin{equation}
\widehat{B}_K =0.738 \pm 0.026
\end{equation}
%
based on two lattice spacings ($a= 0.86$ and $1.16\;$fm) 
and updated fitting~\cite{Kelly:2009fp}.
Our result is also consistent with the previous calculation
using staggered fermions on the coarse ($a\approx 0.12\;$fm)
MILC lattices, which finds 
$\widehat{B}_K= 0.83 \pm 0.02 \pm 0.18$~\cite{gamiz-06}.
We view it as a significant success of lattice QCD that
results with valence staggered and domain-wall
fermions are consistent. Although our calculation shares
lattices with that of Ref.~\cite{ALV-09}, it is completely
independent of that of Refs.~\cite{rbc-uk-08-1,Kelly:2009fp}.

Compared to the other two calculations, our result has the advantages
of pushing closer to the continuum limit,
and being based on three, rather than two lattice spacings.
It also is significantly computationally cheaper.
A clear disadvantage is that we must deal with taste-breaking,
although this is much less of an issue with SU(2) fitting than
with SU(3). Even with SU(2) fitting, however, we do rely on
ChPT to a greater extent than the other calculations.

Our total error (6\%) is significantly larger than that
from the other calculations---4.1\% for Ref.~\cite{ALV-09} and
3.5\% for Ref.~\cite{Kelly:2009fp}. This difference is due almost 
entirely to our use of perturbative (one-loop) matching rather than the
non-perturbative renormalization (NPR) used by the other calculations.
Quantitatively, our matching error is 5.5\% to be compared
to 3.3\%~\cite{ALV-09} and 2.4\%~\cite{Kelly:2009fp}.
In addition, our error estimate is less reliable since it is
based on an educated guess about the possible size of two-loop terms.

It is also of interest to compare our result to those obtained
using two flavors of dynamical quarks (but a quenched strange quark).
This may be a useful comparison because
there are indications that the dependence of dimensionless
quantities on the mass of the strange sea quark is weak.
Indeed, our results for $B_K$ show weak dependence
on all sea-quark masses,
at least within the range of masses that we consider.
In fact, our result is consistent with those obtained
from $N_f=2$ calculations.
A recent calculation using twisted-mass fermions,
and with all errors controlled except for
that due to quenching the strange quark, finds
$\widehat{B}_K=0.729\pm 0.025 \pm 0.016$~\cite{BK_tm_10}.
The result obtained with dynamical overlap fermions,
using a single lattice spacing,
is $\widehat{B}_K=0.758\pm 0.006 \pm 0.071$~\cite{Aoki:2008ss}.
With domain-wall fermions on a single lattice spacing,
the result is $\widehat{B}_K = 0.699 \pm 0.025$~\cite{Aoki:2004ht}.
Finally, with improved (untwisted) Wilson fermions, the
best result is $\widehat{B}_K = 0.69 \pm 0.18$~\cite{Flynn:2004au}.

Looking into the future, our result would be competitive
with those of the other $N_f=2+1$ calculations if we could
reduce our matching error down to the the level achieved
in these other calculations.
We are following two approaches to achieve this reduction:
using 2-loop perturbation theory 
(which would reduce our error estimate to 1.3\%, although this would
remain an educated guess)
and using NPR to determine the matching factors
(which we expect to lead to similar errors to those quoted
by the other calculations.)
First results with NPR using improved staggered fermions
and studying bilinear matching factors are encouraging~\cite{Andrew09}.

Other improvements that are underway are the use of a fourth lattice
spacing (the MILC ``ultrafine'' lattices with spacing $a\approx 0.045\;$fm),
SU(2) fitting including the finite-volume effects predicted by NLO SChPT,
simultaneous continuum and chiral fitting, 
the addition of ensembles with other values of the strange quark mass,
and an improvement in statistics for all ensembles.
We are also calculating matrix elements of operators 
with other Dirac structures,
since these are needed to constrain some theories of physics beyond the 
standard model.

\begin{acknowledgments}
C.~Jung is supported by the US DOE under contract DE-AC02-98CH10886.
The research of W.~Lee is supported by the Creative Research
Initiatives program (3348-20090015) of the NRF grant funded by the
Korean government (MEST).
The work of S.~Sharpe is supported in part by the US DOE grant
no. DE-FG02-96ER40956.
Computations for this work were carried out in part on QCDOC computers 
of the USQCD Collaboration at Brookhaven National Laboratory.
The USQCD Collaboration are funded by the Office of
Science of the U.S. Department of Energy.
\end{acknowledgments}

\appendix
\section{Functions appearing in SU(3) fitting}
\label{app:defs}

In this appendix we present explicit forms for
the functions that appear in the NLO SU(3) SChPT result for
$B_K$ with a mixed action.

\subsection{$\CM_{\textrm{conn}}^0$ and $\CM_{\textrm{disc}}^0$
  \label{app:def-M}}
These two functions are unaffected by the use of a mixed action,
and thus can be obtained from the general results
of Ref.~\cite{steve-06}.
Here we give fully explicit results for $2+1$ flavors.
First we have
\begin{eqnarray}
\CM_{\textrm{conn}}^0 & =& \sum_\textrm{B} \tau^\TB F^{(3)}_\TB
\label{eq:app:Mconndef}
\\
F^{(3)}_\TB &=& -\frac{1}{2}
\Big[(G+X_\TB)\ell(X_\TB)+(G+Y_\TB)\ell(Y_\TB)
\nonumber \\ & & 
+2(G-K_\TB)\ell(K_\TB) -2 G K_\TB\tilde\ell(K_\TB)\Big]
\,,
\end{eqnarray}
where the coefficients give the relative
weight of the different tastes in the loop
\begin{align}
\tau^\TI &= 1/16 \,, \qquad
\tau^\TP = 1/16 \,, \qquad
\tau^\TV = 4/16 \,, \nonumber \\
\tau^\TA &= 4/16 \,, \qquad
\tau^\TT = 6/16 \,,
\label{eq:app:tauB}
\end{align}
which are simply proportional to their multiplicity.
Two standard chiral logarithmic functions arise:
\begin{eqnarray}
\ell(X) &=& X \left[\log(X/\mu_{\rm DR}^2) +\delta^{\rm FV}_1(X) \right]\,,
\label{eq:app:l}
\\
\tilde\ell(X) &=& - \frac{d\ell(X)}{dX} 
\nonumber \\ &=&
-\log(X/\mu_{\rm DR}^2) -1 +\delta^{\rm FV}_3(X)\,,
\label{eq:app:tilde-l}
\end{eqnarray}
where $\mu_\textrm{DR}$ is the scale introduced
by dimensional regularization. The $\mu_\textrm{DR}$ dependence
of the chiral logarithms is canceled by a corresponding dependence
of the LECs. When we evaluate $\ell(X)$ and $\tilde\ell(X)$ numerically
we set $\mu_\textrm{DR}=0.77\;$GeV. 
Although we do not use them in this paper, we quote for
completeness the finite volume corrections to the chiral logarithms:
by
\begin{eqnarray}
\delta^{\rm FV}_1(M^2) &=& \frac4{ML} \sum_{\vec n\ne 0}
\frac{K_1(|\vec n| ML)}{|\vec n|}
\\
\delta^{\rm FV}_3(M^2) &=& 2 \sum_{\vec n\ne 0}
{K_0(|\vec n| ML)}\,,
\end{eqnarray}
with $L$ the box size, and $\vec n$ is a
vector labeling the image position.
These formulae assume that the time direction is much longer than the
spatial directions---but this is easily corrected by including images
in the time direction as well.
The image sums converge fairly quickly because the Bessel functions
die off exponentially.  
For further discussion of the properties of these functions
see Refs.~\cite{steve-92} and \cite{milc-04}.

The second function is given by
\begin{widetext}
\begin{eqnarray}
6 \CM_{\textrm{disc}}^0 &=& 
\ell(\eta_\TI) (G + \eta_\TI)
\frac{(X_\TI-Y_\TI)^2 (L_\TI-\eta_\TI)(S_\TI-\eta_\TI)}
     {(X_\TI-\eta_\TI)^2 (Y_\TI-\eta_\TI)^2}
\nonumber\\
&+&
\frac{(L_\TI-X_\TI)(S_\TI-X_\TI)}{(\eta_\TI-X_\TI)}
\left[ \tilde\ell(X_\TI)(G+X_\TI) - \ell(X_\TI) 
+ \ell(X_\TI) (G + X_\TI)
\left(\frac1{L_\TI-X_\TI}+\frac1{S_\TI-X_\TI}
- \frac1{\eta_\TI-X_\TI} - \frac2{Y_\TI-X_\TI} \right)\right]
\nonumber\\
&+&
\frac{(L_\TI-Y_\TI)(S_\TI-Y_\TI)}{(\eta_\TI-Y_\TI)}
\left[ \tilde\ell(Y_\TI)(G+Y_\TI) - \ell(Y_\TI) 
+ \ell(Y_\TI) (G + Y_\TI)
\left(\frac1{L_\TI-Y_\TI}+\frac1{S_\TI-Y_\TI}
- \frac1{\eta_\TI-Y_\TI} - \frac2{X_\TI-Y_\TI} \right)\right]\,.
\label{eq:app:Mdiscdef}
\end{eqnarray}
\end{widetext}
As is clear from this form, this is the taste singlet contribution
to the ``disconnected'' matrix element, i.e. that involving
hairpin vertices in the loop.
The poles in this function are misleading: due to cancellations
between terms, the function vanishes when
$X_I\to Y_I$ and is finite when $X_I$ or $Y_I$
equal $L_I$, $S_I$ or $\eta_I$. These cancellations could lead to 
numerical instability in the evaluation of $\CM_{\textrm{disc}}$,
but we have not found this to be the case.
Finally, we note that $\CM_{\rm disc}^0$ is independent of
the renormalization scale $\mu_{\rm DR}$.

\subsection{Definitions of $F^{(4)}_\TB$, $F^{(5)}$ and
$F^{(6)}$ \label{app:def-F4}}
These three functions describe the contributions to $\CM_{\rm conn}$
that result from discretization and truncation errors.
Their origin is explained in detail in Ref.~\cite{steve-06}.
Their forms are
\begin{eqnarray}
F^{(4)}_{\TB} &=& \frac{3}{8 f_\pi^2 G }
\left[2 G\ \tilde\ell(K_\TB) \right.
\nonumber \\ & & \hspace*{2.5pc} \left. 
+ \left(\ell(X_\TB)+\ell(Y_\TB)-2\ell(K_\TB)\right)
\right]
\label{eq:app:F4def}
\\
F^{(5)} &=& \frac{3}{8 f_\pi^2 G }
\left[ \left(\ell(X_\TV)+\ell(Y_\TV)-2\ell(K_\TV)\right) \right.
\nonumber \\ & & \hspace*{2.5pc} \left.
+\left(\ell(X_\TA)+\ell(Y_\TA)-2\ell(K_\TA)\right) \right]
\label{eq:app:F5def}
\\
F^{(6)} &=& \frac{3}{8 f_\pi^2 G }
\left[\ell(X_\TT)+\ell(Y_\TT)-2\ell(K_\TT)\right]
\label{eq:app:F6def}
\end{eqnarray}
Note that $F^{(5)}$ and $F^{(6)}$ vanish when $m_x=m_y$.

For completeness, we give the relationships between the coefficients
of these functions and those appearing in the operator enumeration
used in Ref.~\cite{steve-06}. The coefficients of the functions
$F^{(4)}_\TB$ are
\begin{equation}
b_j
= \sum_{\TB'} \frac1{64\pi^2 f^4}\left(C_\chi^{1\TB'} g^{\TB'\TB}
- C^{2\TB'}_\chi h^{\TB'\TB}\right)
\end{equation}
where $\TB'$ runs over all five tastes, and B=I, P, V, A and T,
for $j=6$, $7$, $8$, $9$ and $10$, respectively.
The matrices $g$ and $h$ are defined in Ref.~\cite{steve-06},
and $f$ is the chiral limit value of the pion decay constant.
The coefficient of $F^{(5)}$ is
\begin{equation}
b_{11} = - \frac{4}{\pi^2 f^4} \left( C_\chi^{2\TP}+ 6 C_\chi^{2\TT}\right)
\,,
\end{equation}
while that of $F^{(6)}$ is
\begin{equation}
b_{12} = -\frac{48}{\pi^2 f^4} C_\chi^{2\TT}\,.
\end{equation}

We can use these formulae to estimate the
order of magnitude of the coefficients $b_{6-12}$.
One expects that $C_\chi\sim a^2\Lambda_{\rm QCD}^8$ 
and/or $\sim \alpha^2\Lambda_{\rm QCD}^6$, while the
numerical factors vary from $g^{\TB'\TB}/(64\pi^2)\sim 1/10$
to $48/\pi^2\sim 5$. We simply treat them as of $O(1)$,
and $f\sim\Lambda_{\rm QCD}$, leading to 
the crude estimate given in Eq.~(\ref{eq:b6-12est}).

\subsection{Definitions of $F^{(1)}_\textrm{B}$ and $F^{(2)}_\TB$
\label{app:def-F12}}

These functions give the form of the contributions
to $\CM_{\rm disc}$ that arise from discretization and truncation
errors.
The explicit $2+1$ flavor form of $F^{(1)}_\textrm{B}$ can be deduced from
the general result given in Ref.~\cite{steve-06}:
\begin{widetext}
\begin{eqnarray}
F^{(1)}_\TB &=& \frac{3}{8 f_\pi^2 G }  \Bigg[
\ell(\eta_\TB) \frac{(Y_\TB-X_\TB)^2(L_\TB-\eta_\TB)(S_\TB-\eta_\TB)}
{(X_\TB-\eta_\TB)^2(Y_\TB-\eta_\TB)^2(\eta'_\TB-\eta_\TB)}
\nonumber\\
&&\qquad\qquad +\
\ell(\eta'_\TB) \frac{(Y_\TB-X_\TB)^2(L_\TB-\eta'_\TB)(S_\TB-\eta'_\TB)}
{(X_\TB-\eta'_\TB)^2(Y_\TB-\eta'_\TB)^2(\eta_\TB-\eta'_\TB)}
\nonumber\\
&&\qquad\qquad +\
\frac{(L_\TB-X_\TB)(S_\TB-X_\TB)}{(\eta_\TB-X_\TB)(\eta'_\TB-X_\TB)}
\left\{\tilde\ell(X_\TB) +
\ell(X_\TB) \left(\frac1{L_\TB-X_\TB}+\frac1{S_\TB-X_\TB}
\right.\right.
\nonumber\\
&&\qquad\qquad\qquad\qquad\qquad\qquad\qquad\qquad
\left.\left.- \frac{2}{Y_\TB-X_\TB} -\frac1{\eta_\TB-X_\TB} - \frac1{\eta'_\TB-X_\TB}
\right)\right\} 
\nonumber\\
&&\qquad\qquad+\
\frac{(L_\TB-Y_\TB)(S_\TB-Y_\TB)}{(\eta_\TB-Y_\TB)(\eta'_\TB-Y_\TB)}
\left\{\tilde\ell(Y_\TB) +
\ell(Y_\TB) \left(\frac1{L_\TB-Y_\TB}+\frac1{S_\TB-Y_\TB}
\right.\right.
\nonumber\\
&&\qquad\qquad\qquad\qquad\qquad\qquad\qquad\qquad
\left.\left.- \frac{2}{X_\TB-Y_\TB} -\frac1{\eta_\TB-Y_\TB} - \frac1{\eta'_\TB-Y_\TB}
\right)\right\} \Bigg]\,.
\label{eq:app:F1def}
\end{eqnarray}
\end{widetext}
The coefficients of this function are
\begin{equation}
b_j = \frac1{\pi^2 f^4} 
\left(2 C_\chi^{2\TB}+ C_\chi^{3\TB}\right) 
a^2 \delta_\TB^{\rm MA1}\,,
\label{eq:app:b13_14}
\end{equation}
with B$=$V and A for $j=13$ and $14$, respectively.

The function $F^{(2)}_\TB$ occurs only in the mixed-action set-up.
Its form is
\begin{eqnarray}
F^{(2)}_\TB &=& \frac{3}{8 f_\pi^2 G } \Bigg[ 
2\frac{\ell(Y_\TB)-\ell(X_\TB)}{Y_\TB-X_\TB}
\nonumber\\
&&\qquad\qquad +\ \tilde\ell(X_\TB)+\tilde\ell(Y_\TB) \Bigg]\,.
\label{eq:app:F2def}
\end{eqnarray}
with the coefficients being
\begin{equation}
b_j = \frac1{\pi^2 f^4} 
\left(2 C_\chi^{2\TB}+ C_\chi^{3\TB}\right) 
a^2 \delta_\TB^{\rm MA2}\,,
\label{eq:app:b15_16}
\end{equation}
with B$=$V and A for $j=15$ and $16$, respectively.

The functions $F^{(1,2)}_\TB$ are finite---the apparent poles
have vanishing residues---and vanish when $m_x=m_y$.
They are also independent of the renormalization scale.

The size of the coefficients $b_{13-16}$ can be estimated as
for $b_{6-12}$, with the result given in Eq.~(\ref{eq:b13-16est}).

\section{Functions appearing in SU(2) SChPT fitting
\label{app:su2result}}

In this appendix we describe how SU(2) SChPT results
are obtained from SU(3) SChPT results using the
recipe outlined in Sec.~\ref{ssec:su2} and justified in 
Appendix~\ref{app:su2}. The recipe is to first
expand the SU(3) result in powers of $m_x/m_s$ and $m_\ell/m_s$
(where $m_s$ is here meant generically, and thus includes $m_y$),
keeping only terms of the desired order (here NLO) in
SU(2) power-counting. Secondly, one allows the LECs which
are constants in SU(3) SChPT to become arbitrary functions of
$m_s$ and $m_y$, with the exception of $f$.

First consider the analytic terms $H_2-H_5$ of 
Eqs.~(\ref{eq:H2def})---(\ref{eq:H5def}). $H_2$ is proportional
to $(m_x+m_y)/\Lambda_{\rm QCD}$. The $m_x/\Lambda_{\rm QCD}$ part
remains in SU(2) ChPT, while the $m_y/\Lambda_{\rm QCD}$ part
is absorbed into the $m_s$ dependence of the overall constant
$B_0$. Similarly, $H_3 \propto (m_x+m_y)^2$ leads to an analytic
term proportional to $m_x^2$, to an $m_s$ dependence of the 
analytic term linear in $m_x$,
and to an $m_s^2$ dependence of $B_0$. 
$H_4\propto (m_x-m_y)^2/(m_x+m_y)$
leads to analytic terms of all orders in $m_x$, with the expansion
parameter being $m_x/m_y$ rather than
$m_x/\Lambda_{\rm QCD}$ . Finally,
$H_5 \propto (2m_\ell + m_s)$ leads to an analytic term linear in
$m_\ell$ and further $m_s$ dependence of $B_0$.
The net result is that given in 
Eqs.~(\ref{eq:su2-fit-func}-\ref{eq:q_4}),
with the coefficients $d_j$ having an arbitrary dependence
on the strange quark mass(es).

Next consider the continuum-like chiral logarithms in $H_1$
[see Eq.~(\ref{eq:H1def})]. Here a key role is played by
the overall multiplier $1/G=1/m_{xy:\TP}^2$, which arises
from the denominator in the definition of $B_K$.
In order for a term from $\CM_{\rm conn}^0$ or $\CM_{\rm disc}^0$
to remain of NLO in the SU(2) limit, it must come with
a compensating factor of $G$, $m_y$ or $m_s$. If not,
it will become of NNLO in SU(2) SChPT.

To see how this works in more detail,
consider first $\CM_{\rm conn}$, given in Eq.~(\ref{eq:app:Mconndef}).
Its contribution to $H_1$ consists of a sum over tastes $B$
of the expression
\begin{eqnarray}
\lefteqn{\frac{F_B^{(3)}}{8 \pi^2 f_\pi^2 G}
=
-\frac12\frac{1}{8\pi^2 f_\pi^2 }
\left[
\left(1+\frac{X_B}{G}\right)\ell(X_B) 
+ \right.}
\\
&&\left.\!\!\!\!\!\!\!
\left(1\!+\!\frac{Y_B}{G}\right)\ell(Y_B)
+2\left(1\!-\!\frac{K_B}{G}\right) \ell(K_B)
-2 K_B \tilde\ell(K_B) \right]. \nonumber
\end{eqnarray}
On the first line, the term proportional to
$\ell(X_B)$ survives as a chiral log in the SU(2) limit,
but its coefficient simplifies to $1$. This is because
$X_B/G \propto m_x/m_s$ (where ``$m_x$'' can include
taste-breaking $O(a^2)$ corrections) is suppressed in the
SU(2) limit,
so that, given the overall factor of $1/(8\pi^2 f_\pi^2)$,
the $X_B/G$ term is of NNLO in SU(2) power counting.
On the second line, no terms give rise to SU(2) chiral logarithms,
because the argument of the logarithms is finite when either $m_x$
or $m_\ell$ vanishes. Thus one can expand the logarithms about
$m_x=0$, leading to a power series in $m_x$, in which subsequent
terms are suppressed by $m_x/m_y$ or $m_x/\Lambda\propto X_P/\Lambda_\chi^2$,
with $\Lambda=\Lambda_{\rm QCD}$.
The same is true for the factors multiplying the logarithms.
For example, the last term on the second line becomes
\begin{eqnarray}
\lefteqn{\frac{K_B \tilde\ell(K_B)}{8\pi^2 f_\pi^2} \propto 
\frac{K_B \log[K_B]}{\Lambda_\chi^2} }
\\
&\propto& 
\frac{(m_x\!+\!m_y\! +\!a^2 \Lambda^3)}{\Lambda} 
\log\left[m_x\!+\!m_y\! +\!a^2\Lambda^2\right]
\\
&\propto& f_1(m_y/\Lambda)  
+ f_2(m_y/\Lambda) \frac{m_x}{\Lambda} + \dots
\,.
\end{eqnarray}
The analytic dependence on $m_y$ in $f_1$ can then be 
absorbed into the (unknown) $m_y$ dependence of
the overall coefficient multiplying $H_1$
[which becomes $d_1$ in the SU(2) SChPT result, Eq.~(\ref{eq:su2-fit-func})].
The dependence on $m_x$, which is of NLO, 
can be absorbed into the coefficient of the analytic term $Q_3$.
In this way, one finds that the entire second line
can be absorbed into the coefficients $d_1$ and $d_2$.
The overall result is that
\begin{equation}
\frac{\CM_{\rm conn}^0}{G} \stackrel{SU(2)}{\longrightarrow}
-\frac12 \sum_\TB \tau^\TB \ell(X_\TB)\,.
\end{equation}

Turning next to the contribution of
$\CM_{\rm disc}^0$, Eq.~(\ref{eq:app:Mdiscdef}), we find
two new features.
First, there are logarithms (and other functions) of $\eta_I$.
But since $\eta_I=L_I/3+ 3 S_I/3$ does not vanish when
$m_\ell$ and $a^2$ vanish, these functions can be expanded just
as logarithms of $Y_B$ and $K_B$ were above.
Second, we need to use the result\footnote{%
One might be concerned about the legitimacy of the SU(2) limit, since
ChPT does not control the $m_s$ dependence when $m_s$ becomes large.
If one looks back at the source of this ratio in the derivation, however,
one finds that taking the $m_s\to\infty$ limit in this way amounts to
projecting onto the generators of the appropriate partially quenched
SU(2) subgroup. This is exactly analogous to the way in which one
projects against the ``super-$\eta'$'' by sending $m_0^2\to\infty$ 
in a PQ calculation~\cite{ShShII}.
In other words, it is a mathematical trick and does not imply control
over the actual $m_s$ dependence.
}
\begin{equation}
\frac{(S_I-X_I)}{(\eta_I-X_I)}
\ \stackrel{SU(2)}{\longrightarrow} \ \frac32 \,.
\end{equation}
Third, one can have contributions proportional to 
$(m_\ell/\Lambda)f(m_s/\Lambda)$
(e.g. from the $L_I \widetilde\ell(Y_I)G$ term)
which can be absorbed into the coefficient $d_4$.

Given these results, combined with the arguments used above for
$\CM_{\rm conn}^0$, we find that
the only terms in $\CM_{\rm disc}^0$ which cannot be absorbed
into the coefficients $d_1$, $d_2$ and $d_4$, are those
multiplying $\ln X_I$.
Of these, we must pick out
those multiplied by $G$, since all others are suppressed by $m_x/m_s$
and thus become of NNLO.
The result is then the greatly simplified form
\begin{eqnarray}
\frac{6\CM_{\rm disc}^0}{G} &\ \stackrel{SU(2)}{\longrightarrow}\ &
\frac32 \times \Bigg[ 
(L_I-X_I) \tilde\ell(X_I) + \ell(X_I) 
\nonumber\\
&&\qquad\qquad + {\rm analytic} + {\rm NNLO}
\Bigg]\,.
\end{eqnarray}

All the remaining functions which arise in SU(3) SChPT,
$H_6$-$H_{16}$, become of NNLO in the SU(2) limit.
This is because all the functions which appear, $F^{(2)}-F^{(6)}$, contain
a factor of $1/G$ from the definition of $B_K$, but no
counterbalancing factor in the numerator {\em when the
numerator contains a logarithm of $X_\TB$}. 
For example, in $F^{(4)}_\TB$ [see Eq.~(\ref{eq:app:F4def})]
part of the numerator contains a factor of $G$, but this multiplies
$\ln K_\TB$, which gives only terms analytic in $m_x$. The part containing
$\ln X_\TB$ is multiplied by $X_\TB$, leading to the additional
suppression by $X_\TB/G\sim m_x/m_s$.

The final result, given in Eq.~(\ref{eq:su2-fit-func}),
therefore contains no LECs
induced by discretization or truncation errors.

\section{Derivation of recipe for obtaining the SU(2) SChPT result
\label{app:su2}}

In this Appendix we demonstrate the validity of the recipe used to obtain
the SU(2) SChPT result. This recipe has been described in
Sec.~\ref{ssec:su2} and implemented in Appendix~\ref{app:su2result}.
The essential claim is that it is sufficient to take the SU(2) limit
of the NLO SU(3) expression, as long as one allows LECs (except for $f$)
to have an arbitrary dependence on $m_s$ and $m_y$.
This arbitrary dependence is the only impact of working to all orders
in $r_s=m_s/\Lambda_{\rm QCD}$.\footnote{%
Here and throughout this Appendix we use $m_s$
to represent both $m_s$ and $m_y$.}
In other words, higher orders in SU(3) SChPT do not lead to new
types of functional dependence on the small quantities
$X_P/\Lambda_\chi^2$, $L_P/\Lambda_\chi^2$, $a^2\Lambda_{\rm QCD}^2$
and $\alpha^2$.

In particular, looking at the final result of
Eqs.~(\ref{eq:su2-fit-func}-\ref{eq:q_4}),
there are two main features in need of justification:
\begin{enumerate}
\item
There are no terms involving those LECs that are proportional to
$a^2$ or $\alpha^2$.
\item
The chiral logarithm in the function $Q_1$ 
[Eq.~(\ref{eq:q_1})] is fully predicted. In particular,
its coefficient ($\propto 1/f^2$) has no dependence on $m_s$.
\end{enumerate}
Discretization and truncation errors thus enter only through the
coefficient $d_1$ [see Eq.~(\ref{eq:d_1})] and through the masses
of the PGBs in the chiral logarithms.

The approach we use to justify the recipe is to use SU(3) ChPT,
working to all orders in $r_s$,
but only at NLO in light-quark masses, discretization and truncation effects.
This amounts to a ``poor-person's SU(2) SChPT'', and allows us
to piggy-back on the SU(3) SChPT analysis of Ref.~\cite{steve-06}.
We assume that this all-orders perturbative analysis captures all
possible forms of dependence on the small quantities.

The form of the analytic terms in the SU(2) expression,
Eqs.~(\ref{eq:q_2})-{\ref{eq:q_4}), 
follows immediately in this approach.
Indeed, since the SU(2) limit of the NLO SU(3) expression already gives 
the most general NLO analytic dependence on 
$X_P/\Lambda^2$, $L_P/\Lambda^2$, $a^2\Lambda_{\rm QCD}^2$ and $\alpha^2$,
higher order terms can only generate $r_s$ dependence of their coefficients.

We now turn to the first feature noted above.
We start by displaying the form of the contributions proportional to the
discretization/truncation LECs that arise in
the SU(2) limit of the NLO SU(3) expression.
Keeping only the parts which are non-analytic in light-quark
quantities, one finds that they all have the form
\begin{equation}
\delta B_K\big|_{\rm disc/trunc}^{SU(3) @ {\rm NLO}} \sim
\alpha^2 \frac{M_\pi^2 \log(M_\pi^2)}{G}
\,.
\label{eq:discerrorform}
\end{equation}
We use here a compact notation:
$\alpha^2$ stands for discretization and truncation
effects proportional to $\alpha^2$, $\alpha/\pi$ or
$a^2\Lambda_{\rm QCD}^2$;
$M_\pi^2$ stands for any light-meson mass-squared
[$X_\TB$, $\eta_\TB$ (except $\eta_\TI$) and $L_\TB$]
and also (except in the argument of the logarithm) for
the hairpin vertices $a^2\delta_\TB^{\rm MA1,2}$. 
In addition, the scale $\mu_{\rm DR}$ in the logarithm
is kept implicit.
The form of (\ref{eq:discerrorform}) arises as follows:
the $\alpha^2$ comes from the LECs, the $M_\pi^2\log(M_\pi^2)$
from the pion loop, and the $1/G$ from the denominator in the
definition of $B_K$. 
All other factors of $\Lambda_{\rm QCD}$ cancel,
as required by dimensional analysis.
As noted in Appendix~\ref{app:su2result}, the contribution of
Eq.~(\ref{eq:discerrorform}) is of NNLO in SU(2) power-counting,
because it is suppressed both by $\alpha^2$ and by $M_\pi^2/G$.

The issue then is whether this double suppression
continues to hold when one takes the
SU(2) limit of higher order SU(3) SChPT expressions. 
If (\ref{eq:discerrorform}) simply gets multiplied by a
function of $r_s$, then it remains of NNLO in
SU(2) ChPT.
If, however, some higher order terms reduce in the SU(2) limit to 
\begin{equation}
\delta B_K\big|_{\rm unwanted} \sim
r_s F(r_s) \alpha^2 \log(M_\pi^2)
\,,
\label{eq:unwanted}
\end{equation}
then they would contribute at NLO.
Here $F(r_s)$ is an unknown function, which we assume to
be of $O(1)$ in SU(2) power-counting for $r_s\sim 1$.
The overall factor of $r_s$ is included because we know 
from Eq.~(\ref{eq:discerrorform}) that there is no such term when $r_s=0$.

That (\ref{eq:unwanted}) cannot appear follows from two observations.
First, higher orders in SU(3) ChPT cannot introduce an
enhancement by $G/M_\pi^2\propto m_s/m_\ell$.
Second, chiral logarithms always have the form $M_\pi^2\ln(M_\pi^2)$,
rather than simply $\ln(M_\pi^2)$.
Since the factor of $\alpha^2$ must be
present as we are considering LECs related to discretization or
truncation errors, and the $1/G$ arises from the 
definition of $B_K$,\footnote{%
The $1/G$ can be converted into a $1/\Lambda_\chi^2$ by 
a higher-order correction proportional to $r_s\sim G/\Lambda_\chi^2$.
This makes no difference to our arguments since we treat
$G\sim\Lambda_\chi^2$ in the SU(2) limit.}
one necessarily obtains terms of the form (\ref{eq:discerrorform})
rather than (\ref{eq:unwanted}).

The two observations of the previous paragraph follow from
ChPT power counting, specifically the generalization of
staggered ChPT power counting applicable to a calculation of
$B_K$~\cite{steve-06}. This power-counting ensures that
successive orders in SU(3) SChPT are ``suppressed'' by factors
of $m/\Lambda_{\rm QCD}$ or $\alpha^2$ (using the latter generically).
Thus it is not possible that by going to higher order one can enhance
the result by
a factor of $m_s/m_\ell$---the best that one can do is have
a factor of $r_s\sim O(1)$, which does not change the order in
SU(2) ChPT.

As for the pion loop contribution, this can appear as part of
an arbitrary-order SU(3) diagram in which all the other
particles are ``heavy'' (i.e. with mass-squared proportional to
$m_s$). In the SU(2) limit this diagram collapses to give a pion
tadpole. The contribution of the part of the diagram without the
tadpole can be no larger than
$\delta B_K\sim \alpha^2 F(r_s)$, where the $\alpha^2$ comes from the LEC,
and SU(3) power-counting ensures that $F(r_s)\sim O(1)$.
Adding back in the pion loop, and noting that
each pion field comes explicitly with a factor of $1/f$,
one sees that the loop must give the generic form 
\[
\frac{M_\pi^2\log(M_\pi^2)}{(4\pi f)^2} 
=
\frac{M_\pi^2\log(M_\pi^2)}{\Lambda_\chi^2} 
\sim
\frac{M_\pi^2\log(M_\pi^2)}{G} 
\,,
\]
where in the last step we are using the fact that $G\sim\Lambda_\chi$
in SU(2) power counting.
In more detail, the loop can involve either a flavor non-singlet
pion propagator, in which case $M_\pi^2=X_\TB$, or the flavor singlet
propagator [e.g. of the form in Eq.~(\ref{eq:D_MA}) for $\TB=V,A$], 
in which case the $M_\pi^2$ inside the logarithm can be 
$X_\TB$ or $\eta_\TB$,
while the $M_\pi^2$ outside can also be a hairpin vertex
proportional to $a^2$.
In all cases one ends up with an expression of the NNLO form
(\ref{eq:discerrorform}).

\bigskip
We now turn to the second feature of the result Eq.~(\ref{eq:su2-fit-func})
that needs to be explained, namely that the SU(2) chiral logarithm
has a predicted coefficient, despite working to all orders in $r_s$.
In the continuum, this follows from an SU(2) ChPT analysis in which
the kaon is treated as heavy and serves, along with the four-quark
operator, as a source for pions~\cite{rbc-uk-08-1}.
Using the equations of motion, it is shown in Ref.~\cite{rbc-uk-08-1} that
all operators contributing to $B_K$ at one-loop 
order in SU(2) ChPT (which can have any even number of derivatives acting
on the external kaons) collapse to a {\em single} form.
Since the coupling to pions is determined by SU(2)  chiral symmetry,
both the form of the chiral logarithm and its coefficient
relative to LO are predicted. The result found in this way
is that obtained from Eq.~(\ref{eq:su2-fit-func})
by setting taste-breaking in the pion masses to zero.

Our aim in the following is to show that 
continuum argument can be generalized
to staggered ChPT with only minor changes.
Since we have already dealt with the contribution from LECs
arising from discretization and truncation errors in the operator, 
the remaining
such errors are those coming from $O(a^2)$ taste-breaking effects
arising from the action. 
These enter through the taste-breaking
potential, $\CV$, whose explicit form is given in 
Refs.~\cite{wlee-99,bernard-03}.
Thus we must consider SU(3) SChPT diagrams with a single insertion of 
$\CV$ and any number of powers of $r_s$.

The potential $\CV$ leads to two-point, four-point and higher-order
vertices, all multiplied by an explicit factor of $a^2$.
Those involving four-point or higher-order vertices give 
contributions to $B_K$ of the form (\ref{eq:discerrorform}),
which are thus of NNLO in the SU(2) limit. 
The argument for this is exactly as given above for
the discretization/truncation LECs.
Thus we need only account for the two-point $O(a^2)$ vertices,
which only affect propagators. These will only have an impact
on the SU(2) chiral logarithm if these vertices are included
in the ``light'' meson propagators. Including them on ``heavy''
meson propagators will lead only to corrections of the NNLO
form (\ref{eq:discerrorform}).
The upshot is that we should work to all orders in $r_s$ in
{\em continuum} SU(3) ChPT, keeping $O(a^2)$ corrections
only on the single light meson propagator needed to develop the
non-analyticity.
But working to all orders in $r_s$ in SU(3) ChPT can be achieved
by using {\em continuum} SU(2) ChPT---indeed, the latter goes beyond
perturbation theory.
Thus we end up with the following recipe: use the continuum SU(2) ChPT
calculation of Ref.~\cite{rbc-uk-08-1}
to determine the Feynman diagrams and their
overall factors, but evaluate the integrals using
the light-meson propagators including the $O(a^2)$ corrections.

In fact, we must generalize the calculation of Ref.~\cite{rbc-uk-08-1}
to include the extra taste degree of freedom. This is straightforward,
and follows the methodology used in Ref.~\cite{steve-06}.
We find that, just as in SU(3) ChPT, 
adding taste results in the LO chiral operator having two terms:
\begin{equation}
{\cal O}_K^\chi \propto
{\rm Str}\left(F_1 \xi K\right) {\rm Str}\left(F_2 \xi K\right)
+
{\rm Str}\left(F_1 \xi K F_2 \xi K\right)
\,.
\label{eq:OKSU2}
\end{equation}
Here the pion fields are contained
in $\xi=\sqrt\Sigma=\exp(i \Phi/[2f])$,
while $F_{1,2}$ are spurions which pick
out the flavor and taste of the underlying quark operator.
The argument for the equality of the coefficients of the
two terms is identical to that for SU(3) case given in Ref.~\cite{steve-06}.
Expanding the operator (\ref{eq:OKSU2}) to quadratic order in pion fields,
removing contributions which are common to the factor of
$f_K^2$ in the denominator of $B_K$ and thus cancel, 
and contracting pion fields with staggered ChPT\ propagators leads
to Eq.~(\ref{eq:q_1}).

\section{Tables of $a m_\pi$ and $B_K$}

In this appendix we present the most useful subset
of our results for the ``pion'' masses and $B_K$.
We give results only for the three ensembles, C3, F1 and S1, which
we use to do our continuum extrapolation.
We also quote only the degenerate pion masses, since
$m_\pi^2\propto m_x+m_y$ to very good approximation.
Finally, for $B_K$ we show the subset of our data used
in the SU(2) fits. These are the results that lead our
central value for $B_K$. 
Table~\ref{tab:pion-mass} reports
the average pion masses obtained as described in the main text.
Table~\ref{tab:bk-C3-F1-S1}
reports the values of $B_K$ obtained using
1-loop matching at the scale $\mu=1/a$.

\begin{table}[htb]
\caption{ $a m_\pi$ (Goldstone taste, $\xi_5$)
for the C3, F1 and S1 ensembles.
  \label{tab:pion-mass}}
\begin{ruledtabular}
\begin{tabular}{c  l  l  l }
quark comb. & C3 & F1 & S1 \\
\hline
1-1   & 0.1342(2) & 0.0902(3) & 0.0596(2) \\
2-2   & 0.1861(2) & 0.1244(2) & 0.0819(2) \\
3-3   & 0.2258(2) & 0.1507(2) & 0.0992(1) \\
4-4   & 0.2593(2) & 0.1728(2) & 0.1139(1) \\
5-5   & 0.2888(2) & 0.1924(2) & 0.1269(1) \\
6-6   & 0.3155(2) & 0.2101(2) & 0.1387(1) \\
7-7   & 0.3402(2) & 0.2266(2) & 0.1496(1) \\
8-8   & 0.3634(2) & 0.2419(2) & 0.1599(1) \\
9-9   & 0.3852(2) & 0.2565(2) & 0.1695(1) \\
10-10 & 0.4060(2) & 0.2703(2) & 0.1788(1) \\
\end{tabular}
\end{ruledtabular}
\end{table}
%

\begin{table}[htb]
\caption{Results for $B_K(1/a)$ on the C3, F1, and S1 ensembles.
  \label{tab:bk-C3-F1-S1}}
\begin{ruledtabular}
\begin{tabular}{c  l  l  l  }
quark comb. & C3 & F1 & S1 \\
\hline
1-1   & 0.438(12)  & 0.360(33)  & 0.357(23) \\
2-2   & 0.4815(46) & 0.434(14)  & 0.4147(73) \\
3-3   & 0.5132(27) & 0.4699(77) & 0.4455(41) \\
4-4   & 0.5363(19) & 0.4928(53) & 0.4689(27) \\
5-5   & 0.5547(15) & 0.5118(40) & 0.4878(21) \\
6-6   & 0.5701(12) & 0.5283(33) & 0.5036(17) \\
7-7   & 0.5835(10) & 0.5426(27) & 0.5173(14) \\
8-8   & 0.5954(9)  & 0.5553(23) & 0.5294(13) \\
9-9   & 0.6061(8)  & 0.5667(20) & 0.5403(12) \\
10-10 & 0.6159(7)  & 0.5771(18) & 0.5502(11) \\
\hline
1-8   & 0.5735(38) & 0.5243(97) & 0.4908(58) \\
2-8   & 0.5674(21) & 0.5215(55) & 0.4915(30) \\
3-8   & 0.5690(16) & 0.5248(42) & 0.4978(22) \\
4-8   & 0.5732(13) & 0.5303(35) & 0.5045(18) \\
5-8   & 0.5784(11) & 0.5365(31) & 0.5111(16) \\
\hline
1-9   & 0.5811(38) & 0.5329(95) & 0.4992(57) \\
2-9   & 0.5759(20) & 0.5304(53) & 0.5000(29) \\
3-9   & 0.5772(15) & 0.5332(40) & 0.5058(21) \\
4-9   & 0.5809(12) & 0.5382(33) & 0.5119(17) \\
5-9   & 0.5855(11) & 0.5439(29) & 0.5180(15) \\
\hline
1-10  & 0.5879(38) & 0.5410(93) & 0.5070(56) \\
2-10  & 0.5835(20) & 0.5386(51) & 0.5078(29) \\
3-10  & 0.5848(14) & 0.5410(38) & 0.5132(20) \\
4-10  & 0.5881(12) & 0.5455(31) & 0.5188(17) \\
5-10  & 0.5922(10) & 0.5507(27) & 0.5245(15) \\
\end{tabular}
\end{ruledtabular}
\end{table}
%

\bibliographystyle{apsrev} 
\bibliography{ref} 

\end{document}